\documentclass{IEEEtran}
\usepackage{cite,graphicx,url,amssymb,amsthm,epstopdf,threeparttable,multirow,algorithm,algorithmic}
\usepackage[tight,footnotesize]{subfigure}
\usepackage[usenames]{color}
\usepackage[normalem]{ulem}
\usepackage[cmex10]{amsmath}
\IEEEoverridecommandlockouts

\usepackage{amsfonts,relsize,bm,amsbsy,stmaryrd,verbatim,booktabs,pifont}
\usepackage[colorlinks=true, allcolors=blue]{hyperref}
\usepackage{balance}
\DeclareMathOperator*{\argmin}{arg\,min}
\DeclareMathOperator*{\argmax}{arg\,max}

\theoremstyle{plain}

\newtheorem{theorem}{Theorem}
\newtheorem{corollary}{Corollary}

\theoremstyle{definition}

\theoremstyle{remark}
\newtheorem{remark}{Remark}

\newcommand{\revise}[1]{#1}

\begin{document}

\title{Detection Theory for Union of Subspaces}

\author{Muhammad Asad Lodhi,~\textit{Graduate Student Member, IEEE,} and Waheed U. Bajwa,~\textit{Senior Member, IEEE}
\thanks{Some of the results presented in this paper have appeared in Proceedings of 2018 IEEE Workshop on Statistical Signal
Processing (SSP'18)~\cite{LodhiBajwa.ConfSSP18}. MAL and WUB are with the Department of Electrical and Computer Engineering, Rutgers, The State University of New Jersey, Piscataway, NJ 08854 (Emails: {\tt \{masad.lodhi,~waheed.bajwa\}@rutgers.edu}).}
\thanks{This research is supported in part by the NSF under award CCF-1453073 and by the ARO under award W911NF-14-1-0295.}
}


\maketitle

\begin{abstract}
The focus of this paper is on detection theory for union of subspaces (UoS). To this end, generalized likelihood ratio tests (GLRTs) are presented for detection of signals conforming to the UoS model and detection of the corresponding ``active" subspace. One of the main contributions of this paper is bounds on the performances of these GLRTs in terms of geometry of subspaces under various assumptions on the observation noise. The insights obtained through geometrical interpretation of the GLRTs are also validated through extensive numerical experiments on both synthetic and real-world data.
\end{abstract}



\section{Introduction}
Detection theory has a long history in the signal processing literature. Classical detection theory is often based on the \emph{subspace model}, in which the signal to be detected is assumed to come from a low-dimensional subspace embedded in a high-dimensional ambient space \cite{louis1991statistical,Kay.Book1998}. However, recently a nonlinear generalization of the subspace model, termed the \emph{union of subspaces} (UoS) model \cite{lu2008theory,WuBajwa.ConfICASSP14,WuBajwa.ConfICASSP15,WuBajwa.ITSP15}, has gained attention in the literature due to its ability to better model real-world signals. \revise{Indeed, data in many real-world scenarios tend to be generated by processes that switch/operate in different modes. In such instances, data generated through each mode of the process can be modeled as lying on a subspace, in which case the entire data generated during the process as a whole can be best described as coming from a union of subspaces~\cite{ho2003clustering,hong2006multiscale,yang2008unsupervised,Elhamifar.Vidal.ITPAMI2013,bajwa2014multiple,gini2004radar,wu2015hierarchical}. Some specific instances of such processes include: ($i$) radar target detection involving multiple targets, with only one target being present at a time and each target being characterized by its own specific spectral signature; ($ii$) user detection in a wireless network, with only one user transmitting at a time and each user having its own transmit codebook; and ($iii$) image-based verification of employees in an organization, with the verification system using a database of employees' facial images collected under varying lighting conditions.

Broadly speaking, and under the assumption of processes following the UoS model, we focus on the following questions in this work: ($i$) whether an observed signal (e.g., spectral data, radio frequency (RF) observations, or an image) corresponds to a known generation mechanism (e.g., spectral signatures of known targets, RF transmissions from known users, or faces of known employees); and ($ii$) which mode (e.g., which known target, which known user, or which existing employee) from the known generation mechanism gave rise to the observed signal.} In this context, we revisit in this paper the problem of detection of signals under various additive noise models for the case when the signal conforms to the UoS model. Our goals in this regard are: ($i$) derivation of tests for detection of both the signal and the underlying active subspace \revise{(mode)}, and ($ii$) characterization of the performance of these tests in terms of geometry of the subspaces.

\begin{table*}[!t]
	\centering
	\caption{A brief comparison of this work with related prior works in the literature}
	\label{table}
	\revise{
	\begin{tabular}{|c|c|c|c|c|c|c|}
		\hline
		\textit{\textbf{Work}} & \textit{\textbf{Framework}} & \textit{\textbf{Gaussian Noise Model}}                                                                                                                                     & \textit{\textbf{\begin{tabular}[c]{@{}c@{}}Signal\\ Detection\end{tabular}}} & \textit{\textbf{\begin{tabular}[c]{@{}c@{}}Active Subspace\\ Detection\end{tabular}}} & \textit{\textbf{\begin{tabular}[c]{@{}c@{}}Impact of \\ Geometry\end{tabular}}} \\ \hline
		\textbf{\cite{davenport2010signal}, \cite{yap2014false},\cite{joneidi2016union}}     & \begin{tabular}[c]{@{}c@{}}compressive sensing\end{tabular}                 & white, w/ known variance & \ding{51}                                                                          & \ding{55}                                                                                    & \ding{55}                                                                                 \\ \hline
		\textbf{\cite{gini2004radar}}          & general UoS                                                                       & colored, w/ known cov.~and unknown var.                                                                                                            & \ding{51}                                                                          & \ding{51}                                                                                   & \ding{55}                                                                                 \\ \hline
		\textbf{\cite{wimalajeewa2015subspace}}          & linear sampling of UoS                                                                       & white, w/ known var.                                                                                                            & \ding{55}                                                                          & \ding{51}                                                                                   & \ding{55}                                                                                 \\ \hline
		\textbf{This work}     & general UoS                                                                       & \begin{tabular}[c]{@{}c@{}}colored, w/ known statistics\\ colored, w/ partially unknown statistics\\ colored, w/ completely unknown statistics \end{tabular} & \ding{51}                                                                          & \ding{51}                                                                                   & \ding{51}                                                                                \\ \hline
	\end{tabular}}
\end{table*}

\subsection{Prior work}
There exists a rich body of literature concerning detection of signals under the subspace model; see, e.g., \cite{scharf1994matched,kraut2001adaptive,kraut1999cfar,kelly1986adaptive}. The most well-studied method in this regard is the matched subspace detector~\cite{scharf1994matched}, which projects the received signal onto the subspace of interest and compares its energy against a threshold. A na\"{i}ve approach to detection under the UoS model would be to treat it as a subspace detection problem by replacing the union with direct sum and using the resulting subspace within the matched subspace detector. However, such an approach not only results in high false alarm rates (for obvious reasons), but it also does not enable detection of the active subspace. A better alternative is to treat the detection problem as a multiple hypothesis testing problem, as in \cite{bajwa2014multiple}, with each test given by an individual matched subspace detector. We establish in this paper that such an approach will have the same performance as a GLRT for the case of a single active subspace.

Recently, there have been a few works that are directly related to the detection problem under the UoS model~\cite{davenport2010signal,yap2014false,joneidi2016union,gini2004radar,wimalajeewa2015subspace}. \revise{One of the biggest differences between these (and related) works and this paper is that the existing works cannot explain the variability of detection performance under the UoS model for different problems with same problem parameters (e.g., number and dimension of subspaces, and signal-to-noise ratio); see, e.g., Fig.~\ref{fig:num_sims:geometry} and the accompanying discussion. In contrast, we have been able to establish in this paper that such variability is a quantifiable function of the geometry (expressed in terms of principal angles) of individual subspaces in the union and the geometry of the noise.}

\revise{In terms of explicit comparisons with individual works related to this paper}, \cite{davenport2010signal} studies the problem of signal detection under the compressive sensing framework~\cite{johnson1984extensions}, with the final results involving analysis of a GLRT for a binary hypothesis test. These \emph{compressive detection} results can be considered a special instance of those for detection under the UoS model, since a sparse signal can be thought of as lying in a union of (exponentially many) subspaces~\cite{lu2008theory}. The nature of these results, however, does not enable understanding of the general detection problem under the UoS model, especially in relation to geometry of the underlying subspaces. \revise{First, individual subspaces do not explicitly appear in compressive detection; rather, the results are presented in terms of the so-called ``measurement matrix,'' which obfuscates the role of individual subspaces in detection performance. Second, the most useful of compressive detection results involve the use of \emph{random} measurement matrices; translated into the UoS model, this corresponds to randomly generated subspaces. Since random subspaces tend to be equiangular (with high probability), compressive detection literature does not lend itself to understanding the role of subspace geometry in signal detection.} Similar to \cite{davenport2010signal}, \cite{yap2014false} also studies the compressive detection problem, but in the context of radar-based multi-target detection. While the analysis in~\cite{yap2014false} is based on the use of the LASSO~\cite{tibshirani1996regression} for detection, it too does not offer geometric insights into the general UoS-based detection problem. In \cite{joneidi2016union}, the authors extend the original compressive detection framework of \cite{davenport2010signal} to more general settings, but the final results are still couched in terms of the sparsity framework and they fail to bring out the geometric interplay between the different subspaces.

The work that is most closely related to this paper is~\cite{gini2004radar}, in which the authors study the signal and the active subspace detection problems under the UoS framework in the context of radar target detection. The (signal and active subspace) detection schemes proposed in~\cite{gini2004radar} are based on multiple hypothesis testing. The analysis in~\cite{gini2004radar} is for the case of colored Gaussian noise with unknown variance but \emph{known} covariance matrix. Further, since the analysis is in terms of the spectral signatures of targets, it does not help understand the interplay between the detection performance and the geometry of subspaces. Finally,~\cite{gini2004radar} does not investigate invariance properties of the derived test statistics.

Recently, \cite{wimalajeewa2015subspace} has studied both recovery of a signal conforming to the UoS model and detection of the corresponding active subspace in the presence of a linear sampling operator. This work, however, is fundamentally focused on understanding the role of the sampling operator within the active subspace detection problem. Further, it assumes white Gaussian noise with known variance, does not investigate the related problem of signal detection, and does not focus on the geometry of subspaces as an integral component of the detection problem. 

\subsection{Our contributions}
\revise{The major contributions of this paper include derivation, analysis, and understanding of various GLRTs for the signal and the active subspace detection problems under the UoS model for different noise settings. One of our main contributions} in this regard is a comprehensive understanding of the two detection problems in terms of characterization of the performance of the derived GLRTs through the probabilities of detection, classification, and false alarm, geometry of the underlying subspaces, and invariance properties of the test statistics. One of the key insights of this work is that the probability of correct identification of the active subspace increases with increasing principal angles between subspaces in the union. While this makes intuitive sense, our analysis provides theoretical justification for such an assertion. Further, our work also helps understand the relationship between a binary and a multiple hypothesis testing approach to the signal detection problem under the UoS model. Finally, we provide extensive numerical experiments to highlight the usefulness of our analysis and its superiority to prior works such as~\cite{wimalajeewa2015subspace}. We refer the reader to Table~\ref{table} for a brief comparison of our work with existing literature.

\subsection{Notation and Organization}
We use bold lowercase and bold uppercase letters to denote vectors and matrices, respectively. Given a matrix $\mathbf{A}$, $\mathbf{a}_j$ and ${a}_{ij}$ denote its $j$-th column and $(i,j)$-th entry, respectively. Further, $\mathbf{A}^{-1}$ and $|\mathbf{A}|$ denotes its inverse (if it exists) and its determinant, respectively. Given a vector $\mathbf{a}$, $\|\mathbf{a}\|_p$ denotes its $\ell_p$-norm and $|\mathbf{a}|$ denotes its elementwise absolute values. Finally, $Q(\cdot)$, $\Gamma(\cdot)$, and $K_n(\cdot)$ denote the Gaussian $Q$ function, the Gamma function, and the modified Bessel function of the second kind with parameter $n$, respectively.

The rest of the paper is organized as follows. In Sec.~\ref{sec:prob_form}, we formulate the signal and the active subspace detection problems under the UoS model. Sec.~\ref{SD} \revise{derives} and analyzes the GLRTs for these two problems under different noise conditions. Sec.~\ref{sec:disc} provides a discussion of the results obtained in Sec.~\ref{SD}. Sec.~\ref{num_sims} presents the results of numerical experiments on both synthetic and real-world data, while we conclude the paper in Sec.~\ref{summary}. 

\section{Problem Formulation}  \label{sec:prob_form}
We study two interrelated detection problems in this paper. The first one, referred to as \emph{signal detection}, involves deciding between an observation $\mathbf{y} \in \mathbb{R}^m$ being just noise or it being an unknown signal $\mathbf{x} \in \mathbb{R}^m$ embedded in noise. Mathematically, this can be posed as a binary hypothesis test with the null ($\mathcal{H}_{0}$) and the alternate ($\mathcal{H}_{1}$) hypotheses given by:
\begin{align} \label{eq:prob_form:SD}
&\mathcal{H}_{0}: \quad \mathbf{y = n}; \nonumber \\
&\mathcal{H}_{1}: \quad \mathbf{y = \mathbf{x} + n};
\end{align}
where $\mathbf{n} \in \mathbb{R}^m$ denotes noise that is typically assumed Gaussian. Traditionally,~\eqref{eq:prob_form:SD} has been studied under the assumption of $\mathbf{x}$ belonging to a low-dimensional subspace of $\mathbb{R}^m$~\cite{scharf1994matched,kraut2001adaptive,kraut1999cfar,kelly1986adaptive}. In contrast, our focus is on the case of $\mathbf{x}$ belonging to a \emph{union} of low-dimensional subspaces: $\mathbf{x} \in \underset{k = 1}{\overset{K_0} \bigcup} S_{k}$, where $S_{k} \subset \mathbb{R}^m$ denotes a subspace of $\mathbb{R}^m$. We further assume that the subspaces are pairwise disjoint, $S_{k} \cap S_{k'} = \emptyset$ for $k \not= k'$, and they have the same dimension: $\forall k, \mathrm{dim}(S_k) = n \ll m$.\footnote{One can extend this work to the case of different dimensional subspaces in a straightforward manner at the expense of notational complexity.}

The second problem studied in this paper, which does not arise in classical subspace detection literature, is referred to as \emph{active subspace detection}. The goal in this problem is to not only detect whether $\mathbf{y}$ contains an unknown signal $\mathbf{x}$, but also \emph{identify} the subspace $S_k$ to which $\mathbf{x}$ belongs. Mathematically, this can be posed as a multiple hypothesis test with the null ($\mathcal{H}_{0}$) and the alternate ($\{\mathcal{H}_{k}\}_{k=1}^{K_0}$) hypotheses given by:
\begin{align} \label{eq:prob_form:SuD}
&\mathcal{H}_{0}: \quad \mathbf{y = n}; \nonumber \\
&\mathcal{H}_{k}: \quad \mathbf{y = \mathbf{x} + n}, \ \mathbf{x} \in S_k; \quad k=1,\dots,K_0.
\end{align}

Our goal in this paper is to derive statistical tests for \eqref{eq:prob_form:SD} and \eqref{eq:prob_form:SuD}, and provide a rigorous mathematical understanding of the performance of the derived tests. Our analysis is based on the assumption of $\mathbf{n}$ being a colored Gaussian noise that is distributed as $\mathcal{N}(0,\sigma^{2}\mathbf{R})$ with $\mathbf{R}$ being a full-rank covariance. In particular, we focus on the three cases of ($i$) known noise statistics, ($ii$) known variance ($\sigma^2$), but unknown covariance ($\mathbf{R}$), and ($iii$) unknown variance and covariance. In contrast to prior works~\cite{davenport2010signal,yap2014false,joneidi2016union,gini2004radar,wimalajeewa2015subspace}, we are specifically interested in characterizing our results in terms of the geometry of the underlying subspaces. This geometry can be described through the principal angles between the subspaces, where the $i$-th principal angle between subspace $S_j$ and $S_k$, denoted by $\varphi_i^{(j,k)}, i=1,\dots,n$, is recursively defined as \cite{afriat1957orthogonal}:
\begin{align}
\varphi_i^{(j,k)} &= \arccos \Bigg( \underset{\mathbf{u},\mathbf{v}}{\max} \Bigg\{ \frac{\langle \mathbf{u},\mathbf{v} \rangle}{\|\mathbf{u}\|_2 \|\mathbf{v}\|_2} : \mathbf{u} \in S_j, \mathbf{v} \in S_k, \nonumber \\
						& \qquad \qquad \mathbf{u} \,\bot\, \mathbf{u}_\ell, \mathbf{v} \,\bot\, \mathbf{v}_\ell, \ell = 1,\dots,i-1  \Bigg\} \Bigg),
\end{align}
where $(\mathbf{u}_\ell, \mathbf{v}_\ell) \in S_j \times S_k$ denote the principal vectors associated with the $\ell$-th principal angle. It is straightforward to see that $0 \leq \varphi_1^{(j,k)} \leq \varphi_2^{(j,k)} \leq \dots \leq \varphi_n^{(j,k)} \leq \pi/2$.

We conclude by noting that our statistical tests in the following will be expressed in terms of the following ratios for compactness purposes:
\begin{align*}
T_{\mathbf{z}}(\mathbf{P}) = \frac{\mathbf{z}^T \mathbf{P} \mathbf{z}}{\mathbf{z}^T \mathbf{z}} &, \ T_{\mathbf{z}}^{\eta}(\mathbf{P}) = \frac{\mathbf{z}^T \mathbf{P} \mathbf{z}}{\eta}, \nonumber \\
\ T_{\mathbf{z}}(\mathbf{P}, \mathbf{Q}) = \frac{\mathbf{z}^T \mathbf{P} \mathbf{z}}{\mathbf{z}^T \mathbf{Q} \mathbf{z}} &, \ \overline{T}_{\mathbf{z}}^{\eta}(\mathbf{P}) = \frac{\mathbf{z}^T \mathbf{P} \mathbf{z}}{\eta + \mathbf{z}^T \mathbf{z}},
\end{align*}
where $\mathbf{z}$ and $( \mathbf{P}$, $\mathbf{Q} )$ denote a vector and matrices of appropriate dimensions, respectively, while $\eta > 0$ denotes a constant.

\subsection{Performance metrics}
The performances of the statistical tests proposed in this paper will be characterized in terms of the probabilities of detection ($P_D$), classification ($P_C$), and false alarm ($P_{FA})$. Specifically, let $P_{\mathcal{H}_i}(\cdot) = \Pr(\cdot | \mathcal{H}_i)$, and define the event $\widehat{\mathcal{H}}_i = \{\text{Hypothesis $\mathcal{H}_i$ is accepted}\}$. Then, in the case of signal detection, we have $P_D = P_{\mathcal{H}_1}(\widehat{\mathcal{H}}_1)$ and $P_{FA} = P_{\mathcal{H}_0}(\widehat{\mathcal{H}}_1)$. In contrast, in the case of active subspace detection, we have $P_C = \sum_{k=1}^{K_0} P_{\mathcal{H}_k}(\widehat{\mathcal{H}}_k) \Pr(\mathcal{H}_k)$ and $P_{FA} = P_{\mathcal{H}_0}(\cup_{k=1}^{K_0}\widehat{\mathcal{H}}_k)$.

We conclude by pointing out that some of our forthcoming discussion will use the shorthand $P_{S_k}(\cdot) = \Pr(\cdot | \{\mathbf{x} \in S_k\})$ and $\Psi(\eta_0,\alpha) = \frac{\sqrt{2}}{2^n \Gamma(n/2)} (\eta_{0} \alpha)^{(n-1)/2} K_{(n-1)/2}\left(\frac{\eta_{0} \alpha}{2}\right)$, where $\alpha \in \mathbb{R}_+$ and $\eta_0 \in (0,1/2)$. Using this notation, we can also write $P_{D} = \sum_{k=1}^{K_0} P_{S_{k}}(\widehat{\mathcal{H}}_1) \Pr(\mathbf{x} \in S_k)$.

\section{Main Results} \label{SD}
In this section, we present statistical tests for both the detection problems under various noise conditions. In addition, we provide bounds on the performance metrics for these tests.

\subsection{Known noise statistics} \label{MSD:KN}
We begin with the assumption that both the noise variance, $\sigma^2$, and the covariance, $\mathbf{R}$, are known. It is trivial to see that $\mathbf{y} | \mathcal{H}_0 \sim \mathcal{N}(0,\sigma^2\mathbf{R})$ for both detection problems. Further, in the case of signal detection, we have $\mathbf{y} | \mathcal{H}_1 \sim \mathcal{N}(\mathbf{x},\sigma^2\mathbf{R})$. In contrast, the observations $\mathbf{y}$ under the $k$-th alternate hypothesis in the case of active subspace detection can be expressed as $\mathbf{y} | \mathcal{H}_k \sim \mathcal{N}(\mathbf{H}_k \boldsymbol{\theta}_k,\sigma^2\mathbf{R}), k=1,\dots,K_0$, where $\mathbf{H}_k \in \mathbb{R}^{m \times n}$ denotes a basis for subspace $S_k$ and $\boldsymbol{\theta}_k \in \mathbb{R}^n$ denotes representation coefficients of $\mathbf{x}$ under basis $\mathbf{H}_k$. Since $\mathbf{x}$ and $\boldsymbol{\theta}_k$ are unknown for the signal and the active subspace detection problems, respectively, we resort to the \emph{generalized likelihood ratio tests} (GLRTs) for the two detection problems. Our results in this regard are based on the following definitions: let $\mathbf{z} = \mathbf{R}^{-\frac{1}{2}} \mathbf{y}$ denote the \emph{whitened} observations, $\mathbf{w} = \mathbf{R}^{-\frac{1}{2}} \mathbf{n}$ denote the \emph{whitened} noise, $\mathbf{G}_k = \mathbf{R}^{-\frac{1}{2}} \mathbf{H}_k, k=1,\dots,K_0$, denote the whitened subspace bases, and $\mathbf{P}_{\bar{S}_k} = \mathbf{G}_k (\mathbf{G}_k^T \mathbf{G}_k)^{-1} \mathbf{G}_k^T$ and $\mathbf{P}_{\bar{S}_k}^{\bot} = \mathbf{I} - \mathbf{P}_{\bar{S}_k}$, respectively, denote the projection matrix for the $k$-th whitened subspace and its orthogonal complement.

\begin{theorem}\label{th:MSD:KN:test}
Let $\bar{\gamma} > 0$ denote the test threshold and define $\widehat{k} = \argmax_k (\mathbf{z}^T \mathbf{P}_{\bar{S}_{{k}}} \mathbf{z})$. The GLRT for the signal detection and the active subspace detection problem is, respectively, given by
\begin{align} \label{eq:MSD:KN:test}
T_{\mathbf{z}}^{2 \sigma^2}\left(\mathbf{P}_{\bar{S}_{\widehat{k}}}\right) \underset{\mathcal{H}_{0}}{\overset{\mathcal{H}_{1}}{\gtrless}} \bar{\gamma} \qquad \text{and} \qquad
T_{\mathbf{z}}^{2 \sigma^2}\left(\mathbf{P}_{\bar{S}_{\widehat{k}}}\right) \underset{\mathcal{H}_{0}}{\overset{\mathcal{H}_{\widehat{k}}}{\gtrless}} \bar{\gamma}\,.
\end{align}
\end{theorem}
The proof of this theorem is given in Appendix~\ref{th:MSD:KN:test:proof}, while its interpretation as well as its relationship to the classical test for subspace detection are provided in Sec.~\ref{sec:disc}. We now characterize the performance of the statistical tests in~\eqref{eq:MSD:KN:test} in terms of bounds on $P_{FA}$, $P_D$, and $P_C$. Note that we have to resort to bounds, as opposed to exact expressions, because of the complicated joint distributions that arise in our context; we refer the reader to Appendix~\ref{th:MSD:KN:prob:proof} for further discussion.

\begin{theorem} \label{th:MSD:KN:prob}
The GLRTs in~Theorem~\ref{th:MSD:KN:test} for the signal and the active subspace detection problems result in probability of false alarm that is upper bounded by:
\begin{align} \label{eq:MSD:KN:prob}
P_{FA} 
	   &\leq \min \Big\{ 1 \;,\; \underset{k = 1}{\overset{K_0} \sum} \Pr\Big(T_{{\mathbf{w}}}^{2 \sigma^2}(\mathbf{P}_{\bar{S}_k}) > \bar{\gamma} \Big) \Big\}.
\end{align}

Further, in the case of signal detection, the probability of detection $P_{D} = \sum_{k=1}^{K_0} P_{S_{k}}(\widehat{\mathcal{H}}_1) \Pr(\mathbf{x} \in S_k)$ can be upper and lower bounded by the fact that
\begin{align} \label{eq:MSD:KN:prob:detect}
P_{S_k}(\widehat{\mathcal{H}}_1) &\leq \min \Big\{ 1 \;,\; {\underset{i = 1}{\overset{K_0} \sum}} {P}_{S_{k}} \Big( T_{\mathbf{z}}^{2 \sigma^2}(\mathbf{P}_{\bar{S}_i}) > \bar{\gamma} \Big) \Big\}, \ \text{and}\nonumber \\
P_{S_{k}}(\widehat{\mathcal{H}}_1) &\geq {\underset{i = 1}{\overset{K_0} \sum}}\frac{\left[{P}_{S_{k}} \Big( T_{\mathbf{z}}^{2 \sigma^2}(\mathbf{P}_{\bar{S}_i}) > \bar{\gamma} \Big)\right]^2 }{ \underset{j = 1}{\overset{K_0} \sum} {P}_{S_{k}} \Big( T_{\mathbf{z}}^{2 \sigma^2}(\mathbf{P}_{\bar{S}_i}) > \bar{\gamma} , T_{\mathbf{z}}^{2 \sigma^2}(\mathbf{P}_{\bar{S}_j}) > \bar{\gamma} \Big) }.
\end{align}

Finally, the probability of classification $P_C$ for active subspace detection can be lower bounded by the fact that
\begin{align} \label{eq:MSuD:KN:prob}
P_{\mathcal{H}_k}(\widehat{\mathcal{H}}_k) &\geq  \max \Big\{ 0 \;,\; {P}_{S_k} (  T_{\mathbf{z}}^{2 \sigma^2}(\mathbf{P}_{\bar{S}_k}) > \bar{\gamma} ) +  \nonumber \\
&{{\underset{j = 1,j \neq k}{\overset{K_0} \sum}}} {P}_{S_k} ( T_{\mathbf{z}}( \mathbf{P}_{\bar{S}_k} , \mathbf{P}_{\bar{S}_j} ) > 1) - (K_0 - 1) \Big\}.
\end{align}
\end{theorem}
The proof of this theorem is given in Appendix~\ref{th:MSD:KN:prob:proof}. It is worth noting that probabilities of the form ${P}_{S_{k}}( T_{\mathbf{z}}^{2 \sigma^2}(\mathbf{P}_{\bar{S}_j}) > \bar{\gamma})$ correspond to tail probabilities of chi-squared random variables, whereas the probabilities ${P}_{S_k} ( T_{\mathbf{z}}( \mathbf{P}_{\bar{S}_k} , \mathbf{P}_{\bar{S}_j} ) > 1)$ involve ratios of \emph{dependent} chi-squared variables whose distributions can be numerically computed.

\begin{remark} \label{remark:MSuD:KN}
It is noted in Appendix~\ref{th:MSD:KN:prob:proof} that \eqref{eq:MSuD:KN:prob} can be further lower bounded using \cite[Lemma~1]{wimalajeewa2015subspace} as $P_{\mathcal{H}_k}(\widehat{\mathcal{H}}_k) \geq \max \big\{0 , {P}_{S_k}(T_{\mathbf{z}}^{2 \sigma^2}(\mathbf{P}_{\bar{S}_k}) > \bar{\gamma}) - \sum\limits_{j:j\not=k} Q \big( \frac{1}{2} (1-2\eta_{0}) \sqrt{\lambda_{j \backslash k}}\big) - \sum\limits_{j:j\not=k} \Psi(\eta_{0},\lambda_{j \backslash k}) \big\}$, where $\lambda_{j \backslash k} = \mathbf{z}^T \mathbf{P}_{\bar{S}_j}^{\bot} \mathbf{z} / \sigma^2$ when $\mathbf{z} \in \bar{S}_k$. This bound, however, depends further on $\eta_0$. Numerical experiments reported in Sec.~\ref{num_sims} show the looseness of this bound for the case of $\eta_0 = 0.25$, the value advertised in~\cite{wimalajeewa2015subspace}.
\end{remark}

\begin{remark}
A heuristic approach to detecting signals under the UoS model would be to use the multiple hypothesis tests of \cite{bajwa2014multiple}, where each test is an individual matched subspace detector. The final decision can then be made by taking the union of binary outputs from each matched detector and declaring detection if any one of them has detected a signal. It is straightforward to see however that this final decision rule coincides with the decision rule in \eqref{eq:MSD:KN:test}. Thus, in the event that only one subspace is active, the testing procedure in \cite{bajwa2014multiple} effectively reduces to a GLRT.
\end{remark}

\subsection{Unknown noise covariance} \label{ASD:UC}
Next, we consider the case of colored noise with unknown covariance matrix $\mathbf{R}$. In this case, we also assume access to $N_0$ noise samples $\boldsymbol{\xi}_p \sim \mathcal{N}(0,\mathbf{R}), p = 1,\dots,N_0$ ($N_0 > m$ to obtain a non-singular estimate of $\mathbf{R}$), which is a standard assumption in the detection literature~\cite{kraut2001adaptive,kraut1999cfar,kelly1986adaptive}. As before, we use GLRTs to obtain decision rules for the two detection problems. Our results make use of the following definitions: let $\boldsymbol{\Sigma} = \frac{1}{N_0} \sum_{p = 1}^{N_0} {\boldsymbol{\xi}_p} {\boldsymbol{\xi}_p}^T$ denote sample covariance of noise samples, $\widehat{\mathbf{z}} = \boldsymbol{\Sigma}^{-\frac{1}{2}} \mathbf{y}$ denote the \emph{empirically whitened} observations, $\widehat{\mathbf{w}} = \boldsymbol{\Sigma}^{-\frac{1}{2}} \mathbf{n}$ denote the \emph{empirically whitened} noise, $\widehat{\mathbf{G}}_k = \boldsymbol{\Sigma}^{-\frac{1}{2}} \mathbf{H}_k, k=1,\dots,K_0$, denote the empirically whitened subspace bases, and $\widehat{\mathbf{P}}_{\bar{S}_k} = \widehat{\mathbf{G}}_k (\widehat{\mathbf{G}}_k^T \widehat{\mathbf{G}}_k)^{-1} \widehat{\mathbf{G}}_k^T$ denote the projection matrix for the $k$-th empirically whitened subspace.
\begin{theorem}\label{th:ASD:UC:test}
Let $\bar{\gamma} > 0$ denote the test threshold and define $\widehat{k} = \argmax_k (\widehat{\mathbf{z}}^T \widehat{\mathbf{P}}_{\bar{S}_k} \widehat{\mathbf{z}})$. The GLRT for the signal detection and the active subspace detection problem is, respectively, given by:
\begin{align} \label{eq:ASD:UC:test}
\overline{T}_{\widehat{\mathbf{z}}}^{N_0 \sigma^2}(\widehat{\mathbf{P}}_{\bar{S}_{\widehat{k}}}) \underset{\mathcal{H}_{0}}{\overset{\mathcal{H}_{1}}{\gtrless}} \bar{\gamma} \qquad \text{and} \qquad
\overline{T}_{\widehat{\mathbf{z}}}^{N_0 \sigma^2}(\widehat{\mathbf{P}}_{\bar{S}_{\widehat{k}}}) \underset{\mathcal{H}_{0}}{\overset{\mathcal{H}_{\widehat{k}}}{\gtrless}} \bar{\gamma}.
\end{align}
\end{theorem}
The proof of this theorem is provided in Appendix \ref{th:ASD:UC:test:proof}, while some discussion on interpretation and relationship to the classical test for subspace detection is provided in Sec.~\ref{sec:disc}. We now characterize the performance of the statistical tests in~\eqref{eq:ASD:UC:test} in terms of bounds on $P_{FA}$, $P_D$, and $P_C$.
\begin{theorem}\label{th:ASD:UC:prob}
The GLRTs for the signal and the active subspace detection problems in Theorem~\ref{th:ASD:UC:test} result in the probability of false alarm that is upper bounded by:
\begin{align} \label{eq:ASD:UC:prob}
P_{FA} &\leq \min \Big\{ 1 \;,\; \underset{k = 1}{\overset{K_0} \sum} \Pr \Big( \overline{T}_{{\widehat{\mathbf{w}}}}^{N_0 \sigma^2}(\widehat{\mathbf{P}}_{\bar{S}_k}) > \bar{\gamma} \Big) \Big\}.
\end{align}

Further, in the case of signal detection, the probability of detection $P_{D} = \sum_{k=1}^{K_0} P_{S_{k}}(\widehat{\mathcal{H}}_1) \Pr(\mathbf{x} \in S_k)$ can be upper and lower bounded by the fact that
\begin{align}
P_{S_{k}} \Big( \widehat{\mathcal{H}}_1 \Big) &\leq \min \Big\{ 1 \;,\; {\underset{i = 1}{\overset{K_0} \sum}} {P}_{S_{k}} \Big( \overline{T}_{\widehat{\mathbf{z}}}^{N_0 \sigma^2}(\widehat{\mathbf{P}}_{\bar{S}_i}) > \bar{\gamma} \Big) \Big\}, ~and \nonumber \\
P_{S_{k}} \Big( \widehat{\mathcal{H}}_1 \Big) 	&\geq {\underset{i = 1}{\overset{K_0} \sum}} \frac{ \left[ {P}_{S_{k}} \Big( \overline{T}_{\widehat{\mathbf{z}}}^{N_0 \sigma^2}(\widehat{\mathbf{P}}_{\bar{S}_i}) > \bar{\gamma} \Big) \right]^2 }{ \underset{j = 1}{\overset{K_0} \sum} {P}_{S_{k}} \Big( \overline{T}_{\widehat{\mathbf{z}}}^{N_0 \sigma^2}(\widehat{\mathbf{P}}_{\bar{S}_i}) > \bar{\gamma} , \overline{T}_{\widehat{\mathbf{z}}}^{N_0 \sigma^2}(\widehat{\mathbf{P}}_{\bar{S}_j}) > \bar{\gamma} \Big) }.
\end{align}
Finally, the probability of classification $P_C$ for active subspace detection can be lower bounded by the fact that
\begin{align} \label{eq:ASuD:UC:prob}
P_{\mathcal{H}_k}&(\widehat{\mathcal{H}}_k) \geq  \max \Big\{ 0 \;,\; {P}_{S_k} (  \overline{T}_{\widehat{\mathbf{z}}}^{N_0 \sigma^2}(\widehat{\mathbf{P}}_{\bar{S}_k}) > \bar{\gamma} ) + \nonumber \\
&  {{\underset{j = 1,j \neq k}{\overset{K_0} \sum}}} {P}_{S_k} ( T_{\widehat{\mathbf{z}}}( \widehat{\mathbf{P}}_{\bar{S}_k} , \widehat{\mathbf{P}}_{\bar{S}_j} ) > 1) - (K_0 - 1) \Big\}.
\end{align}
\end{theorem}
The proof of this theorem follows along similar lines as for the proof of Theorem \ref{th:MSD:KN:prob} and is omitted due to space constraints. In contrast to Theorem~\ref{th:MSD:KN:prob}, the terms of the form ${P}_{S_k} (  \overline{T}_{\widehat{\mathbf{z}}}^{N_0 \sigma^2}(\widehat{\mathbf{P}}_{\bar{S}_k}) > \bar{\gamma} )$ and ${P}_{S_k} ( T_{\widehat{\mathbf{z}}}( \widehat{\mathbf{P}}_{\bar{S}_k} , \widehat{\mathbf{P}}_{\bar{S}_j} ) > 1)$ involve probabilities of the ratios of \emph{dependent} chi-squared variables and have to be computed numerically.
\begin{remark} \label{remark:ASuD:UC}
One can again further lower bound \eqref{eq:ASuD:UC:prob} using \cite[Lemma~1]{wimalajeewa2015subspace} as: $P_{\mathcal{H}_k}(\widehat{\mathcal{H}}_k) \geq \max \Big\{ 0 , {P}_{S_k} (  \overline{T}_{\widehat{\mathbf{z}}}^{N_0 \sigma^2}(\widehat{\mathbf{P}}_{\bar{S}_k}) > \bar{\gamma} ) - {\underset{j:j\not=k}{\overset{K_0} \sum}} Q \big( \frac{1}{2} (1-2\eta_{0}) \sqrt{\widehat{\lambda}_{j \backslash k}}\big) - {\underset{j:j\not=k}{\overset{K_0} \sum}} \Psi(\eta_{0},\widehat{\lambda}_{j \backslash k}) \Big\}$, where $\widehat{\lambda}_{j \backslash k} = \frac{1}{\sigma^2} \widehat{\mathbf{z}}^T \widehat{\mathbf{P}}_{\bar{S}_j}^{\bot} \widehat{\mathbf{z}}$ when $\mathbf{z} \in \bar{S}_k$.
\end{remark}

\subsection{Unknown noise statistics} \label{ASD:UN}
We now address adaptive detection in settings where the covariance matrix $\mathbf{R}$ and variance $\sigma^2$ are both unknown. Once again assuming access to $N_{0}$ noise samples and using the notation of Sec.~\ref{ASD:UC}, the GLRTs lead to the following decision rules.
\begin{theorem} \label{th:ASD:UN:test}
Let $\bar{\gamma} > 0$ denote the test threshold and define $\widehat{k} = \argmax_k (\widehat{\mathbf{z}}^T \widehat{\mathbf{P}}_{\bar{S}_k} \widehat{\mathbf{z}})$. The GLRT for the signal detection and the active subspace detection problem is, respectively, given by:
\begin{align} \label{eq:ASD:UN:test}
T_{\widehat{\mathbf{z}}}(\widehat{\mathbf{P}}_{\bar{S}_{\widehat{k}}}) \underset{\mathcal{H}_{0}}{\overset{\mathcal{H}_{1}}{\gtrless}} \bar{\gamma} \quad \text{and} \quad
T_{\widehat{\mathbf{z}}}(\widehat{\mathbf{P}}_{\bar{S}_{\widehat{k}}}) \underset{\mathcal{H}_{0}}{\overset{\mathcal{H}_{\widehat{k}}}{\gtrless}} \bar{\gamma}.
\end{align}
\end{theorem}
The proof of this theorem is given in Appendix \ref{th:ASD:UN:test:proof}, with corresponding discussion in Sec.~\ref{sec:disc}. The performance of the statistical tests in~\eqref{eq:ASD:UN:test} is given by the following theorem.
\begin{theorem} \label{th:ASD:UN:prob}
The GLRTs for the signal and the active subspace detection problems in Theorem~\ref{th:ASD:UN:test} result in the probability of false alarm that is upper bounded by:
\begin{align} \label{eq:ASD:UN:prob}
P_{FA} 
	   &\leq \min \Big\{ 1 \;,\; \underset{k = 1}{\overset{K_0} \sum} \Pr \Big( T_{\widehat{\mathbf{w}}}(\widehat{\mathbf{P}}_{\bar{S}_k}) > \bar{\gamma} \Big) \Big\}.
\end{align}

Further, in the case of signal detection, the probability of detection $P_{D} = \sum_{k=1}^{K_0} P_{S_{k}}(\widehat{\mathcal{H}}_1) \Pr(\mathbf{x} \in S_k)$ can be upper and lower bounded by the fact that
\begin{align}
P_{S_{k}} \Big( \widehat{\mathcal{H}}_1 \Big) &\leq \min \Big\{ 1 \;,\; {\underset{i = 1}{\overset{K_0} \sum}} {P}_{S_{k}} \Big( T_{\widehat{\mathbf{z}}}(\widehat{\mathbf{P}}_{\bar{S}_i}) > \bar{\gamma} \Big) \Big\}, ~and \nonumber \\
P_{S_{k}} \Big( \widehat{\mathcal{H}}_1 \Big) 	&\geq {\underset{i = 1}{\overset{K_0} \sum}} \frac{ \left[ {P}_{S_{k}} \Big( T_{\widehat{\mathbf{z}}}(\widehat{\mathbf{P}}_{\bar{S}_i}) > \bar{\gamma} \Big) \right]^2 }{ \underset{j = 1}{\overset{K_0} \sum} {P}_{S_{k}} \Big( T_{\widehat{\mathbf{z}}}(\widehat{\mathbf{P}}_{\bar{S}_i}) > \bar{\gamma} , T_{\widehat{\mathbf{z}}}(\widehat{\mathbf{P}}_{\bar{S}_j}) > \bar{\gamma} \Big) }.
\end{align}
Finally, the probability of classification $P_C$ for active subspace detection can be lower bounded by the fact that
\begin{align} \label{eq:ASuD:UN:prob}
P_{\mathcal{H}_k}&(\widehat{\mathcal{H}}_k) \geq  \max \Big\{ 0 \;,\; {P}_{S_k} (  T_{\widehat{\mathbf{z}}}(\widehat{\mathbf{P}}_{\bar{S}_k}) > \bar{\gamma} ) + \nonumber \\
&{{\underset{j = 1,j \neq k}{\overset{K_0} \sum}}} {P}_{S_k} ( T_{\widehat{\mathbf{z}}}( \widehat{\mathbf{P}}_{\bar{S}_k} , \widehat{\mathbf{P}}_{\bar{S}_j} ) > 1) - (K_0 - 1) \Big\}.
\end{align}
\end{theorem}
The proof of this theorem is also similar to the proof of Theorem \ref{th:MSD:KN:prob} and is thus omitted. Similar to Theorem~\ref{th:ASD:UC:prob}, the terms of the form ${P}_{S_k} (  T_{\widehat{\mathbf{z}}}(\widehat{\mathbf{P}}_{\bar{S}_k}) > \bar{\gamma} )$ and ${P}_{S_k} ( T_{\widehat{\mathbf{z}}}( \widehat{\mathbf{P}}_{\bar{S}_k} , \widehat{\mathbf{P}}_{\bar{S}_j} ) > 1)$ need to be computed numerically.
\begin{remark}
Similar to Remark~\ref{remark:ASuD:UC}, a looser lower bound can be derived here as well, with the only difference being that $\overline{T}_{\widehat{\mathbf{z}}}^{N_0 \sigma^2}(\widehat{\mathbf{P}}_{\bar{S}_k})$ is replaced by $T_{\widehat{\mathbf{z}}}(\widehat{\mathbf{P}}_{\bar{S}_k})$.
\end{remark}

\section{Discussion} \label{sec:disc}
In this section we discuss some characteristics of the various test statistics obtained in Sec.~\ref{SD}. We also describe the impact of geometry of the subspaces in the union and the geometry of the colored noise on the detection performances.

\revise{
\subsection{UoS detection versus classical subspace detection}
First, we compare the test statistics for signal detection under the UoS model (\eqref{eq:MSD:KN:test},\eqref{eq:ASD:UC:test} and \eqref{eq:ASD:UN:test}) with their counterparts under the subspace model \cite{scharf1994matched,kraut2001adaptive,kraut1999cfar,kelly1986adaptive}. Under the subspace observation model, the signal $\mathbf{x}$ is assumed to belong to a single subspace, $\mathbf{x} = \mathbf{H} \boldsymbol{\theta}$, where $\mathbf{H}$ contains the subspace bases. The corresponding test statistics for known noise statistics, unknown noise covariance and unknown noise statistics are, respectively, given by~\cite{scharf1994matched,kraut2001adaptive,kraut1999cfar,kelly1986adaptive}:
\begin{align} \label{eq:subspace:tests}
T_{\mathbf{z}}^{2 \sigma^2}\left(\mathbf{P}_{\bar{S}}\right) \underset{\mathcal{H}_{0}}{\overset{\mathcal{H}_{1}}{\gtrless}} \bar{\gamma}
,
\overline{T}_{\widehat{\mathbf{z}}}^{N_0 \sigma^2}(\widehat{\mathbf{P}}_{\bar{S}}) \underset{\mathcal{H}_{0}}{\overset{\mathcal{H}_{1}}{\gtrless}} \bar{\gamma}
, \text{ and} \,\,\,
T_{\widehat{\mathbf{z}}}(\widehat{\mathbf{P}}_{\bar{S}}) \underset{\mathcal{H}_{0}}{\overset{\mathcal{H}_{1}}{\gtrless}} \bar{\gamma}.
\end{align}
At a first glance, the test statistics for the UoS model and the subspace model look similar. However, the numerator of the statistics for the subspace model corresponds to the energy of the observed signal after projection onto the relevant subspace. In contrast, since we deal with multiple subspaces, we have to rely on projection onto the subspace that captures the most energy of the observed signal.

\begin{figure} [!h]
	\centering
	\includegraphics[width=\columnwidth] {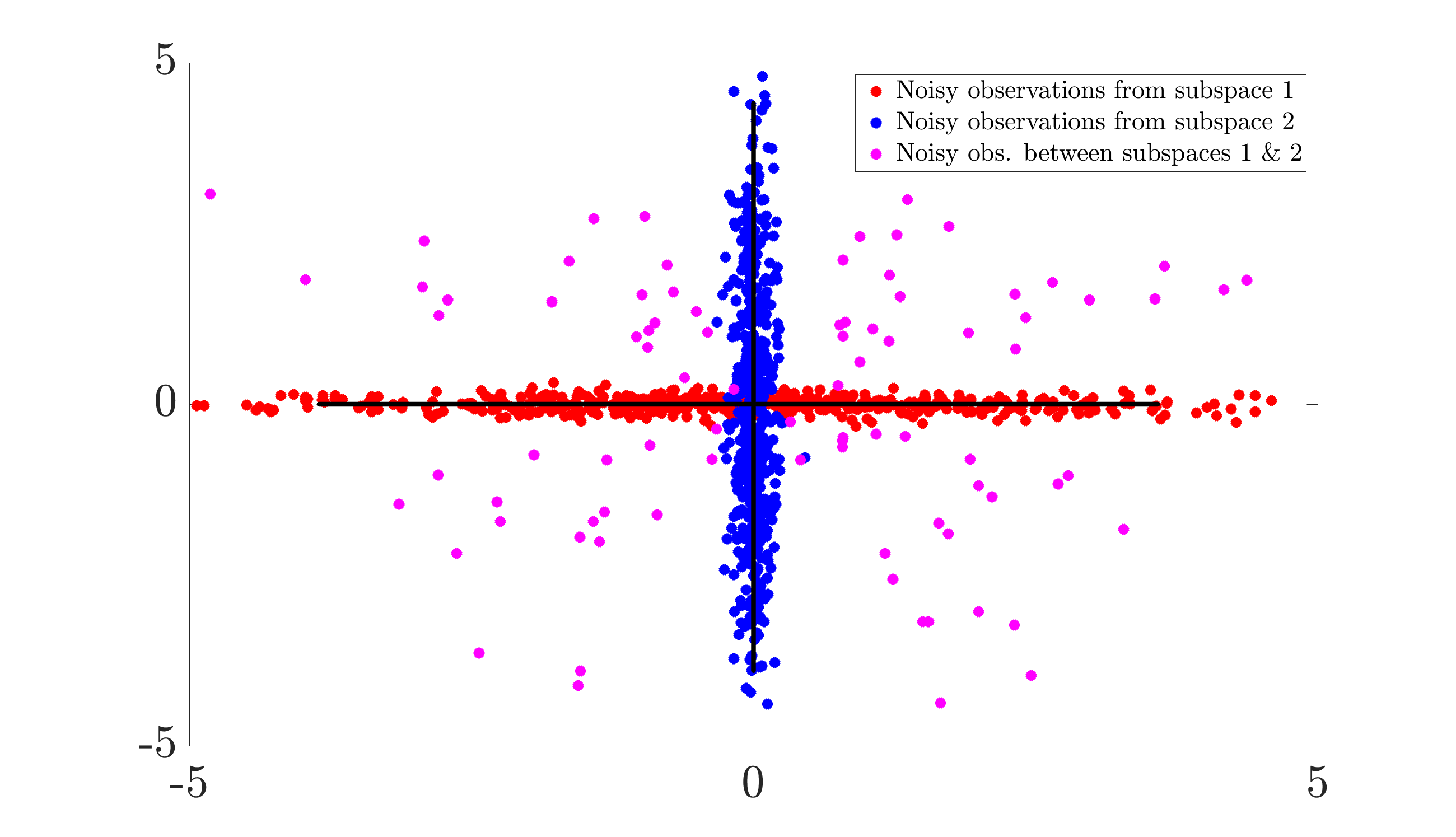}
	\caption{ \label{fig:SD:discuss:UoS_vs_Sub}
		{\revise{This figure highlights the difference between UoS- and classical subspace-based detection of signals generated under the UoS model. The red and blue dots correspond to noisy signals generated from a union of two subspaces, while the magenta dots represent observations that do not belong to the union. UoS-based detection would be able to reject the magenta observations, whereas subspace-based detection would accept them as signals since they belong to the direct sum of the underlying subspaces.}}}
\end{figure}

Next, we discuss advantages of the UoS-based test statistics over the respective subspace-based test statistics for signal detection. Under the assumption of the (noisy) signal being generated under the UoS model, the test statistics derived in this paper reject observations that correspond to the ``gaps'' between the individual subspaces; see, e.g., Fig.~\ref{fig:SD:discuss:UoS_vs_Sub}, in which observations in the gaps correspond to magenta-colored dots. In contrast, subspace-based detection needs to resort to direct sum of the underlying subspaces in the union. This, in turn, leads to higher false alarm rates since observations in gaps that belong to the direct sum are falsely accepted as signals; in Fig.~\ref{fig:SD:discuss:UoS_vs_Sub}, e.g., subspace-based detection will accept all magenta observations as signals. We also refer the reader to Sec.~\ref{num_sims} for numerical validation of this fact.

Finally, the presence of multiple subspaces in the union also results in the problem of active subspace detection, which does not arise in the context of classical subspace detection as it only considers one underlying subspace.
}

\subsection{Signal detection versus active subspace detection}
Notice that the test statistics for active subspace detection have forms similar to those for signal detection. The main difference lies in the performance of these statistics when detecting either the signal or the active subspace. The detection performance for active subspace detection is lower than that for signal detection. This is due to the fact that for signal detection, the detector is not concerned with detecting the true subspace from which the observed signal is coming and can afford to confuse one subspace with another as long as it detects the presence of a signal. That is not the case with active subspace detection, where this confusion matters, and thus we observe the loss in performance. This fact was also highlighted by Gini et al. in \cite{gini2004radar}.

\subsection{Invariance properties of the test statistics}
We now examine the invariance properties of our test statistics for signal detection. Since our test statistics for active subspace detection are similar to those for signal detection under UoS model, they exhibit similar invariance properties.

From the expressions in \eqref{eq:MSD:KN:test}, \eqref{eq:ASD:UC:test} and \eqref{eq:ASD:UN:test}, notice that the statistics are invariant to the rotations in $\bar{S}_{\widehat{k}}$. This means all rotated versions of the relevant signal (for rotations in $\bar{S}_{\widehat{k}}$) will result in same detection performance. Moreover, the statistics also exhibit invariance with respect to the translations in $\bar{S}_{\widehat{k}}^{\bot}$ (which is the orthogonal subspace of $\bar{S}_{\widehat{k}}$). This implies that any additive interference from $\bar{S}_{\widehat{k}}^{\bot}$ is unnoticeable to the detectors since they only measure the energy of $\mathbf{z}$ in the subspace $\bar{S}_{\widehat{k}}$.
Additionally, the test statistic for detection in unknown noise statistics \eqref{eq:ASD:UN:test} is also invariant to the scaling of the observed signals, i.e., scaled versions of a signal will result in same detection performance with this test. This is due to the fact that both numerator and denominator in \eqref{eq:ASD:UN:test} are quadratic forms of the \emph{whitended/empirically whitened} observations $\mathbf{z}$, without any additive terms.

\subsection{Influence of geometry between whitened subspaces on detection probability} \label{MSD:KN:influence:subspace}
The detection performance of our detector decreases only slightly as the angles between the subspaces increase. This can be seen from an alternate expression for the probability of union of events. For example, the probability of union of two events, $A$, and $B$, can be written as: $P(A \cup B) 	= P(A) + P(B) - P(A B) = P({A} B') + P(A' {B}) + P({A} {B})$ where $A'$, and $B'$ are the complements of the corresponding events. One can thus see that the probability of union of events is directly proportional to the probability of the intersection of events (and their complements). For the case of detection probability, these intersections are $k$-tuples of the form $\underset{j = 1}{\overset{k} \cap} \big\{ T_{\widehat{\mathbf{z}}}^{2 \sigma^2}(\widehat{\mathbf{P}}_{\bar{S}_j}) > \bar{\gamma} \big\}$ (and their complements). When a pair (or more) of subspaces are close to each other, i.e., the principal angles between whitened/empirically whitened subspaces are small, the probability of these $k$-tuples is larger compared to when the subspaces are far apart.

Intuitively, since signal detection problem is not concerned with the detection of the active subspace, confusing a (noisy) signal coming from one subspace as being generated from another subspace does not matter significantly. In fact, this confusion helps the detection task as long as a signal is actually present. Interestingly, when the subspaces are far apart, i.e., principal angles are large, chances of such confusion are less and the probability of detection is slightly decreased.

\subsection{Influence of geometry between whitened subspaces on correct classification probability} \label{MSuD:KN:influence:subspace}
We now examine the influence of geometry between whitened subspaces on the probability of correct classification. This analysis in particular sets us apart from other related works such as~\cite{davenport2010signal,yap2014false,joneidi2016union,gini2004radar,wimalajeewa2015subspace}, as we make the influence of geometry explicit through the principal angles between subspaces. We start with the case of active subspace detection in known noise statistics. The crux of our analysis is given in the following theorem.
\begin{theorem} \label{th:MSuD:KN:angles}
When the active subspaces are detected using the test in Theorem \ref{th:MSD:KN:test}, the lower bound on the probability of correct classification increases with increasing principal angles between the whitened subspaces.
\end{theorem}
The proof of this theorem is detailed in Appendix \ref{th:MSuD:KN:angles:proof}. The following corollary can also be obtained form Theorem \ref{th:MSuD:KN:angles}.
\begin{corollary}
Suppose the noise is white Gaussian, i.e., $\mathbf{n} \sim \mathcal{N}(0,\sigma^{2}\mathbf{I})$. When the active subspaces are detected using the test in Theorem \ref{th:MSD:KN:test}, the lower bound on the probability of correct classification increases with increasing principal angles between the subspaces in the union.
\end{corollary}

Similarly, in the case of other noise settings (unknown covariance and unknown noise statistics), the probability of correct classification of individual subspaces increases with increasing principal angles between the \emph{empirically whitened subspaces}. This also follows trivially from Theorem \ref{th:MSuD:KN:angles}.
	\begin{figure} [!t]
   	\centering
   	\includegraphics[width=6.5cm] {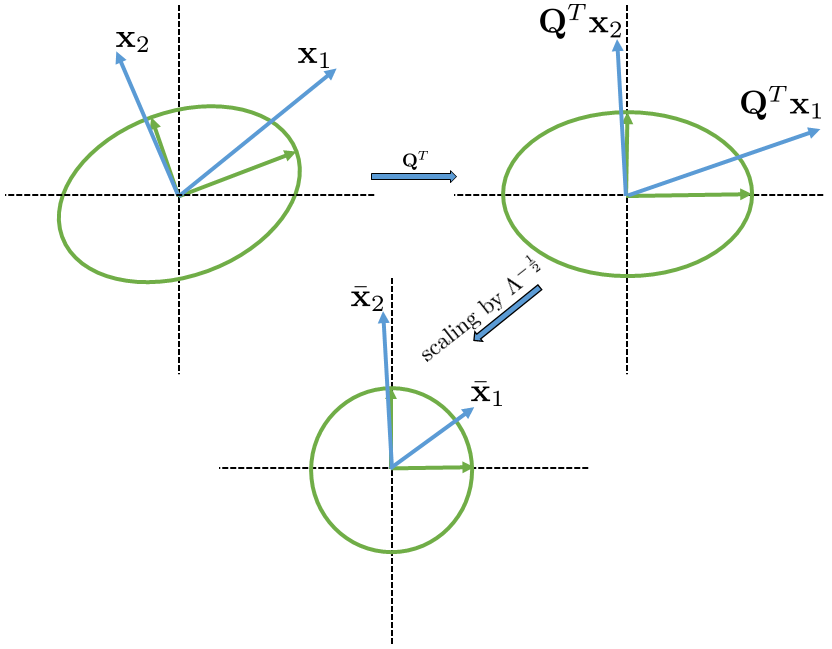}
   	\caption{ \label{fig:ASD:discuss:noise_geometry}
This figure shows the effect of geometry of colored noise on two signals coming from two different subspaces. The ellipse represents the covariance of the colored noise with the green vectors representing the eigenvectors of the covariance. The blue vectors $\mathbf{x}_1$ and $\mathbf{x}_2$ represent signals from the two different subspaces. The first operation during whitening can be seen as rotation by $\mathbf{Q}^T$ to align the canonical bases with the noise eigenvectors. The second operation of scaling by $\Lambda^{-\frac{1}{2}}$ scales each axis by the inverse of the corresponding eigenvalue. Thus, the closer a subspace is to the leading eigenvectors of noise covariance, the lower is its detection probability as it suffers more attenuation during whitening.}
   	\end{figure}

\subsection{Influence of geometry of colored noise} \label{MSD:KN:influence:noise}
To characterize the effect of noise geometry on two detection problems, we focus on the terms $\mathbf{z}^T \mathbf{P}_{\bar{S}_k} \mathbf{z}$ in \eqref{eq:MSD:KN:test}. 
We can see that $\mathbf{z}^T \mathbf{P}_{\bar{S}_k} \mathbf{z} = (\bar{\mathbf{x}} + \mathbf{w})^T \mathbf{P}_{\bar{S}_k} (\bar{\mathbf{x}} + \mathbf{w}) = \bar{\mathbf{x}}^T \mathbf{P}_{\bar{S}_k} \bar{\mathbf{x}} + 2 \mathbf{w}^T \mathbf{P}_{\bar{S}_k} \bar{\mathbf{x}} + \mathbf{w}^T \mathbf{P}_{\bar{S}_k} \mathbf{w},$
where $\bar{\mathbf{x}} = \mathbf{R}^{-\frac{1}{2}} \mathbf{x}$. The norm of $\bar{\mathbf{x}}$ can be expressed as:
\begin{align} \label{x_norm}
\| \bar{\mathbf{x}} \|_2^2 &= \mathbf{x}^T \mathbf{Q} \mathbf{\Lambda}^{-1} \mathbf{Q}^T \mathbf{x} = \| \mathbf{\Lambda}^{-\frac{1}{2}} \mathbf{x}^Q \|_2^2 = \underset{i = 1}{\overset{m}{\sum}} \frac{(x^Q_i)^2}{\lambda_i}
\end{align}
where $\mathbf{x}^Q = \mathbf{Q}^T \mathbf{x}$, and $\mathbf{R} = \mathbf{Q} \mathbf{\Lambda} \mathbf{Q}^T$ is the eigenvalue decomposition of $\mathbf{R}$. The matrix $\mathbf{\Lambda}$ contains the eigenvalues $\lambda_1 \geq \lambda_2 \geq \dots \lambda_m$ on the diagonal and the matrix $\mathbf{Q}$ has the eigenvectors of the covariance as its columns. Note that $\mathbf{Q}^T$ is a rotation matrix that rotates and aligns the canonical bases of the observation space with the eigenvectors of the covariance, i.e., $\mathbf{Q}^T$ performs unscaled whitening. We can see from the last expression in \eqref{x_norm} that $x^Q_i$ for smaller values of $i$ gets attenuated by a larger $\lambda_i$ than $x^Q_i$ for larger values of $i$ (since $\lambda_1 \geq \lambda_2 \geq \dots \lambda_m$). This implies that subspaces (and signals) with more energy in lower indices after unscaled whitening, suffer more attenuation and have a lower $\| \bar{\mathbf{x}} \|_2$.
Thus, subspaces (signals) closer to the higher-order eigenvectors of the covariance (i.e., eigenvectors corresponding to higher eigenvalues) end up having a lower $\| \bar{\mathbf{x}} \|_2$.

With slight algebraic manipulations, we can see also that $\| \bar{\mathbf{x}} \|_2$ (and other terms proportional to it) appears in the numerator of our test statistics. This dictates that for same signal-to-noise ratio (SNR), i.e., SNR $= \frac{\| \mathbf{x} \|_2^2}{\sigma^2}$, a lower $\| \bar{\mathbf{x}} \|_2$ will result in a lower detection probability. Thus we conclude that for the same SNR, subspaces with more energy closer to the higher-order eigenvectors of the covariance have lower detection probability and vice versa. This make intuitive sense: a subspace with more influence of noise (i.e., a subspace that lives closer to the higher-order eigenvectors of the covariance) will have a lower detection rate than a subspace with less influence of noise. A depiction of this observation is shown in Fig.\,\ref{fig:ASD:discuss:noise_geometry}.

Since the same quadratic forms appear in the numerator of the test statistics for active subspace detection, we also conclude from this discussion that the subspaces with more energy near the higher-order eigenvectors of the covariance have lower probability of correct classification. 

\section{Numerical Experiments} \label{num_sims}
In this section, we present numerical experiments to examine the tightness of various bounds derived in this paper and verify the trends of performance metrics with respect to the geometry of the subspaces.

\begin{figure} [!ht]
	\begin{center}
		\begin{tabular}{c}
			{\includegraphics[width=\columnwidth] {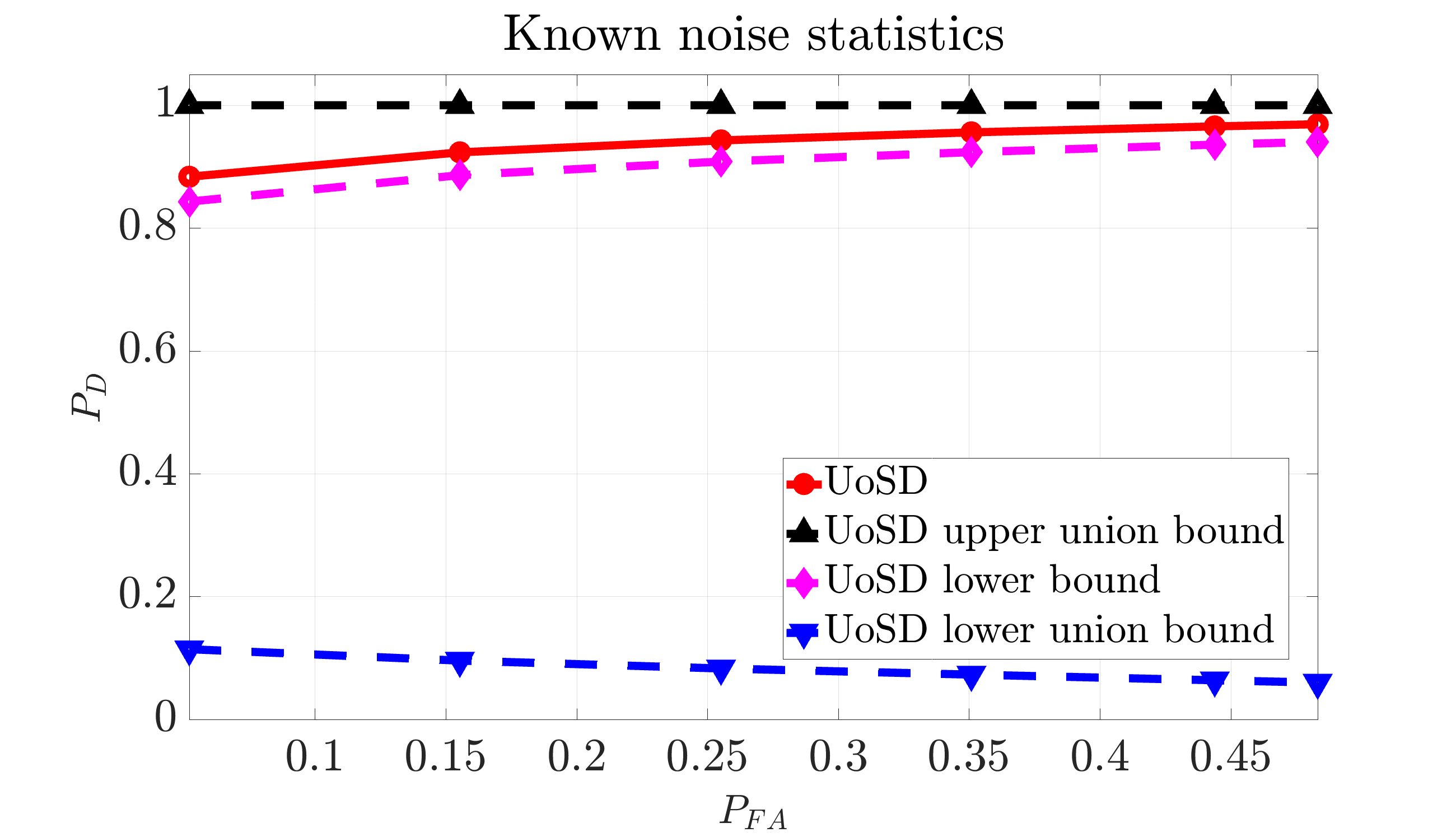}}\\
			{\includegraphics[width=\columnwidth]  {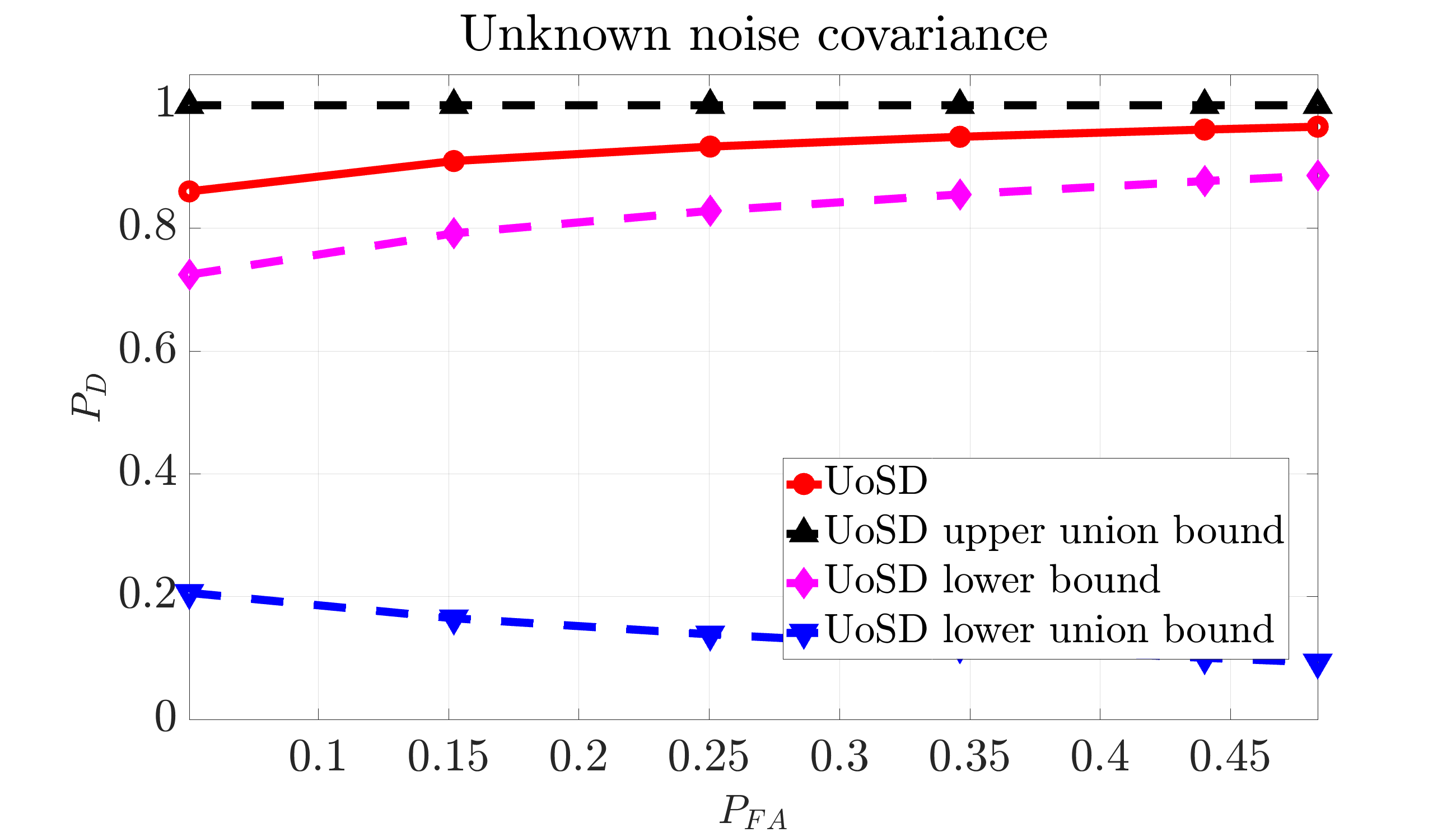}}\\
			{\includegraphics[width=\columnwidth]  {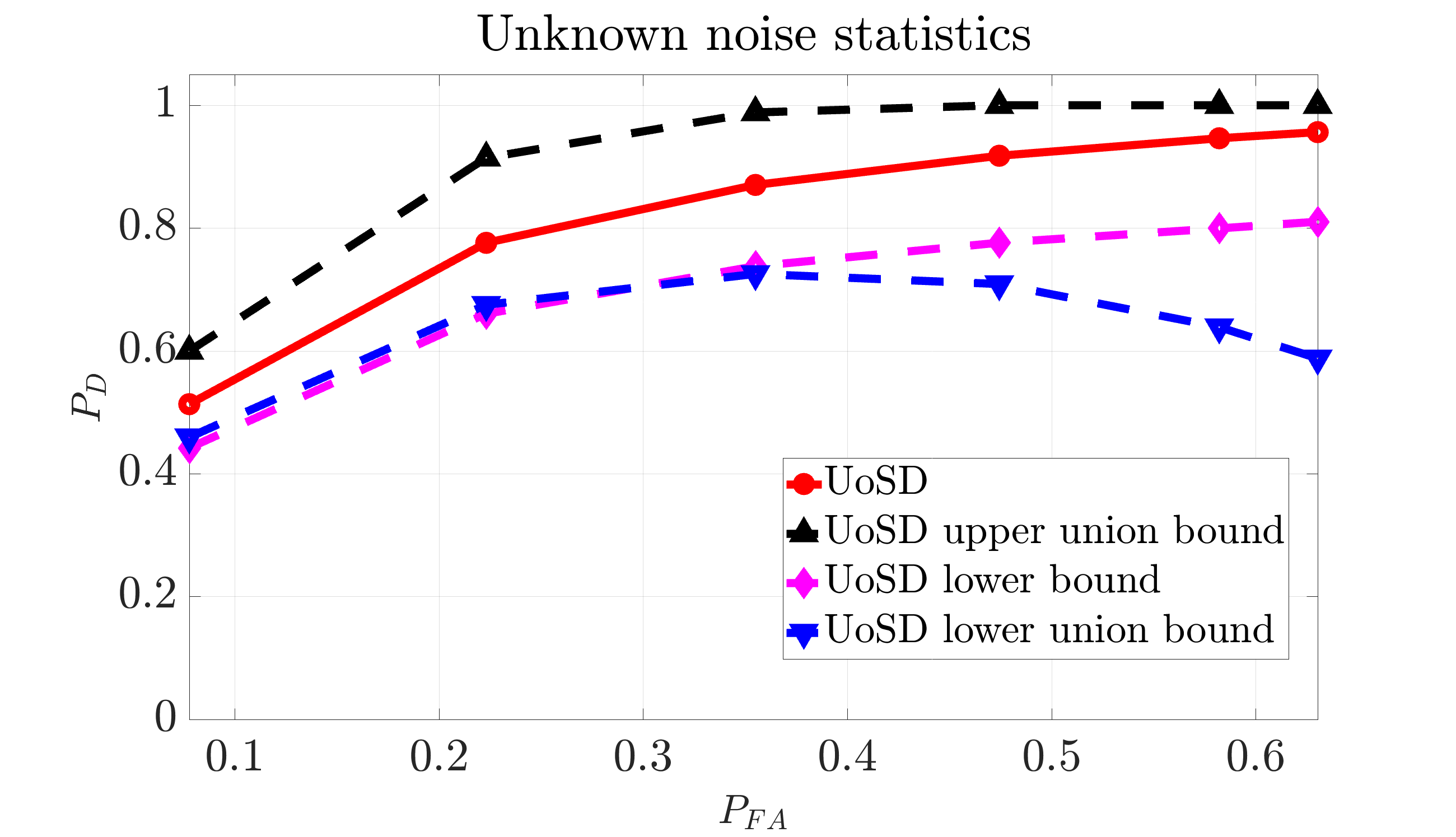}}
		\end{tabular}
	\end{center}
	\caption{ \label{fig:num_sims:MSD:ROC}
		ROC curves for signal detection under the UoS model (labeled UoSD) and the derived bounds. Each subfigure shows four plots under the UoS model: the upper union bound on the detection probability, the true detection probability, the lower bound on the detection probability, and the lower union bound. Starting from the top, the subfigures show the ROC curves under known noise statistics, unknown noise covariance and unknown noise statistics, respectively.}
\end{figure}

\subsection{Synthetic data} \label{num_sims:monte_carlo}
We run Monte-Carlo experiments for signal and active subspace detection problems under different noise settings using synthetic data. Our general procedure for these experiments is as follows: we consider a union of \textit{three} $2$-dimensional subspaces in a 4-dimensional space. The subspaces are structured to highlight the effect of geometry between subspaces. The first and third subspaces are fixed and the angles between them are kept constant. As for the second subspace, we make different realizations of it with increasing principal angles with respect to the first subspace. This process is repeated for different levels of false alarm probabilities and SNR levels. The threshold for each false alarm level is determined numerically. When unmentioned, the false alarm rate is upper bounded at $10^{-1}$ and the SNR is 10 dB. Each experiment is averaged over 10000 trials.

\subsubsection{Signal detection problem}
\begin{figure} [!ht]
	\begin{center}
		\begin{tabular}{c}
			{\includegraphics[width=\columnwidth] {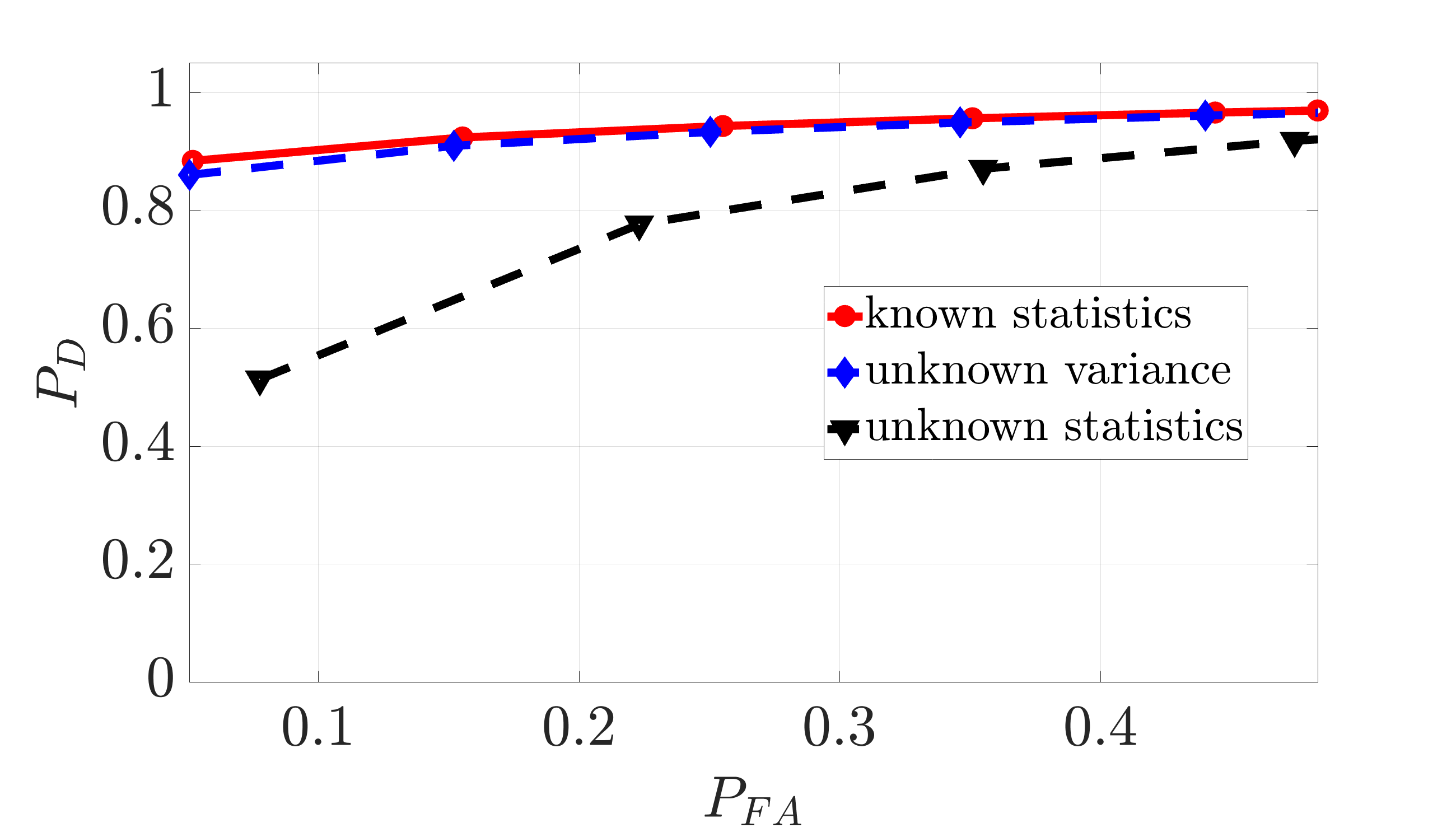}} \\
			{\includegraphics[width=\columnwidth] {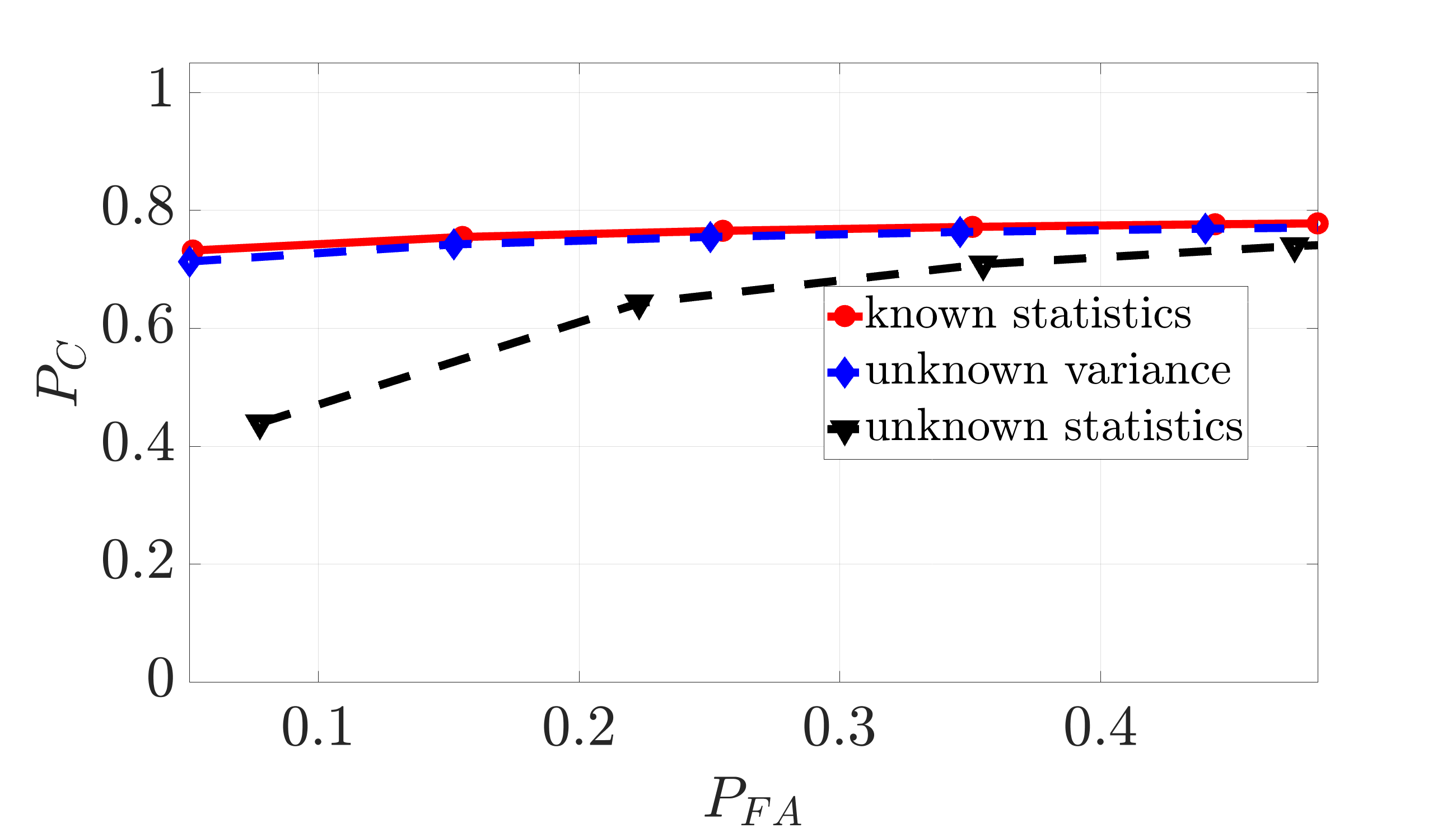}}
		\end{tabular}
	\end{center}
	\caption{ \label{fig:num_sims:MSD:ROC_comparison}
		ROC curves for signal (top) and active subspace (bottom) detection under the UoS model for different noise settings.}
\end{figure}

The receiver operating characteristic (ROC) curve of signal detection for the tests derived in this paper, with their respective upper and lower bounds, is given in Fig.~\ref{fig:num_sims:MSD:ROC}. We can see that the lower union bound is much looser compared to the upper union bound and the lower bound derived in the paper. Moreover, Fig.~\ref{fig:num_sims:MSD:ROC_comparison} provides a comparison of different noise scenarios, from which we conclude that the best performance is given under known noise statistics.

\begin{figure*} [!ht]
	\begin{center}
		\begin{tabular}{c c c}
			{\includegraphics[height=3.43cm] {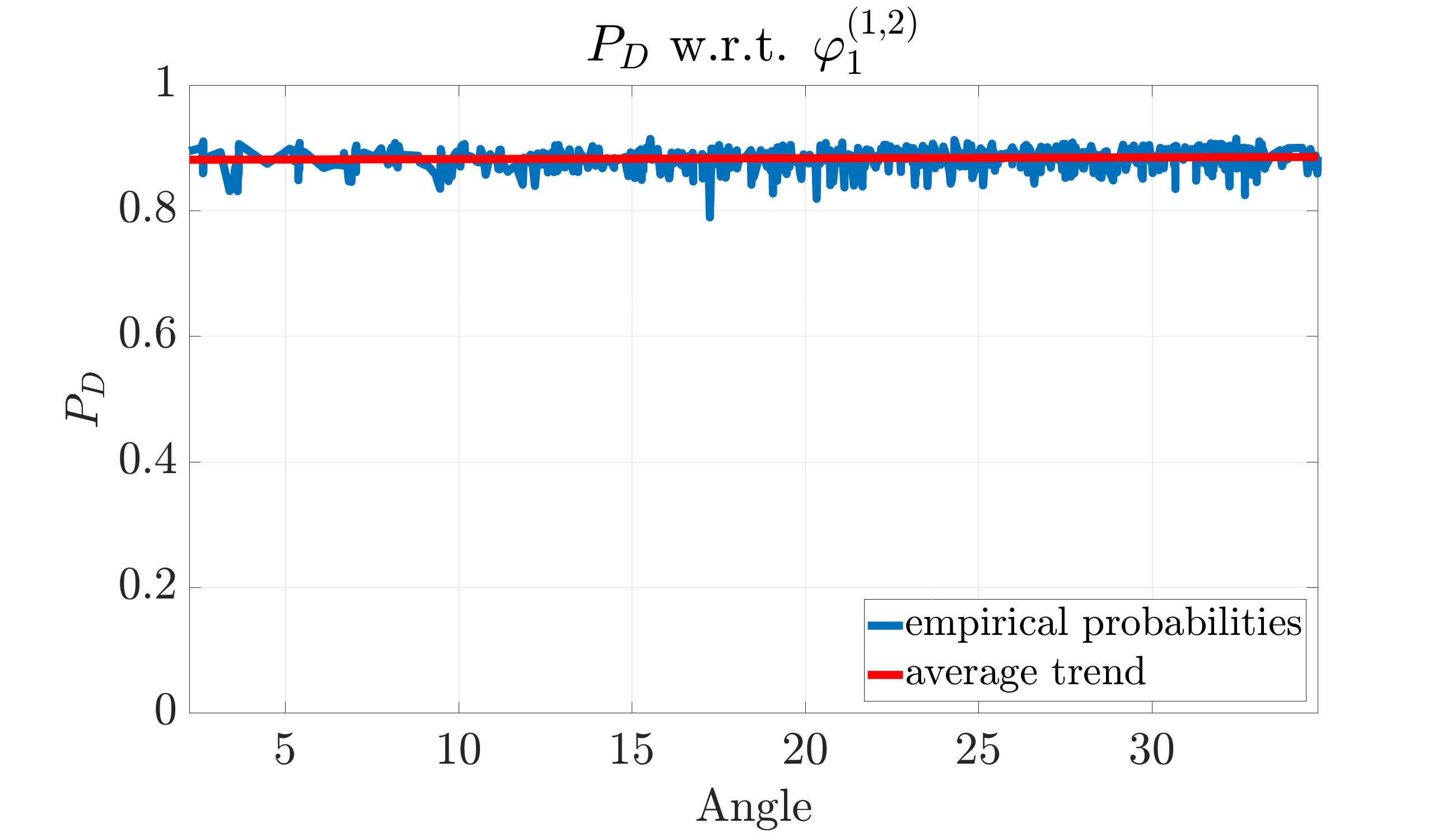}} &
			{\includegraphics[height=3.43cm] {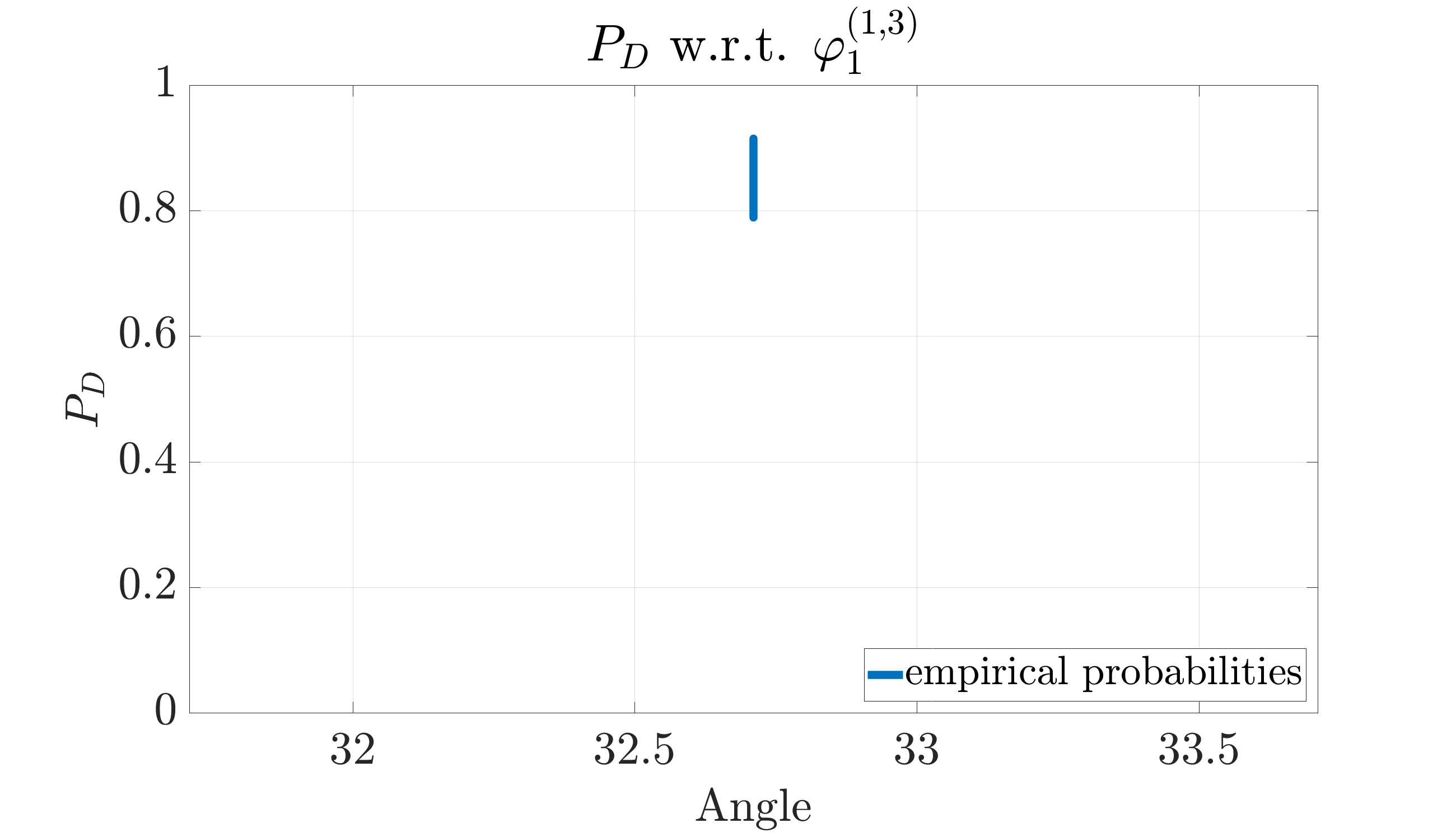}} &
			{\includegraphics[height=3.43cm] {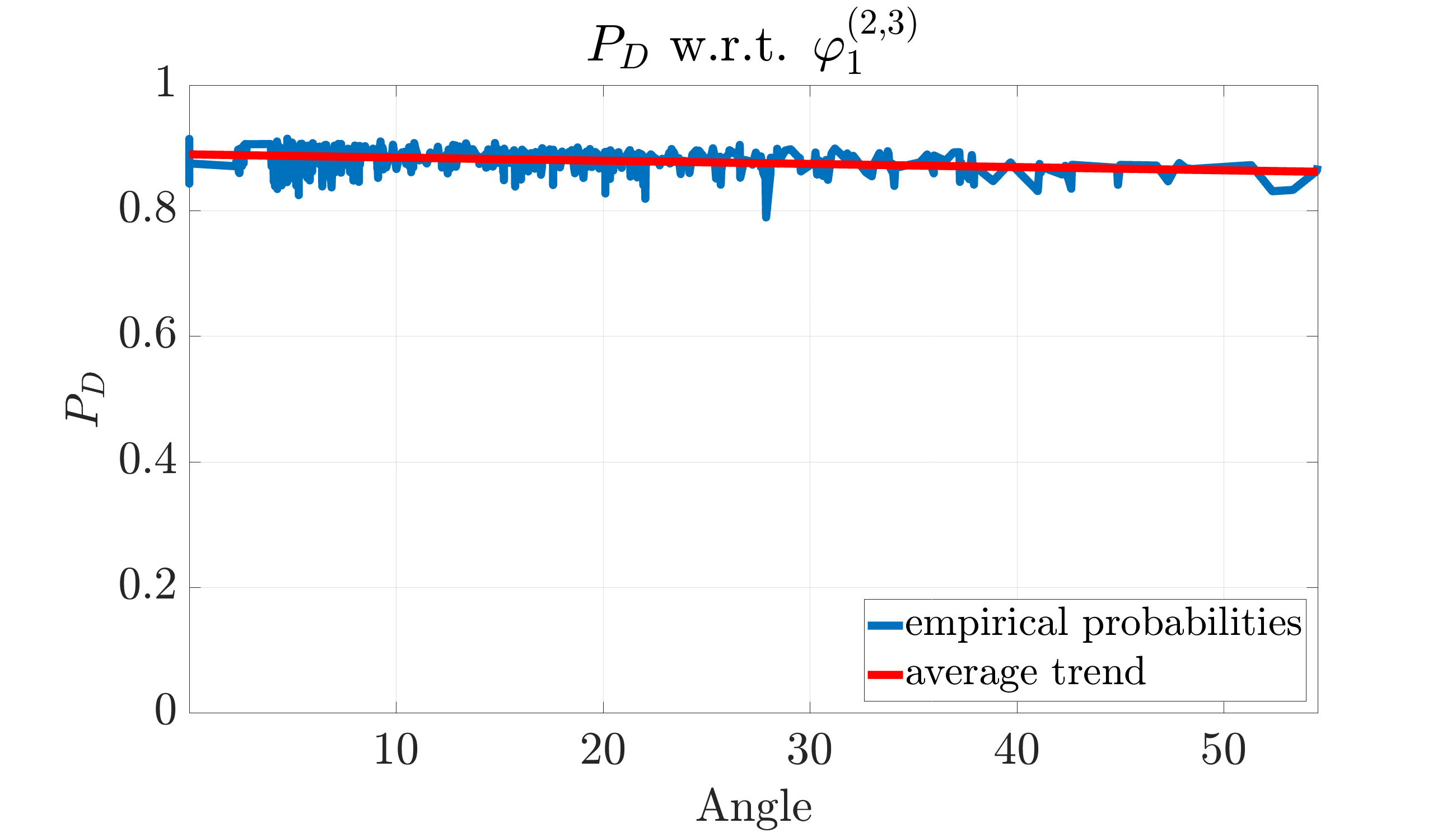}}
			\\
			{\includegraphics[height=3.43cm] {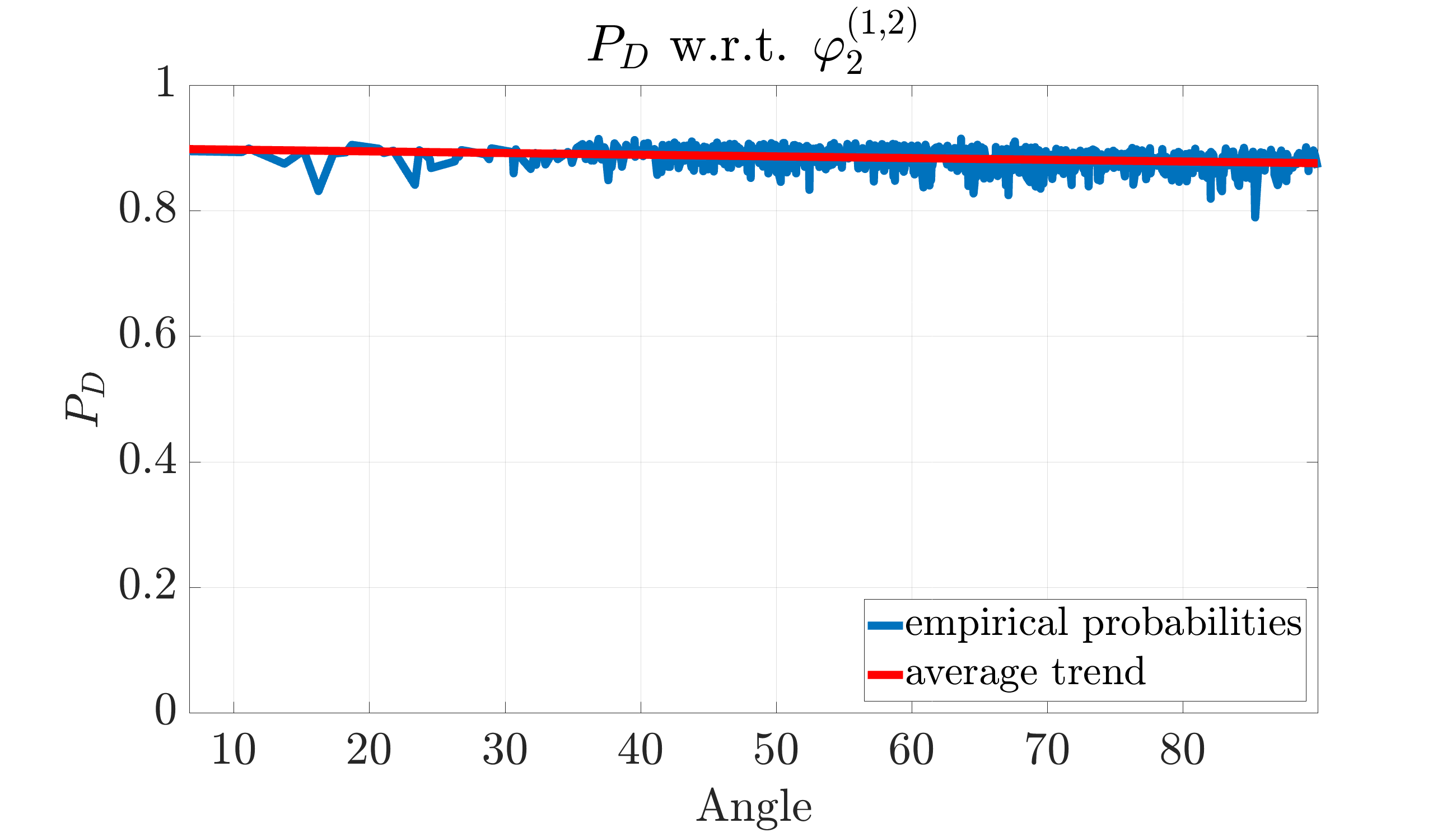}} &
			{\includegraphics[height=3.43cm] {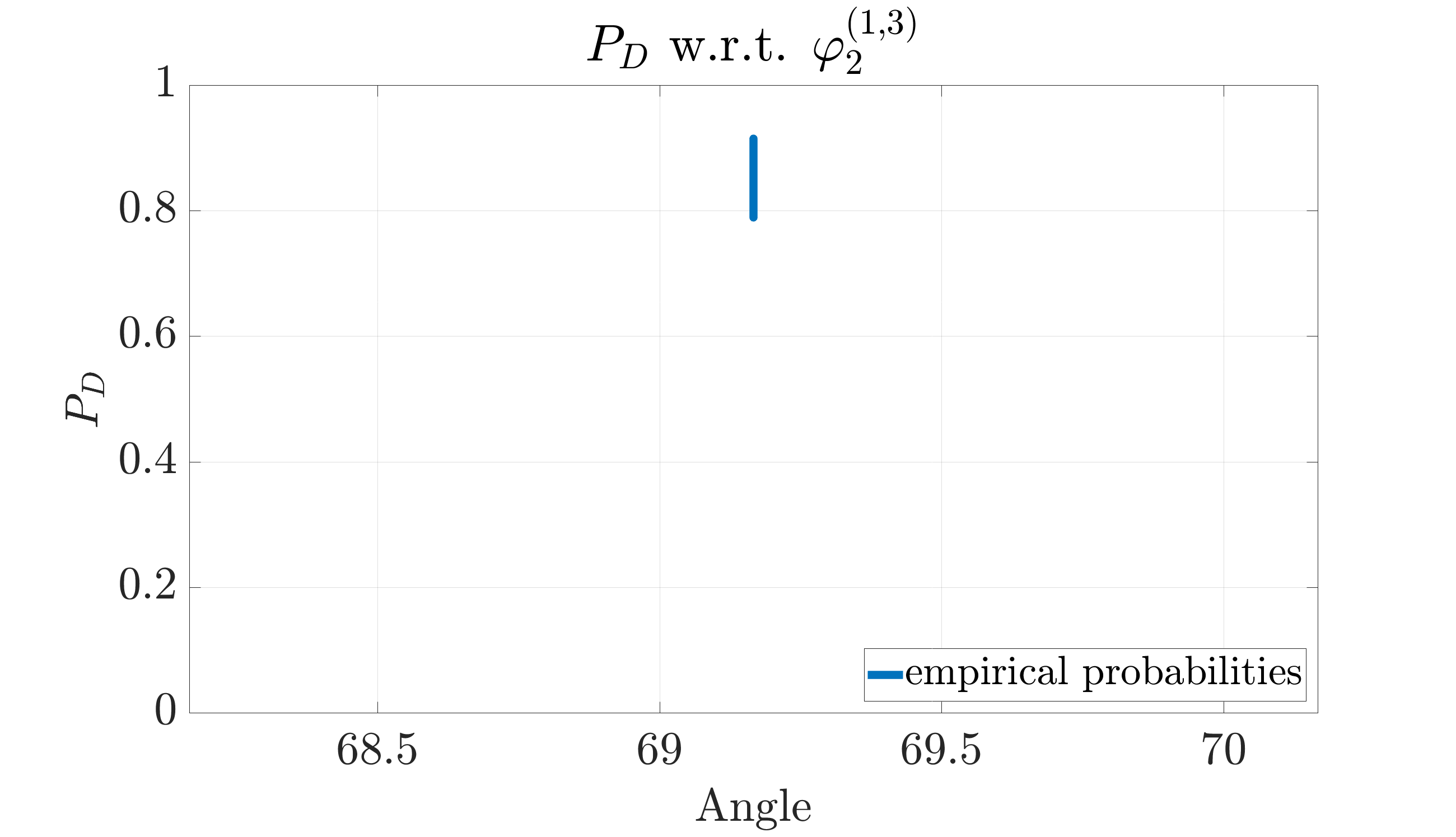}} &
			{\includegraphics[height=3.43cm] {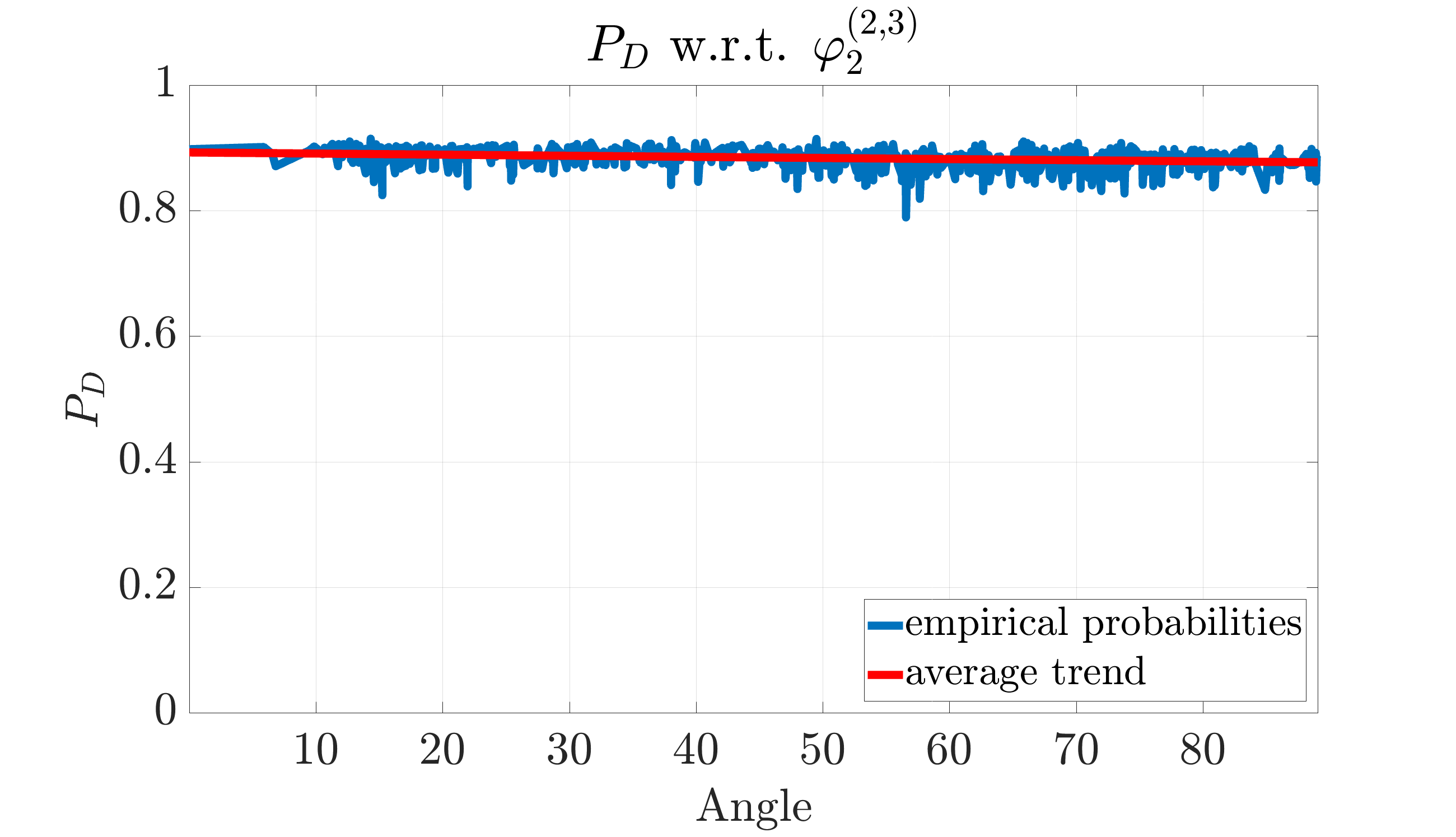}}
		\end{tabular}
	\end{center}
	\caption[example]
	{ \label{fig:num_sims:MSD:KN:P_D}
		The probability of detection with respect to the principal angles between whitened subspaces when the noise statistics are fully known. The angles/whitened angles between subspaces 1 and 3 are fixed, but the probabilities change due to changing angles with subspace 2, and thus we see a vertical line for the detection probability with respect to $\varphi_1^{(1,3)}$ and $\varphi_2^{(1,3)}$. For other angles, we see a minimal decrease in probability as the angles increase.}
\end{figure*}

Next, Fig.~\ref{fig:num_sims:MSD:KN:P_D} shows the effect of subspace angles on the detection probability. We see that the principal angles between whitened subspaces have indeed minimal effect on the detection probability under known noise settings. A similar behavior can also be seen for detection probability under other noise settings, but we omit those plots in the interest of space.

\begin{figure*} [!ht]
   	\begin{center}
   	\begin{tabular}{c c c}
   	{\includegraphics[height=3.43cm] {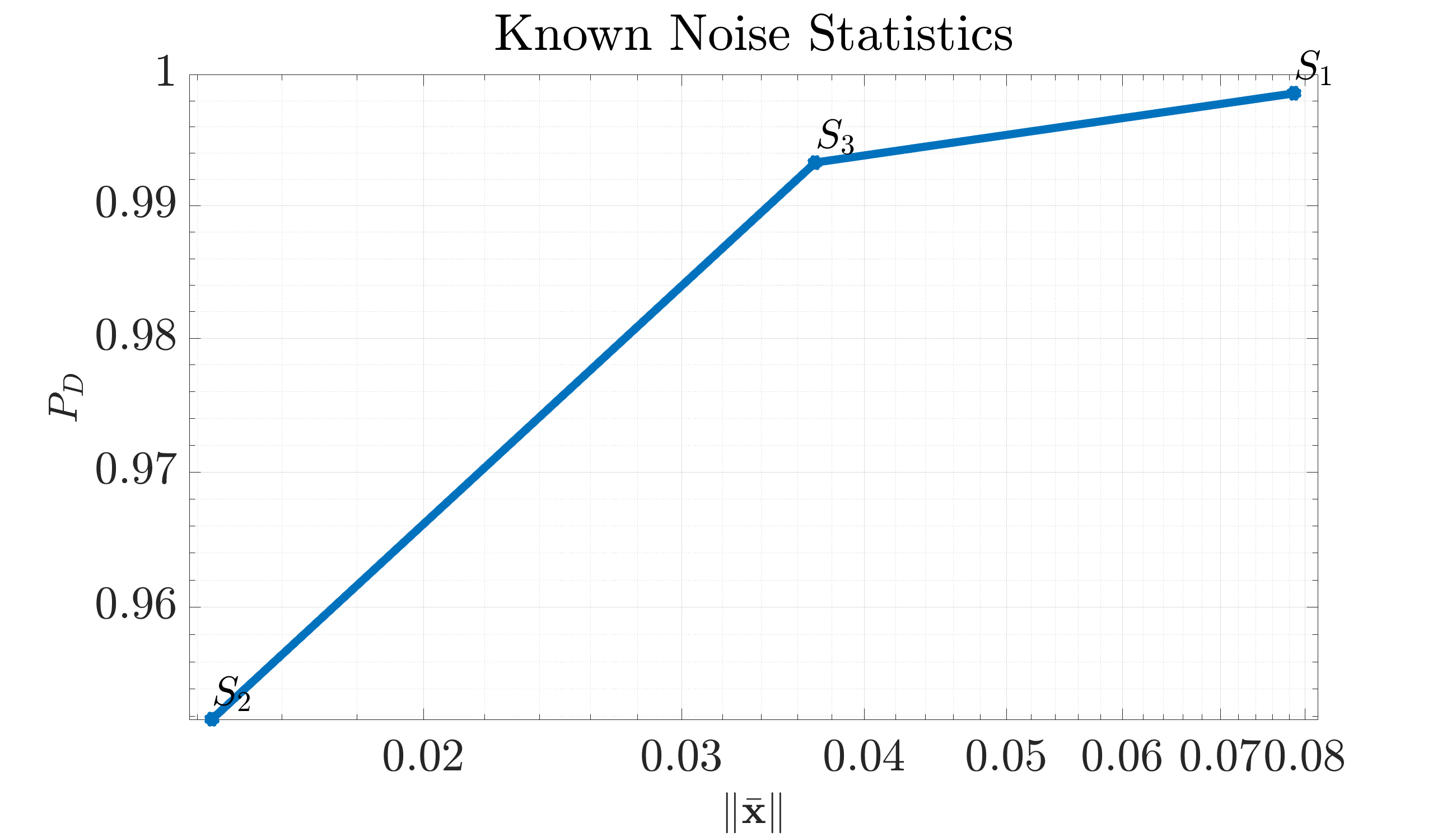}}
   	{\includegraphics[height=3.43cm] {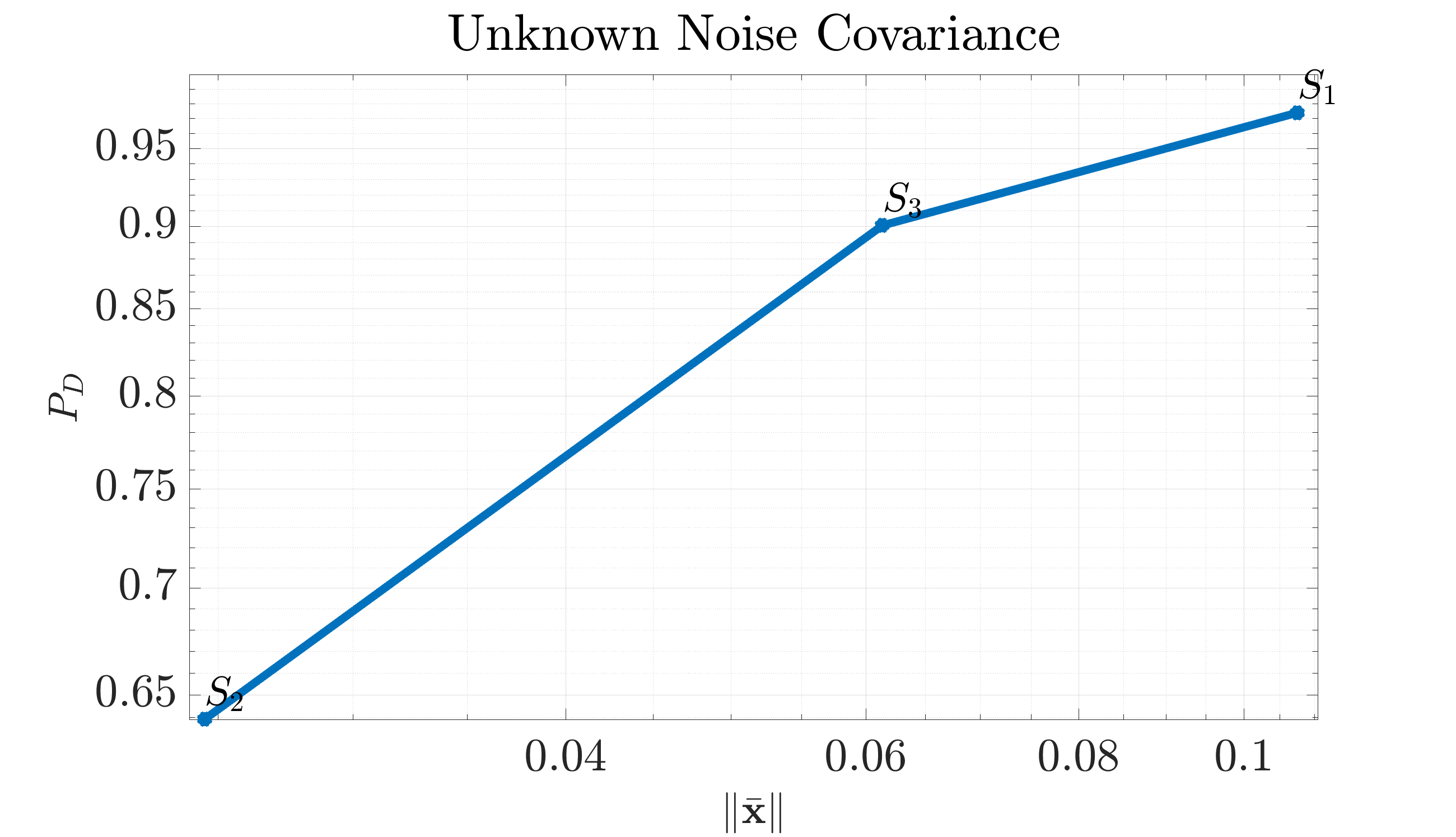}}
   	{\includegraphics[height=3.43cm] {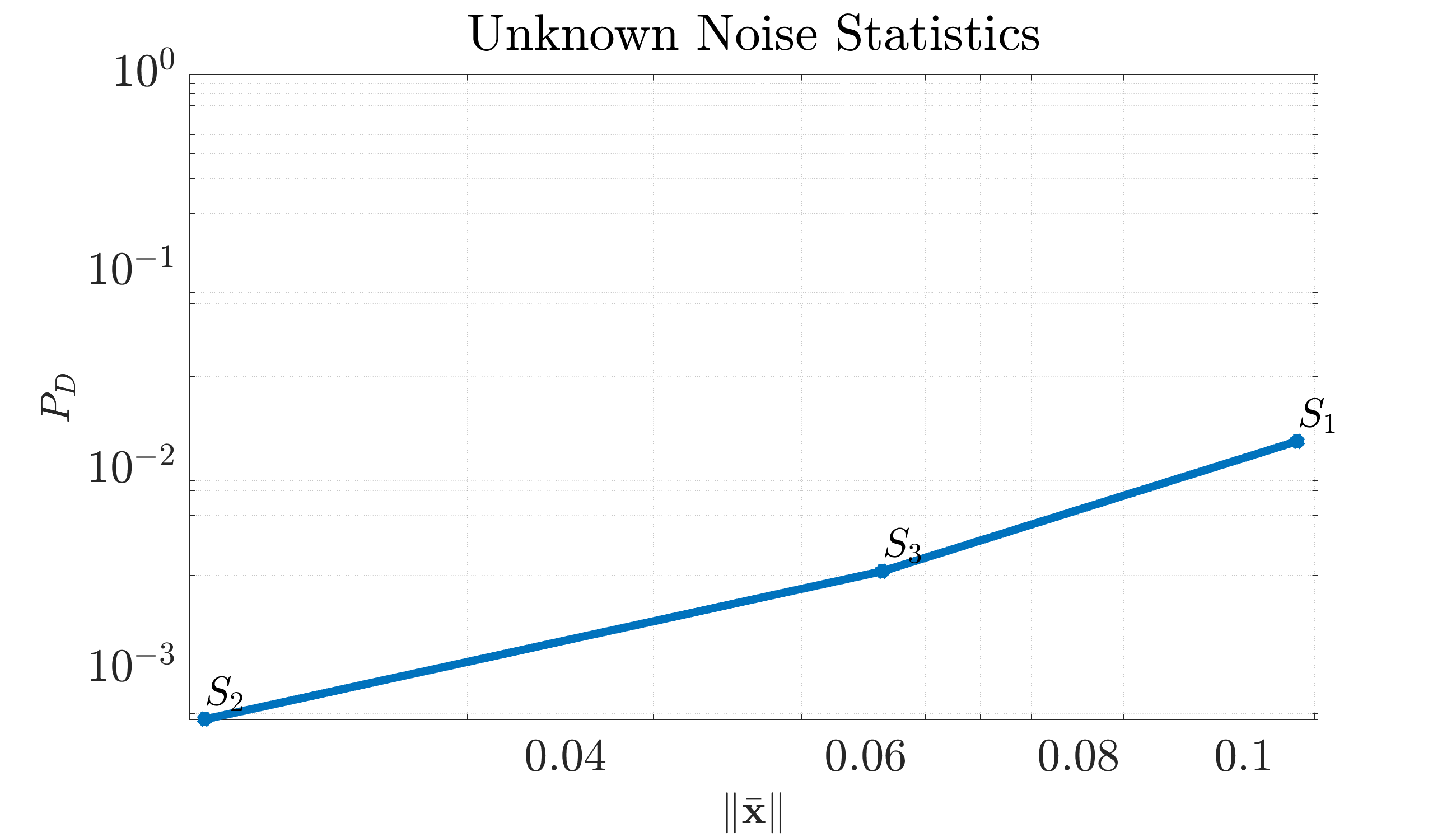}}
   	\end{tabular}
   	\end{center}
   	\caption{ \label{fig:num_sims:MSD:noise_geometry}
	Each subfigure shows that the closer a subspace is to the higher-order eigenvectors of the noise covariance, the lower is its detection probability. On the x-axis we have the average $\| \bar{\mathbf{x}} \|$ over 12500 random signals for each subspace and the on y-axis we have the detection probability. The subspace with bases closer to the higher-order eigenvectors has lower $\| \bar{\mathbf{x}} \|$ and thus lower detection probability.}
   	\end{figure*}

To show the influence of the geometry of noise, we consider three $2$-dimensional subspaces in a $4$-dimensional space and randomly generate a noise covariance matrix. We then add noise to the eigenvectors of the noise covariance matrix and use them as bases for two of our subspaces. Starting from the eigenvectors corresponding to the smallest eigenvalues, we successively pick $n$ noisy eigenvectors for subspaces $S_1$ and $S_2$ in the union. The bases of the third subspace $S_3$ are generated randomly from a standard normal distribution. We noted in Sec.~\ref{MSD:KN:influence:noise} that subspaces with more basis vectors closer to the higher-order eigenvectors of the noise covariance have lower $\| \bar{\mathbf{x}} \|_2$ and thus a lower detection probability, and vice versa. This trend can be clearly seen in Fig.~\ref{fig:num_sims:MSD:noise_geometry} for signal detection under each noise setting.

\subsubsection{Active subspace detection problem}
\begin{figure*} [!ht]
   	\begin{center}
   	\begin{tabular}{c c c}
   	{\includegraphics[height=3.43cm] {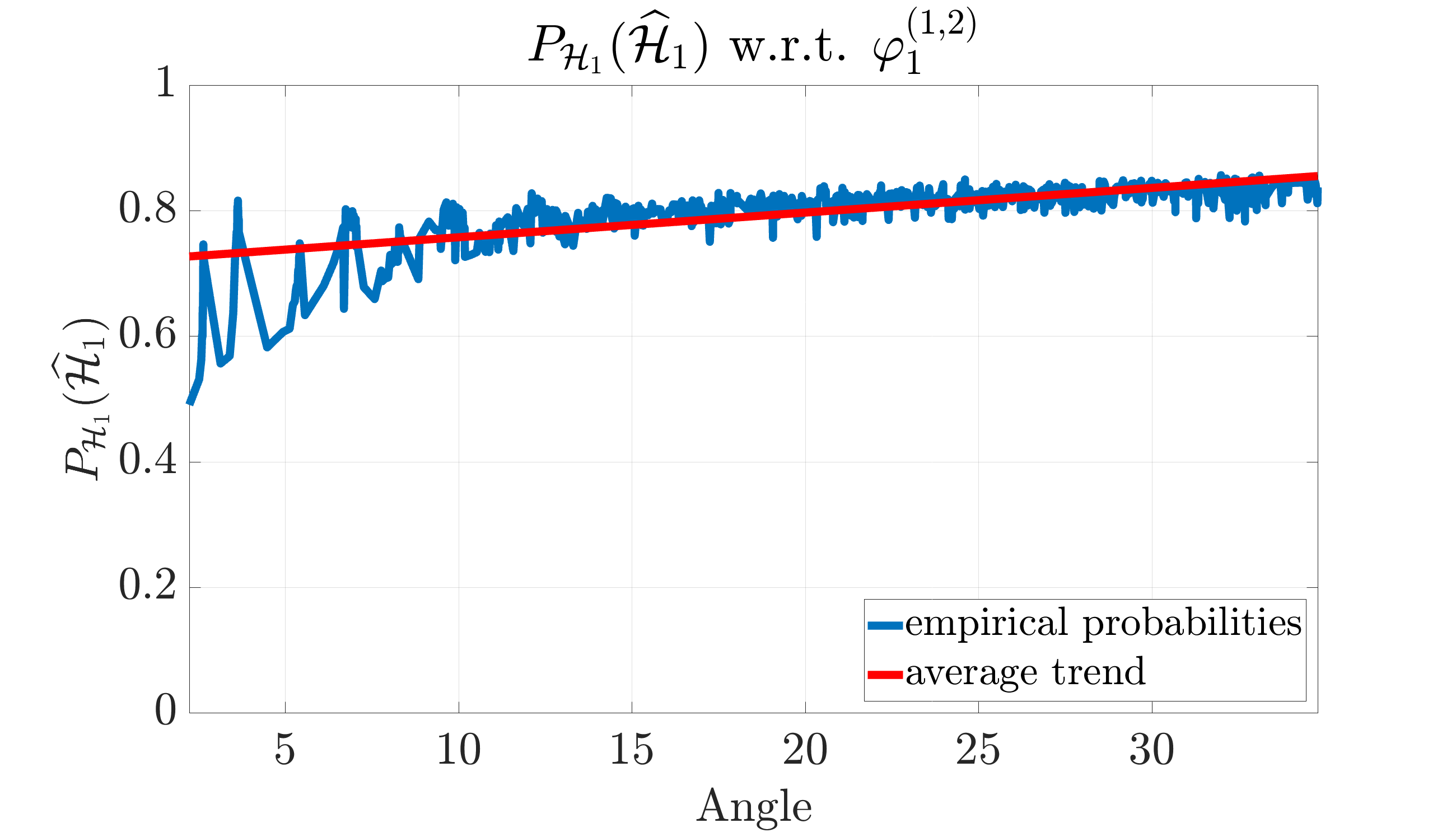}} &
   	{\includegraphics[height=3.43cm]
   	{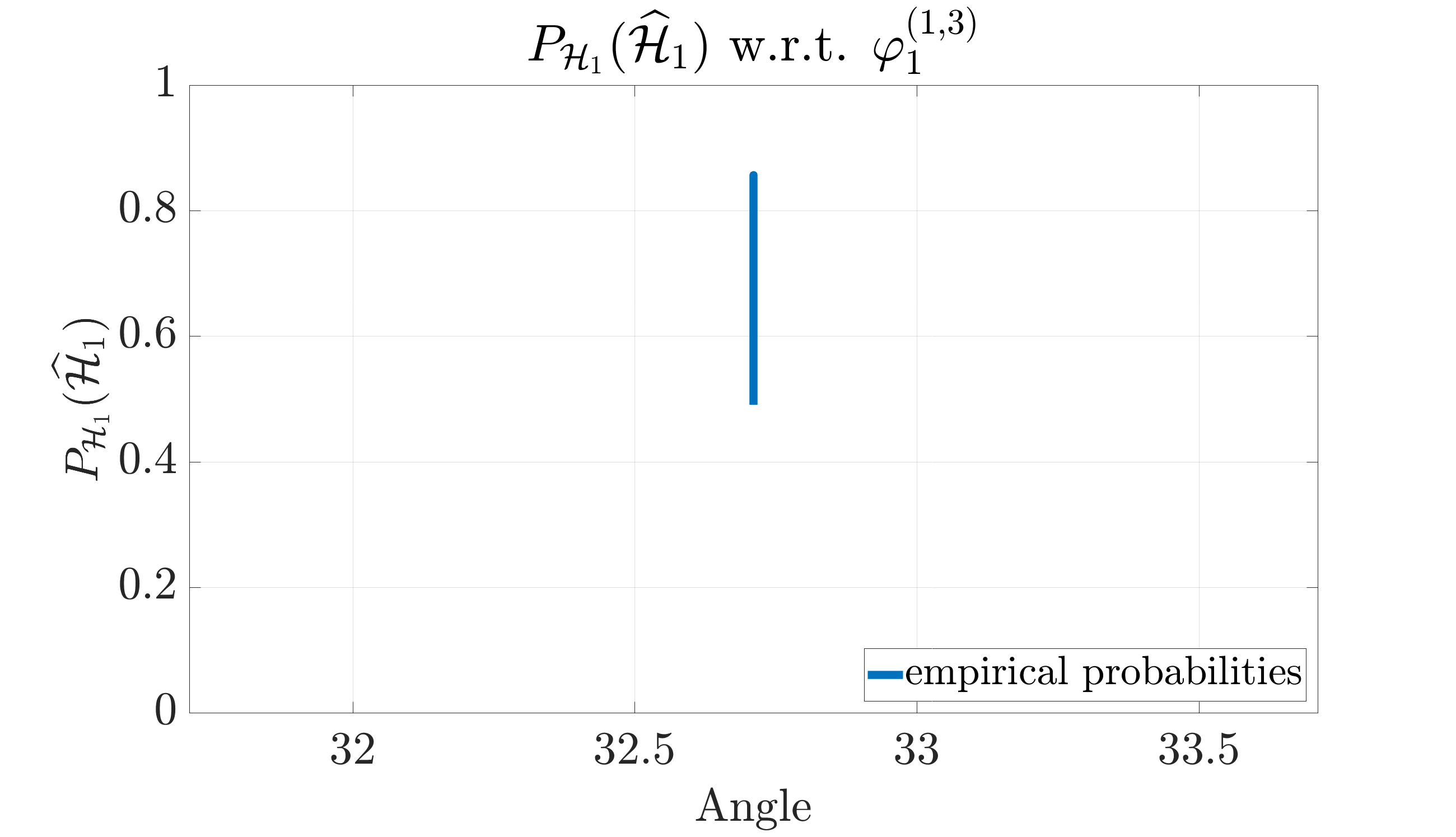}} &
    {\includegraphics[height=3.43cm]
   	{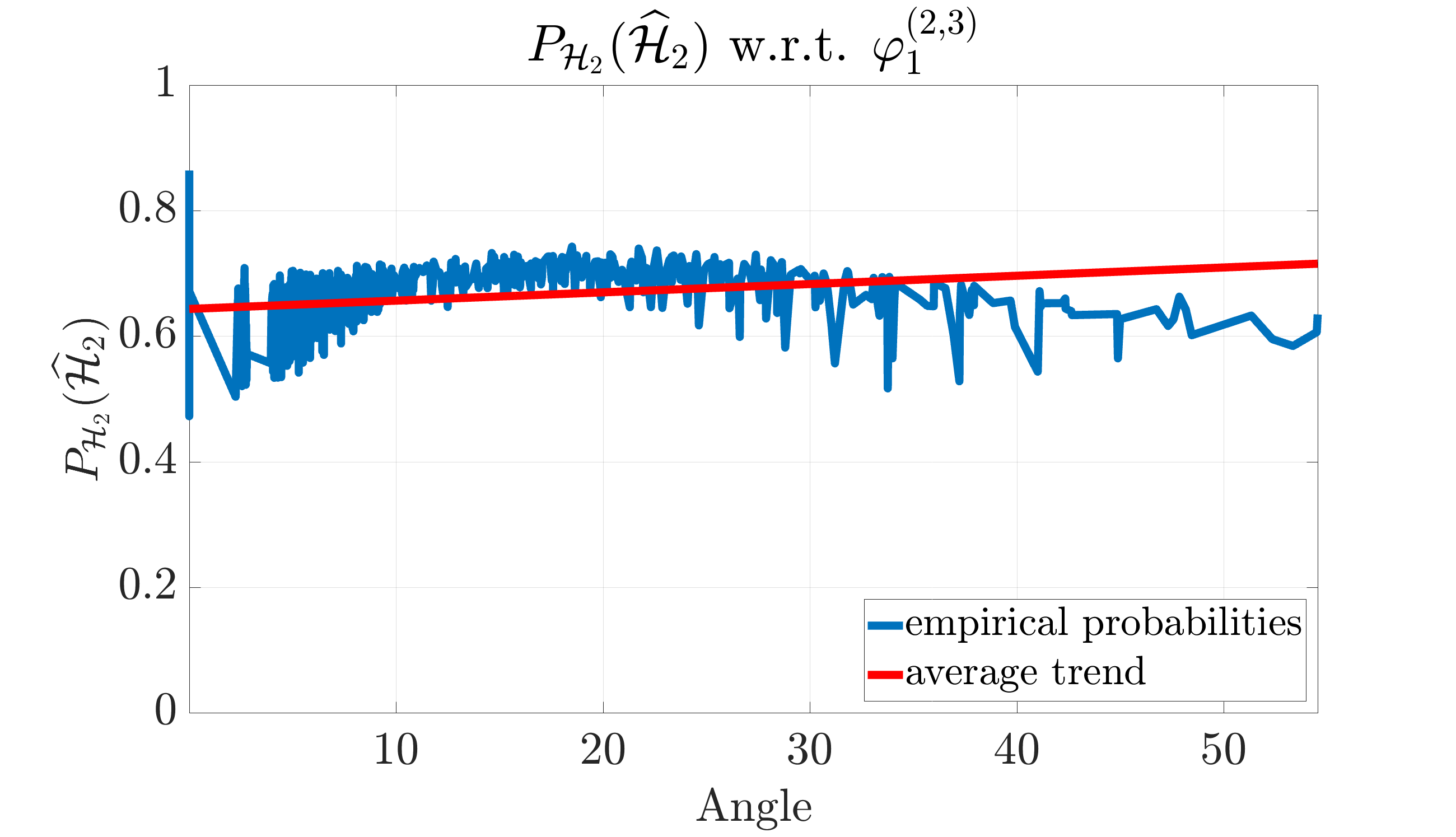}}
   	\\
	{\includegraphics[height=3.43cm] {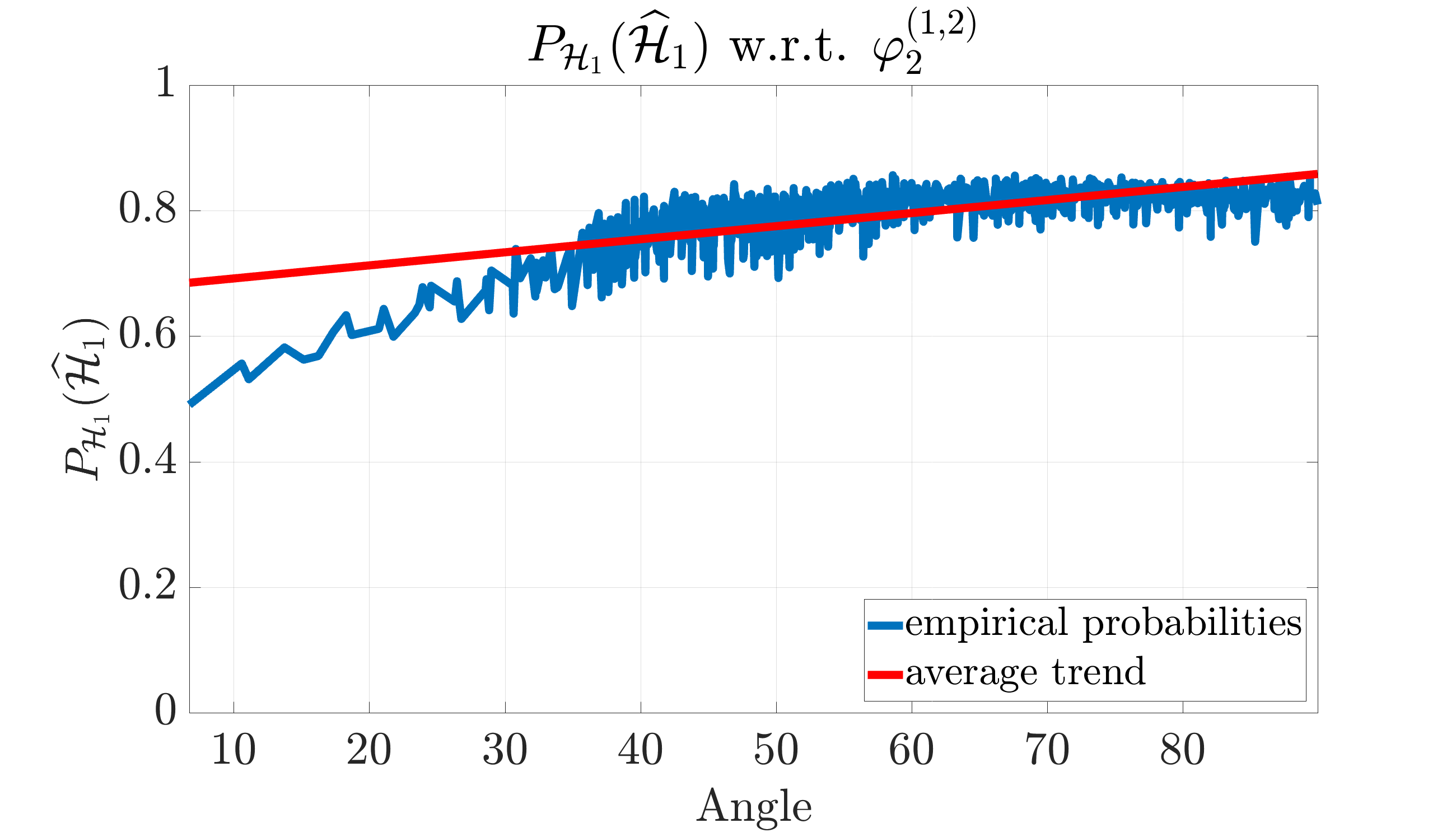}} &
	{\includegraphics[height=3.43cm] {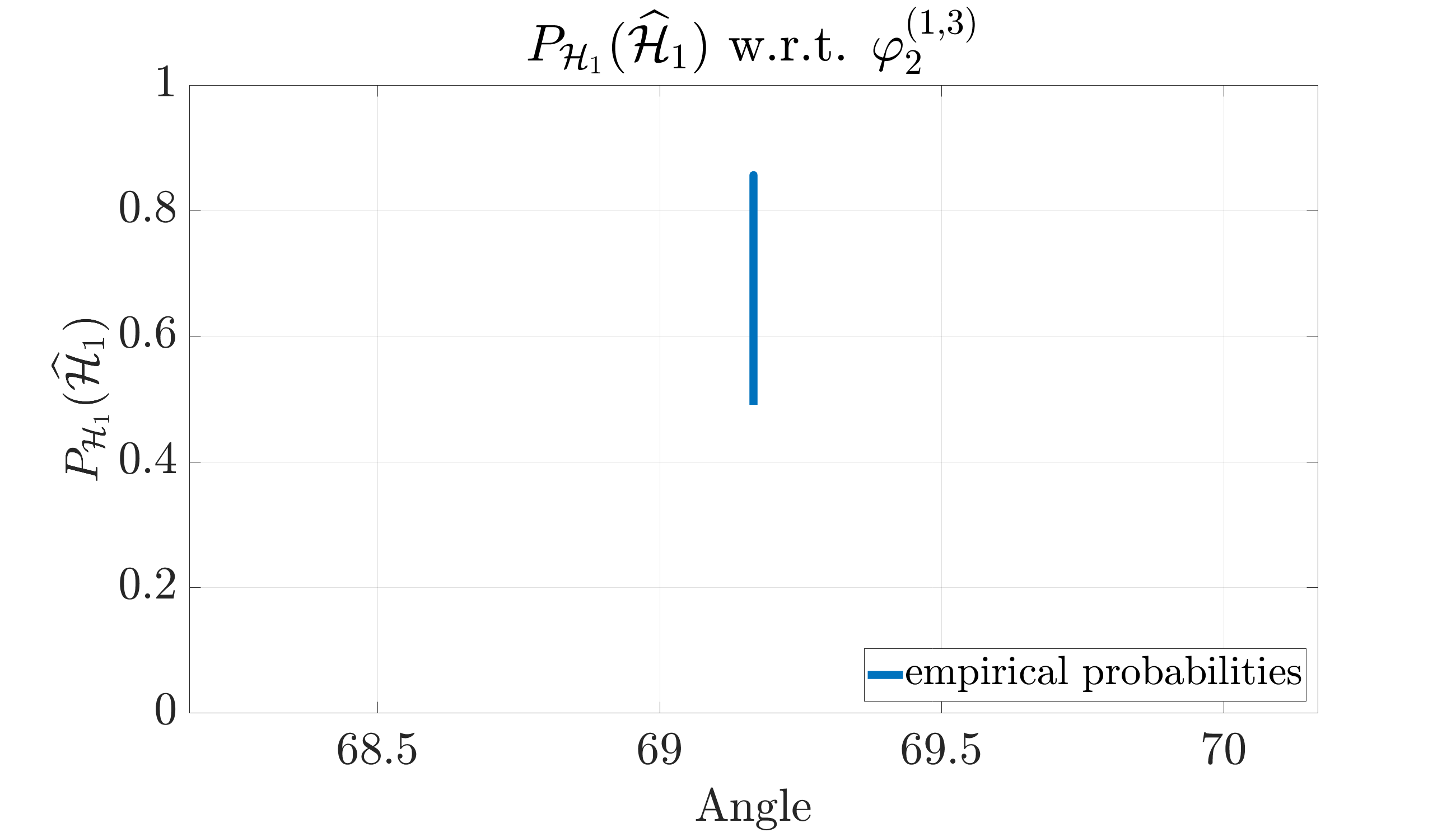}} &
	{\includegraphics[height=3.43cm]
	{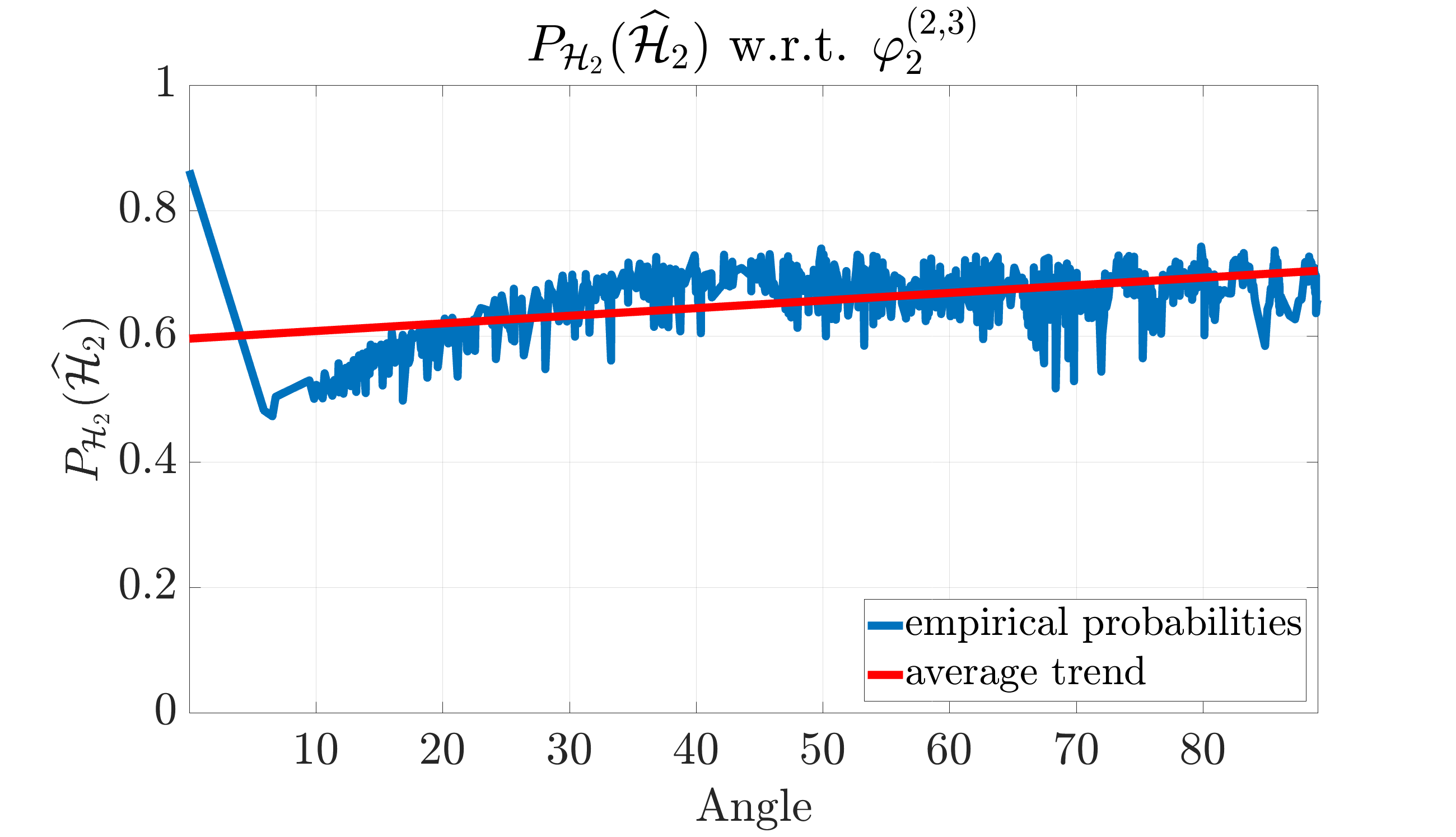}}
	\\
	{\includegraphics[height=3.43cm] {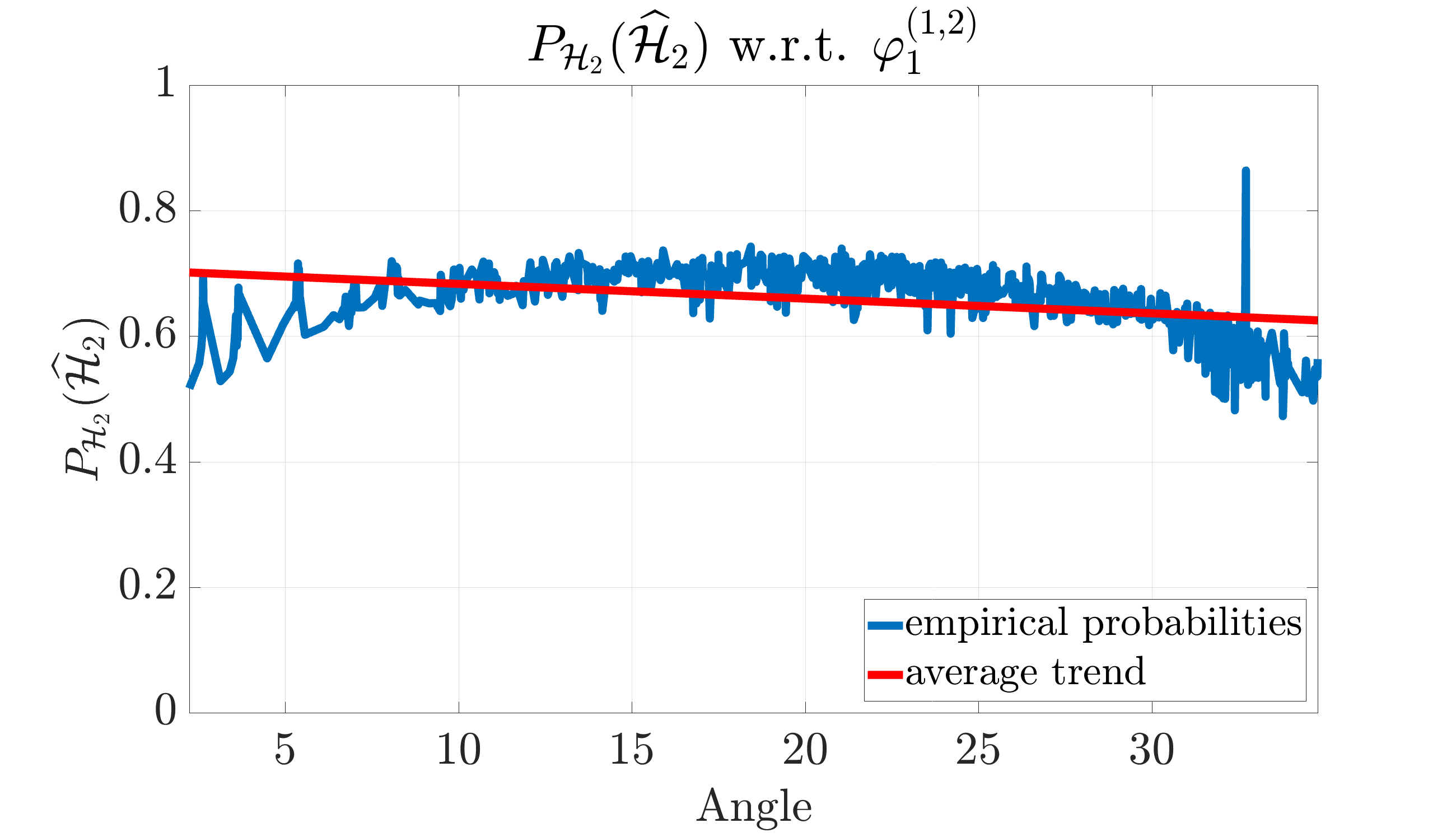}} &
	{\includegraphics[height=3.43cm] {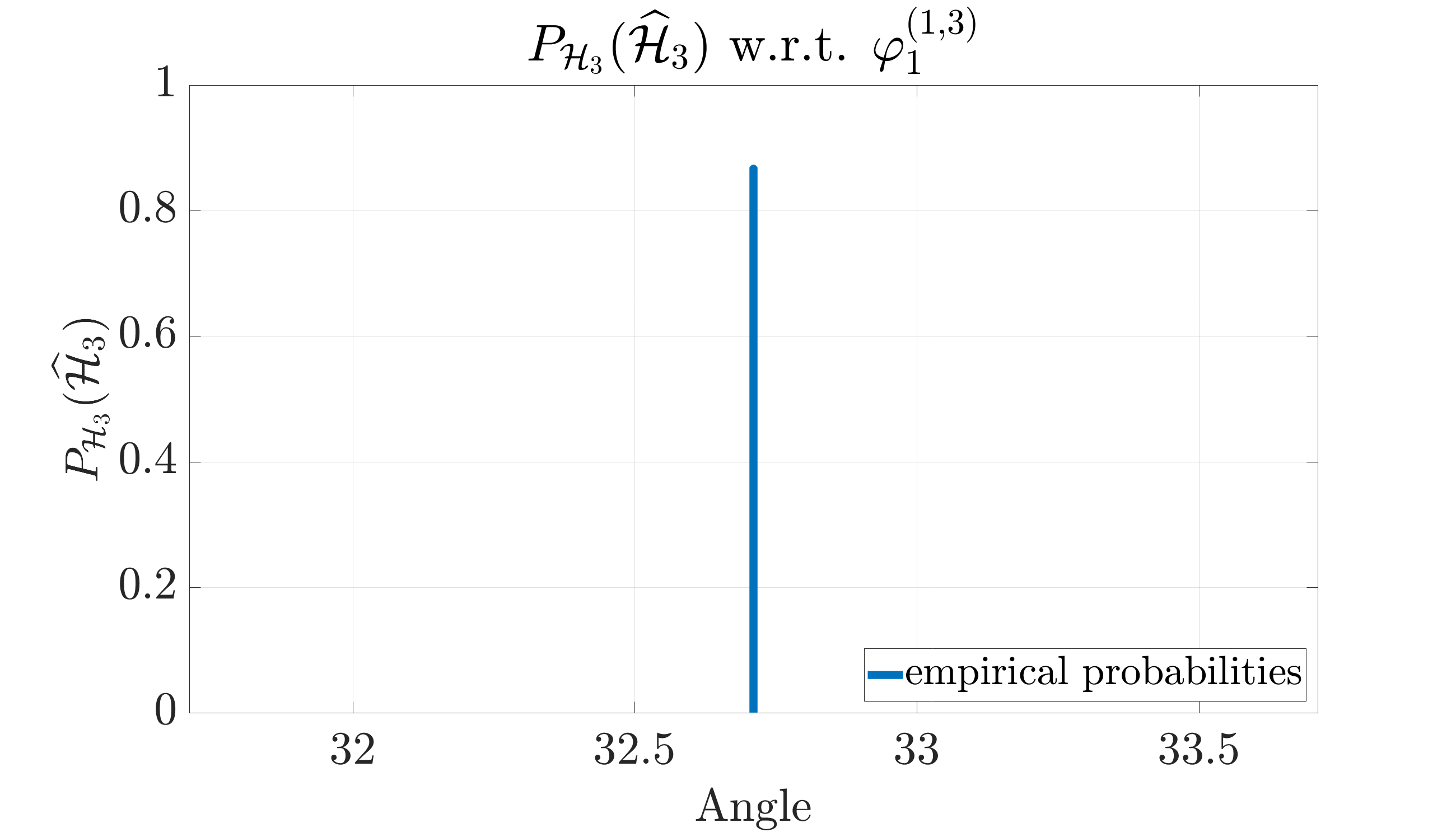}} &
	{\includegraphics[height=3.43cm]
	{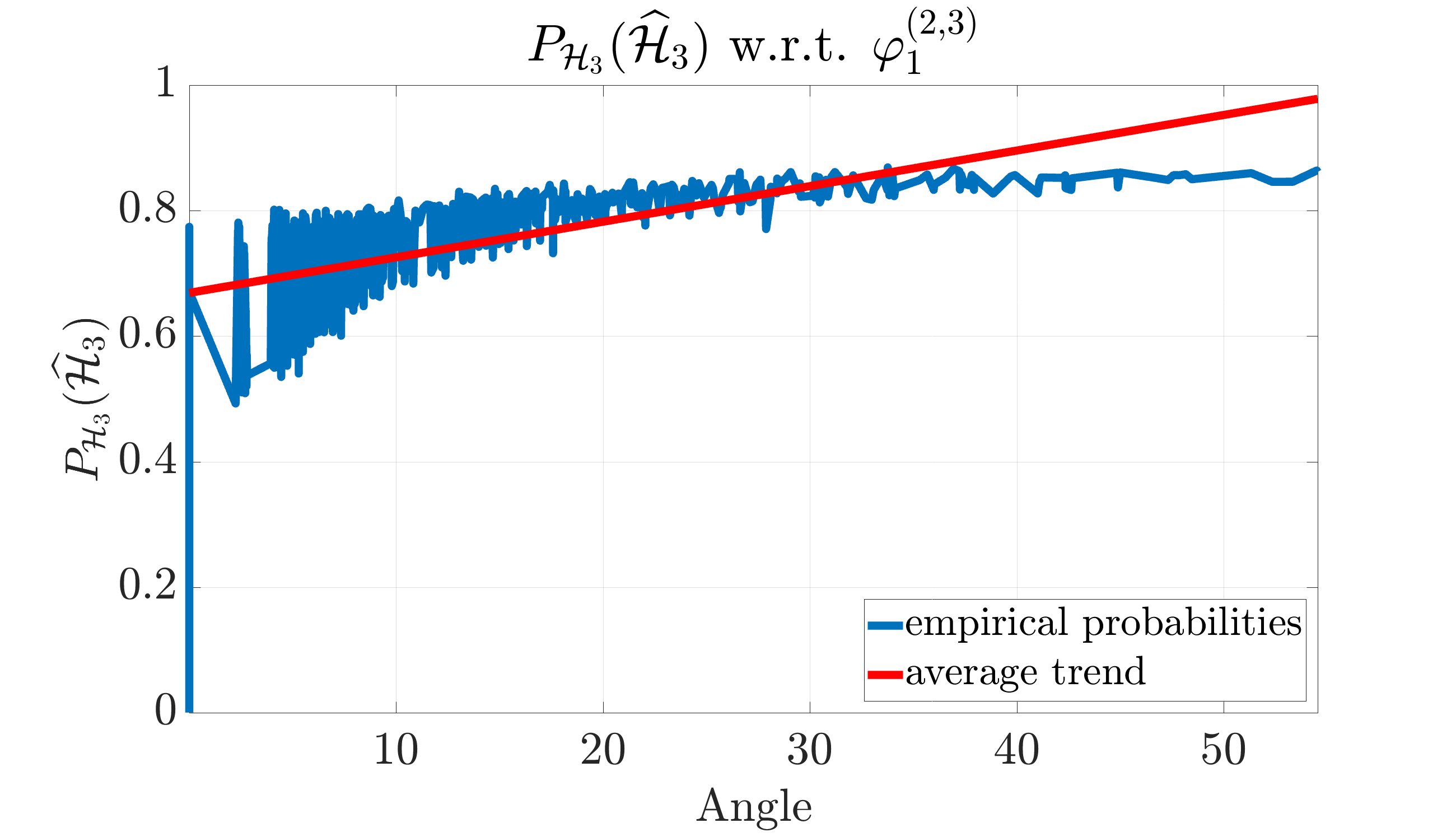}}
	\\
	{\includegraphics[height=3.43cm] {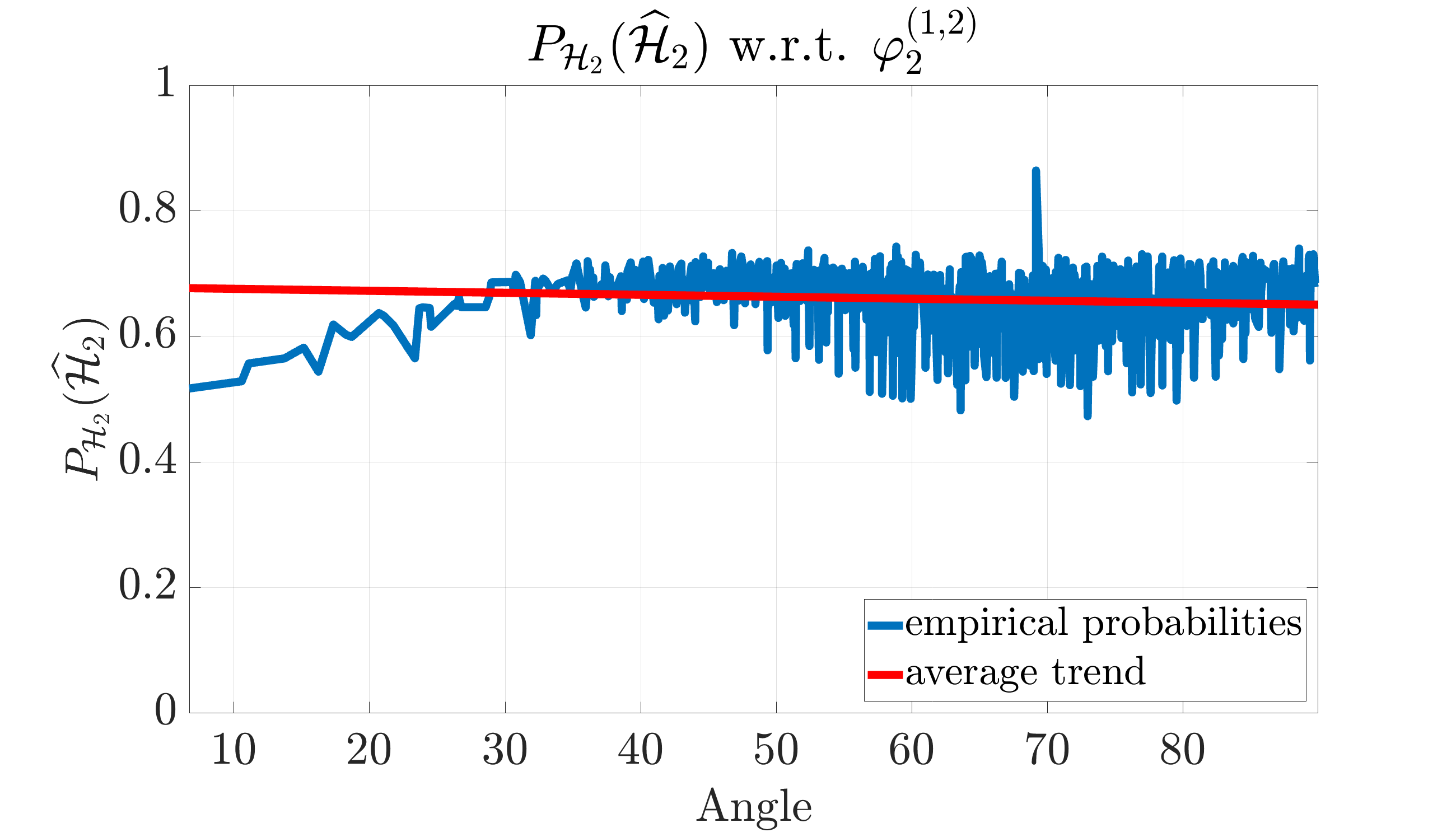}} &
	{\includegraphics[height=3.43cm] {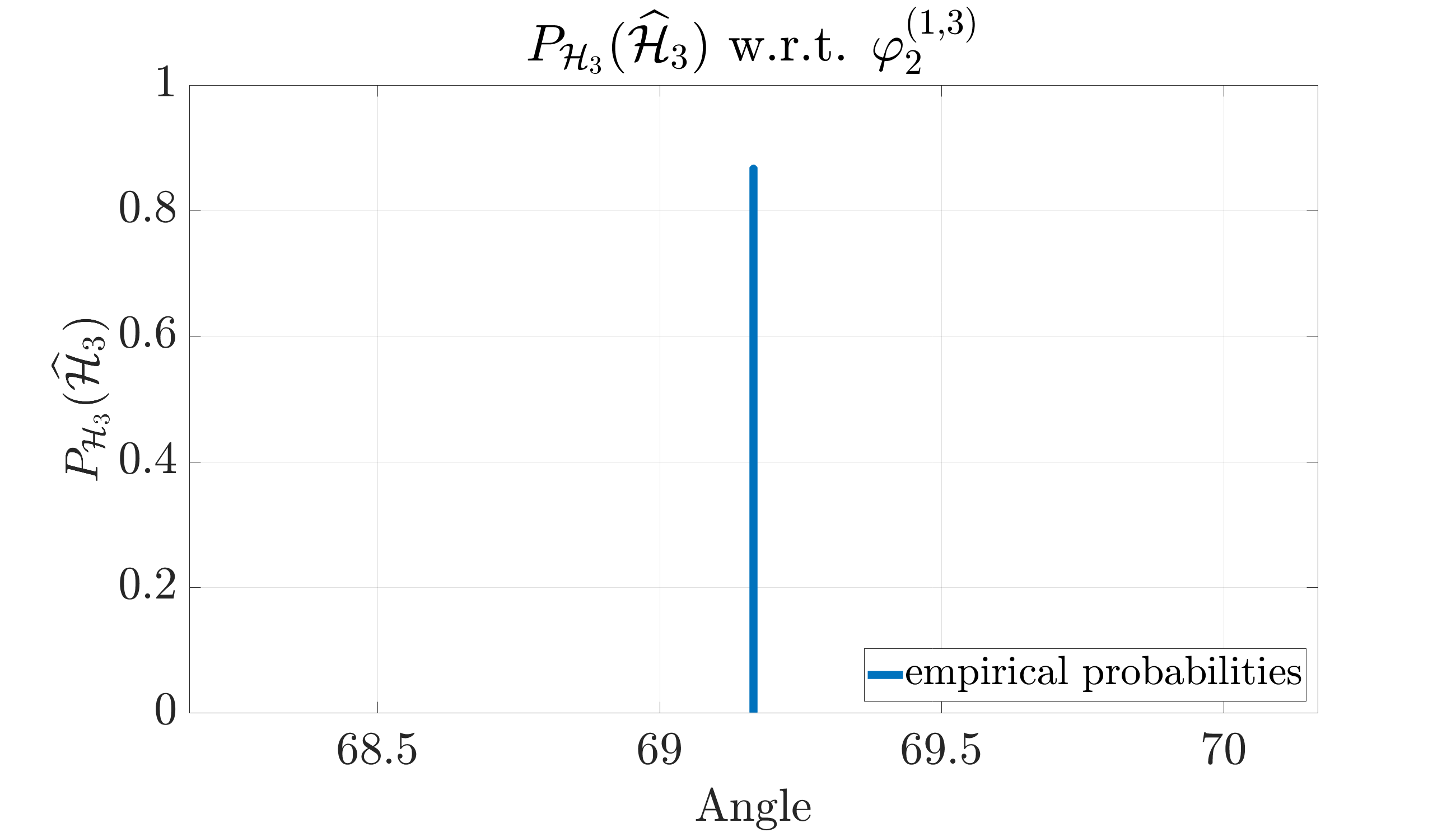}} &
	{\includegraphics[height=3.43cm]
	{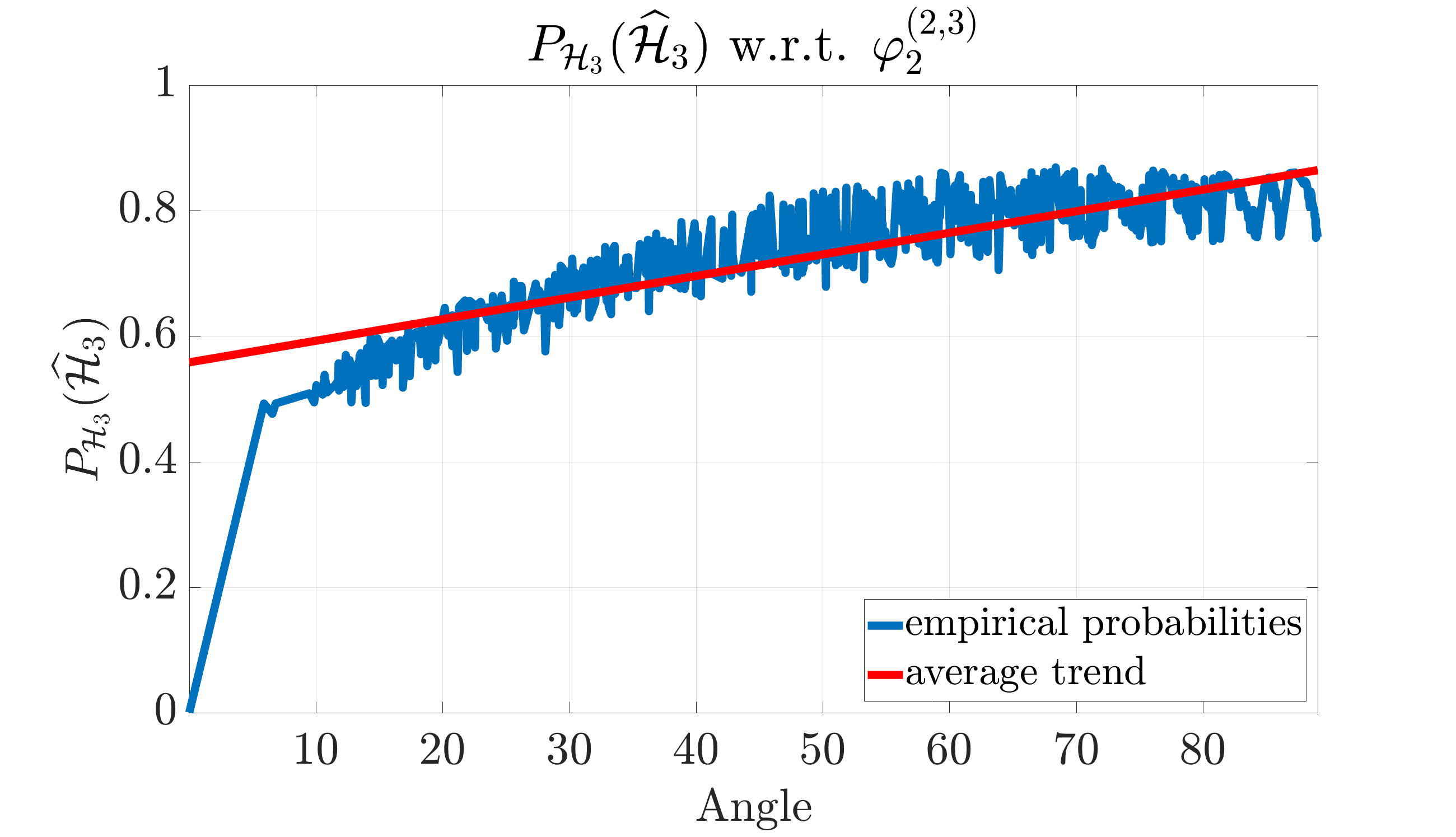}}
   	\end{tabular}
   	\end{center}
   	\caption{ \label{fig:num_sims:ASuD:KN:P_C}
	In known noise settings, the probability of correct classification increases with the increasing principal angles between whitened subspaces.}
   	\end{figure*}

\begin{figure} [!ht]
\begin{center}
\begin{tabular}{c}
	{\includegraphics[width=\columnwidth] {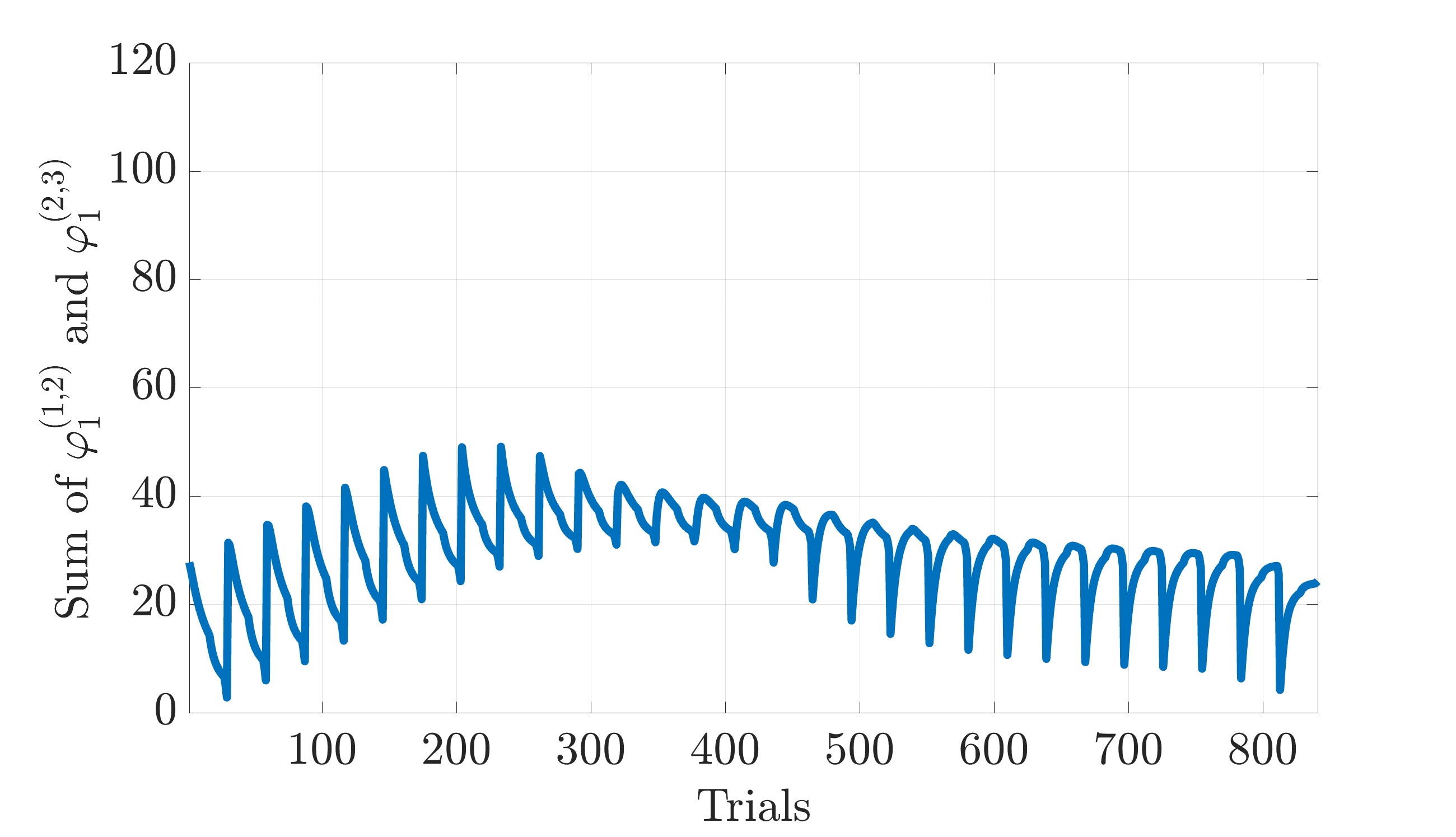}}
\end{tabular}
\end{center}
	\caption{ \label{fig:num_sims:ASuD:KN:subspace2:sum_angles}
		Sum of minimum principal angles subspace $S_2$ makes with subspace $S_1$ and subspace $S_3$. As $S_2$ moves away from $S_1$, the average of this sum increases initially and then decreases. The effect of this on the probability of classification $P_{S_2}(\widehat{\mathcal{H}}_2)$ can be seen in Fig.~\ref{fig:num_sims:ASuD:KN:P_C}.}
\end{figure}

We now demonstrate that the probability of correct classification increases with increasing principal angles between (whitened/empirically whitened) subspaces. This trend can be seen in Fig.~\ref{fig:num_sims:ASuD:KN:P_C} for active subspace detection under known noise settings. Notice that for subspace $S_2$, the probability $P_{S_2}(\widehat{\mathcal{H}}_2)$ first increases then decreases. This is because as we keep increasing the angles between $S_1$ and $S_2$, $S_2$ keeps moving closer to $S_3$. Since $S_1$ and $S_3$ are fixed, the angles that $S_2$ collectively makes with $S_1$ and $S_3$ first increase and then decrease, resulting in the observed behavior for $P_{S_2}(\widehat{\mathcal{H}}_2)$. This insight is verified in Fig.~\ref{fig:num_sims:ASuD:KN:subspace2:sum_angles} in terms of the plot of $\varphi_{1}^{1,2} + \varphi_{1}^{2,3}$ as a function of the number of trials. Similar trends for probabilities are seen under other noise settings, which are omitted due to space constraints.

\begin{figure*} [!ht]
	\begin{center}
		\begin{tabular}{c c c}
			{\includegraphics[height=3.43cm] {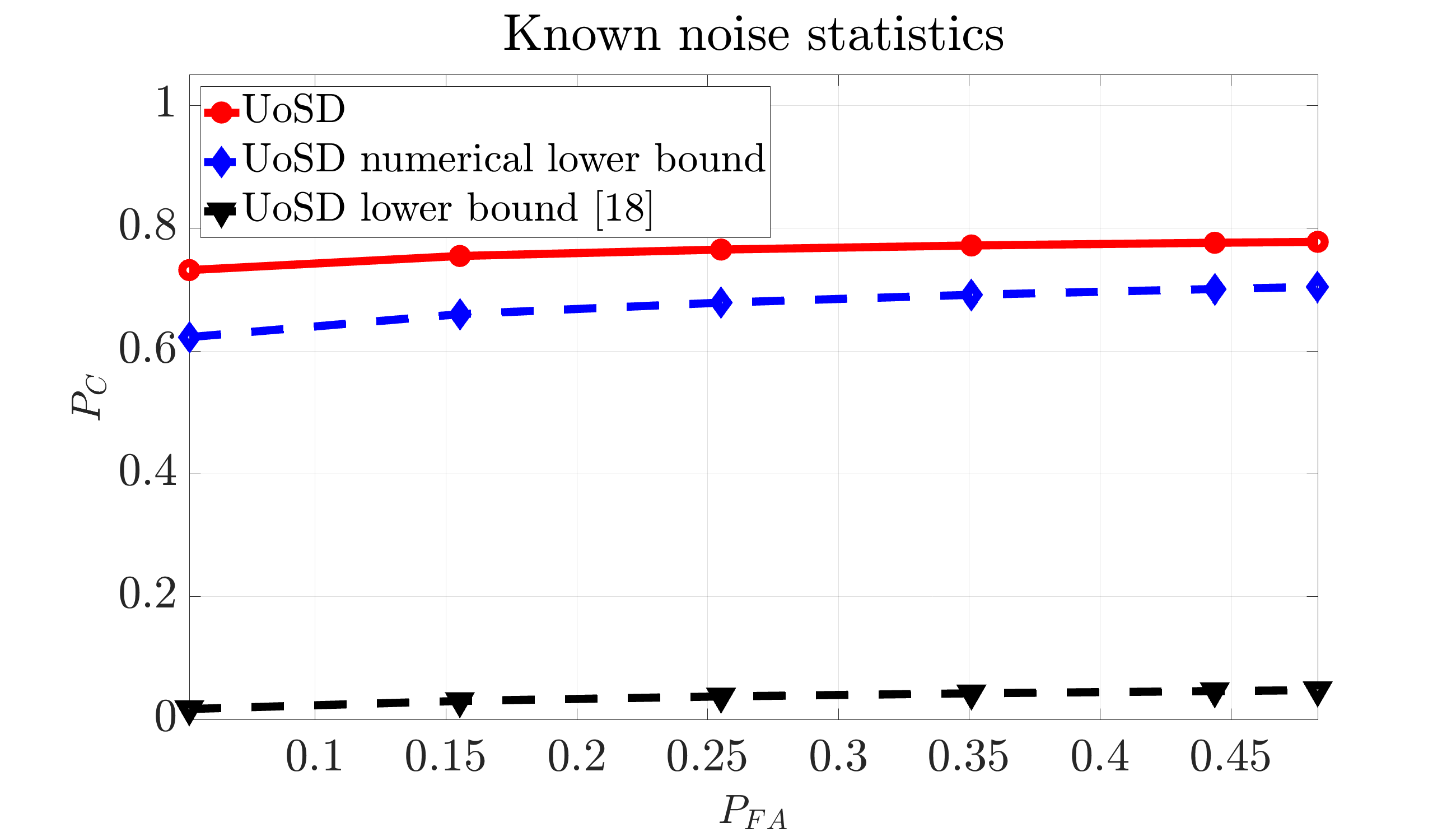}}
			{\includegraphics[height=3.43cm] {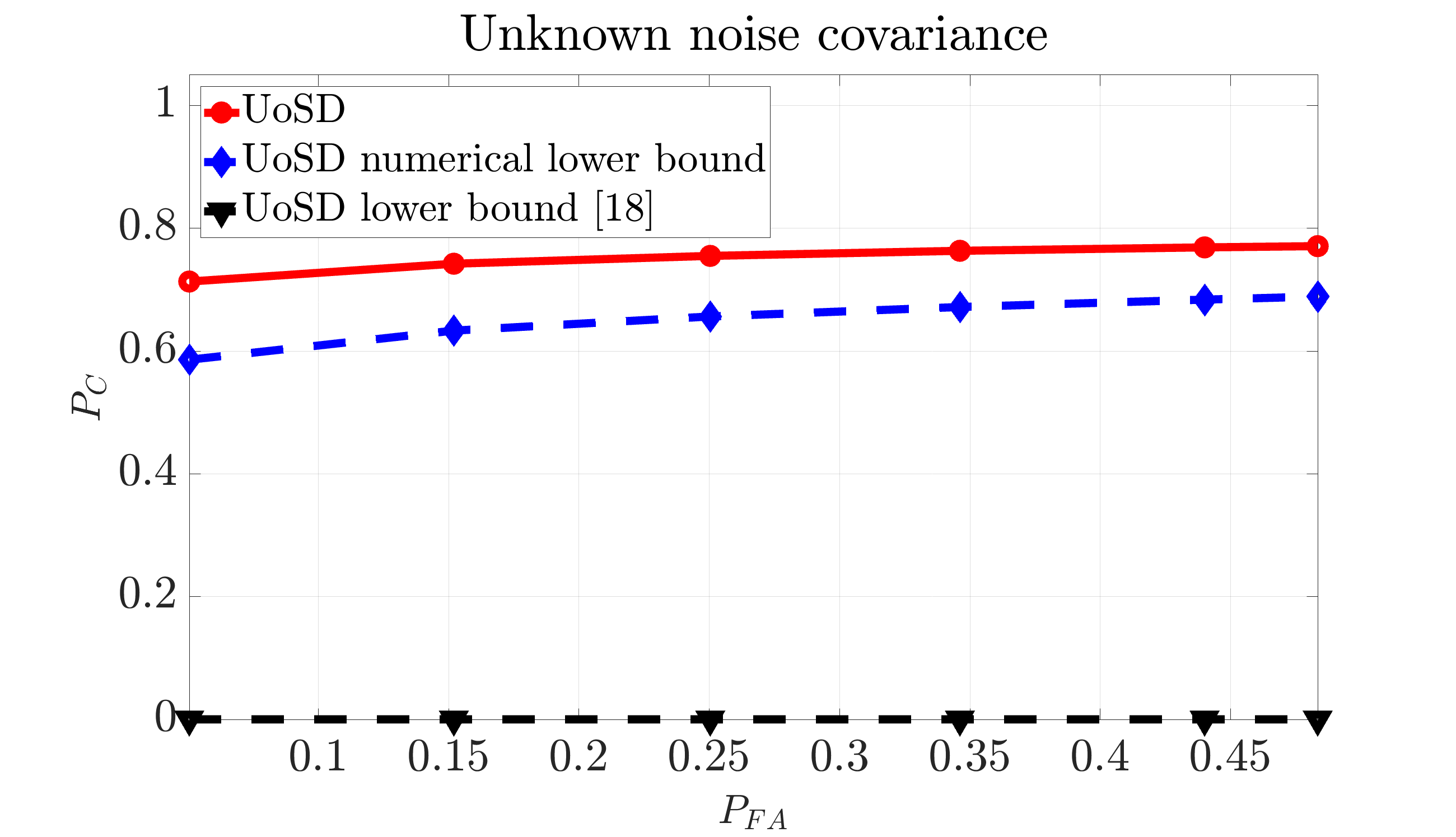}}
			{\includegraphics[height=3.43cm] {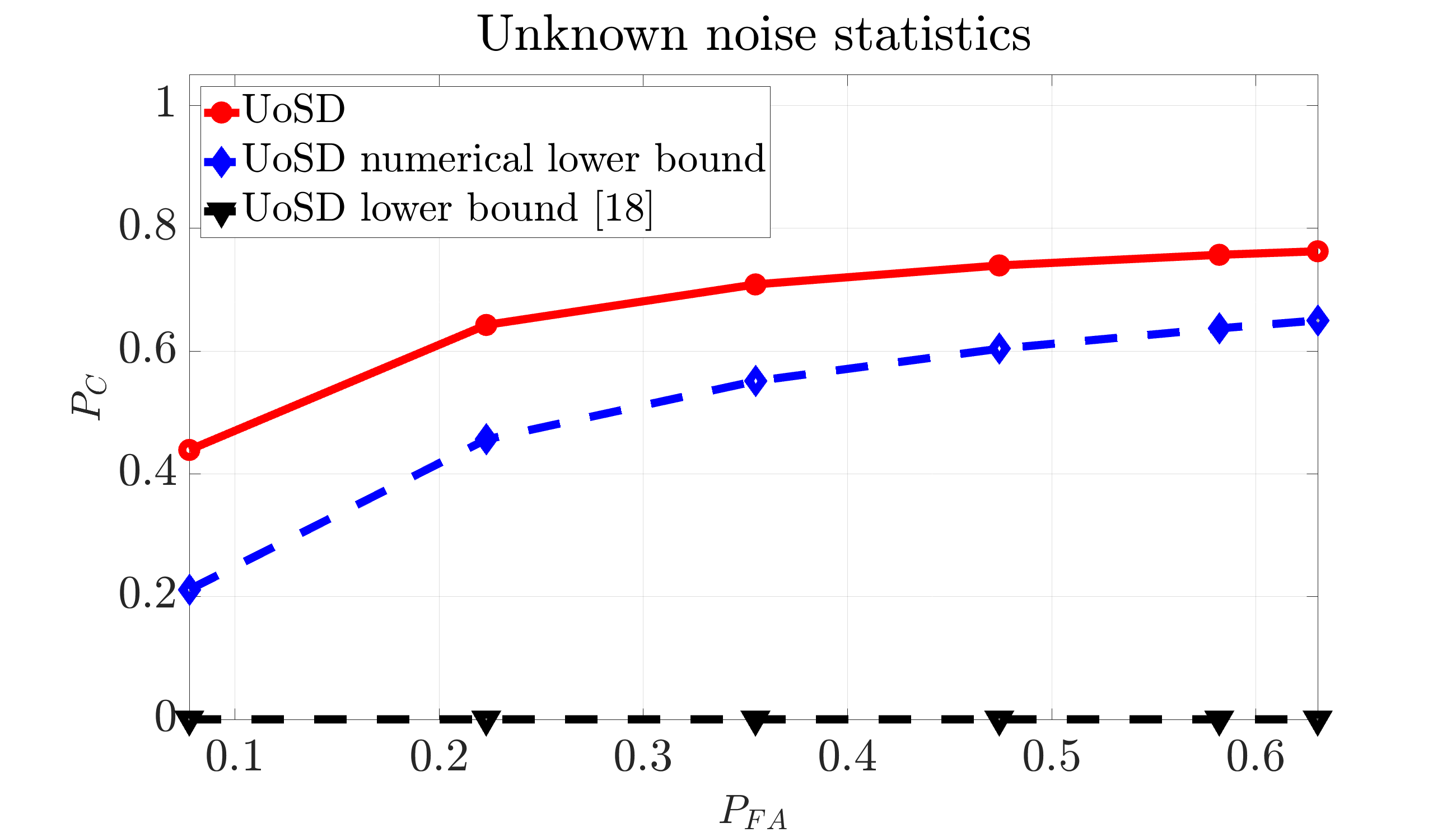}}
		\end{tabular}
	\end{center}
	\caption{ \label{fig:num_sims:MSD:ROC:P_C}
		ROC curves for active subspace detection under the UoS model (labeled UoSD) and the derived bounds. All subfigures show three plots: the true classification probability under UoS, the lower bound on the classification probability computed numerically and the lower bound derived using \cite{wimalajeewa2015subspace}. Starting from the left, the sub-figures show the ROC curves under known noise statistics, unknown noise covariance and unknown noise statistics.}
\end{figure*}

Next, we plot the ROC curves for the probability of correct classification and the various bounds derived under different noise settings in Fig.~\ref{fig:num_sims:MSD:ROC:P_C}. We see that the lower bounds derived from \cite{wimalajeewa2015subspace} are very loose, compared to our lower bounds. A comparison of the probability of correct classification under different noise settings is provided in Fig.~\ref{fig:num_sims:MSD:ROC_comparison}.

\begin{figure*} [!ht]
   	\begin{center}
   	\begin{tabular}{c c c}
   	{\includegraphics[height=3.43cm] {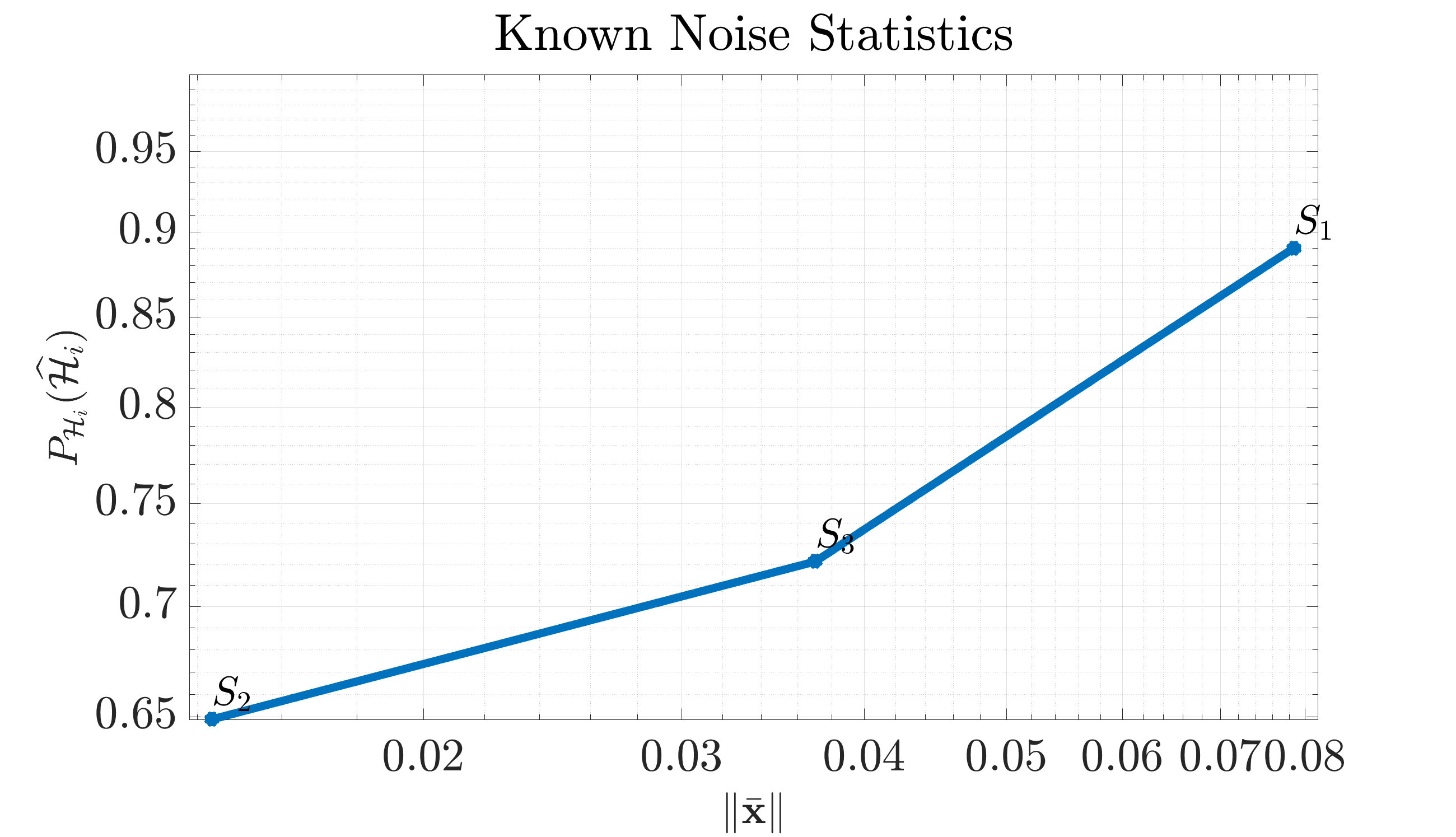}}
   	{\includegraphics[height=3.43cm] {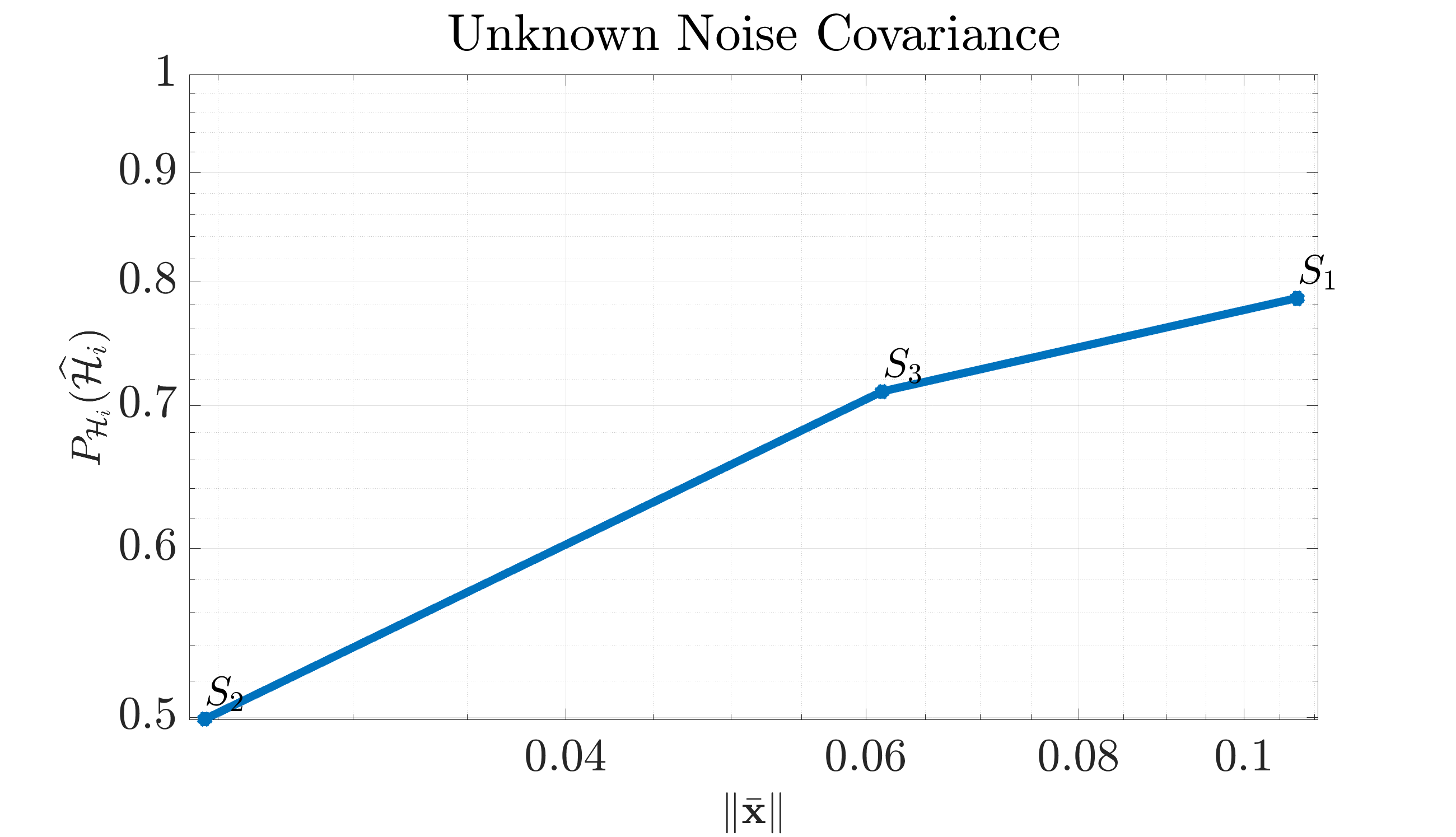}}
   	{\includegraphics[height=3.43cm] {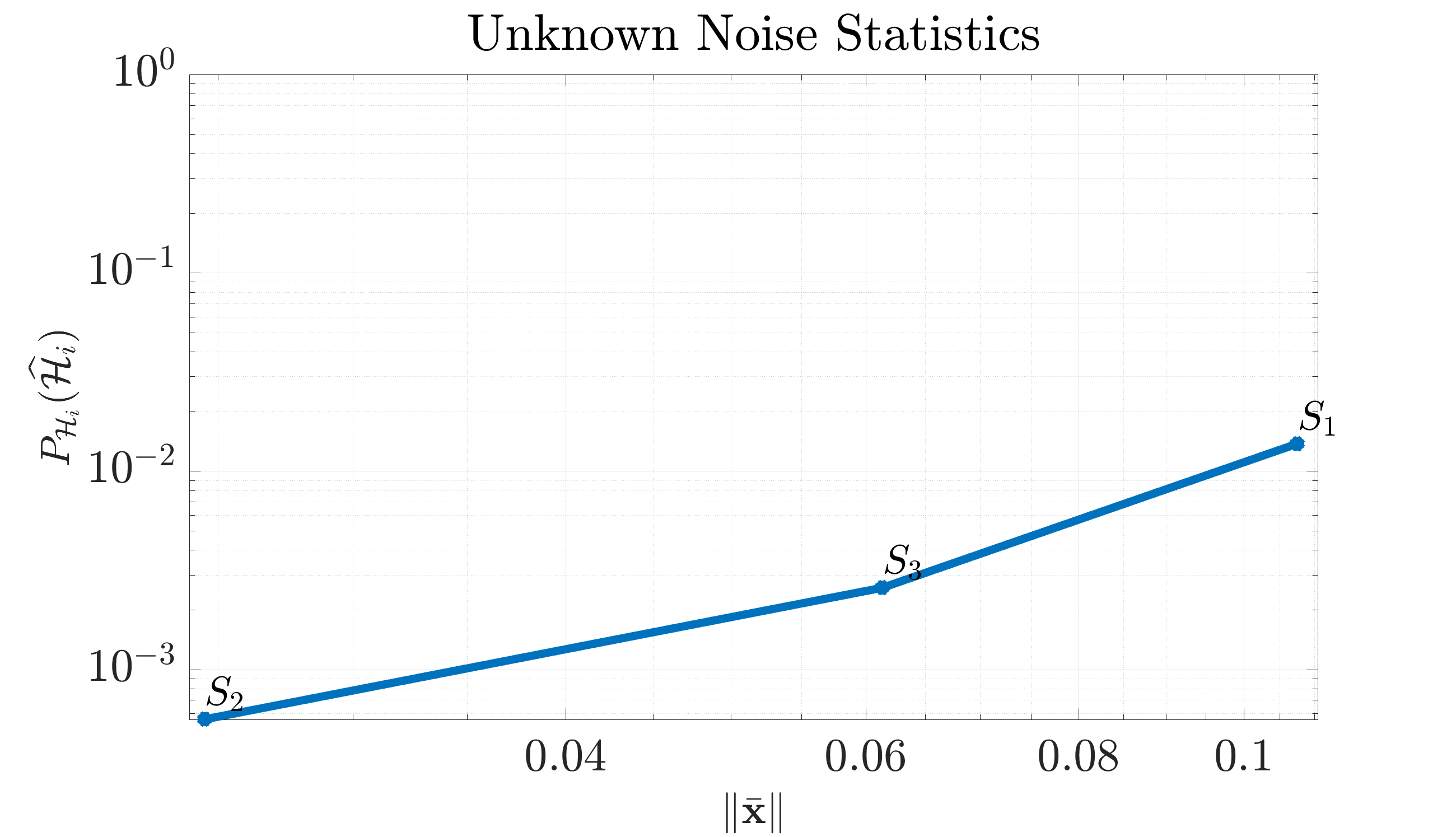}}
   	\end{tabular}
   	\end{center}
   	\caption{ \label{fig:num_sims:ASuD:noise_geometry}
	Each subfigure shows that the closer a subspace is to the higher-order eigenvectors of the noise covariance, the lower is its classification probability $P_{\mathcal{H}_k}(\widehat{\mathcal{H}}_k)$. The setup here is similar to the one for Fig.~\ref{fig:num_sims:MSD:noise_geometry}.}
   	\end{figure*}

We further show the influence of noise geometry on active subspace detection. We use the same setup as for signal detection. We can see from Fig.~\ref{fig:num_sims:ASuD:noise_geometry} that subspaces closer to the higher-order eigenvectors of the noise covariance have lower detection probability, and vice versa.

\begin{figure*} [!ht]
	\begin{center}
		\begin{tabular}{c c c}
			{\includegraphics[height=3.43cm] {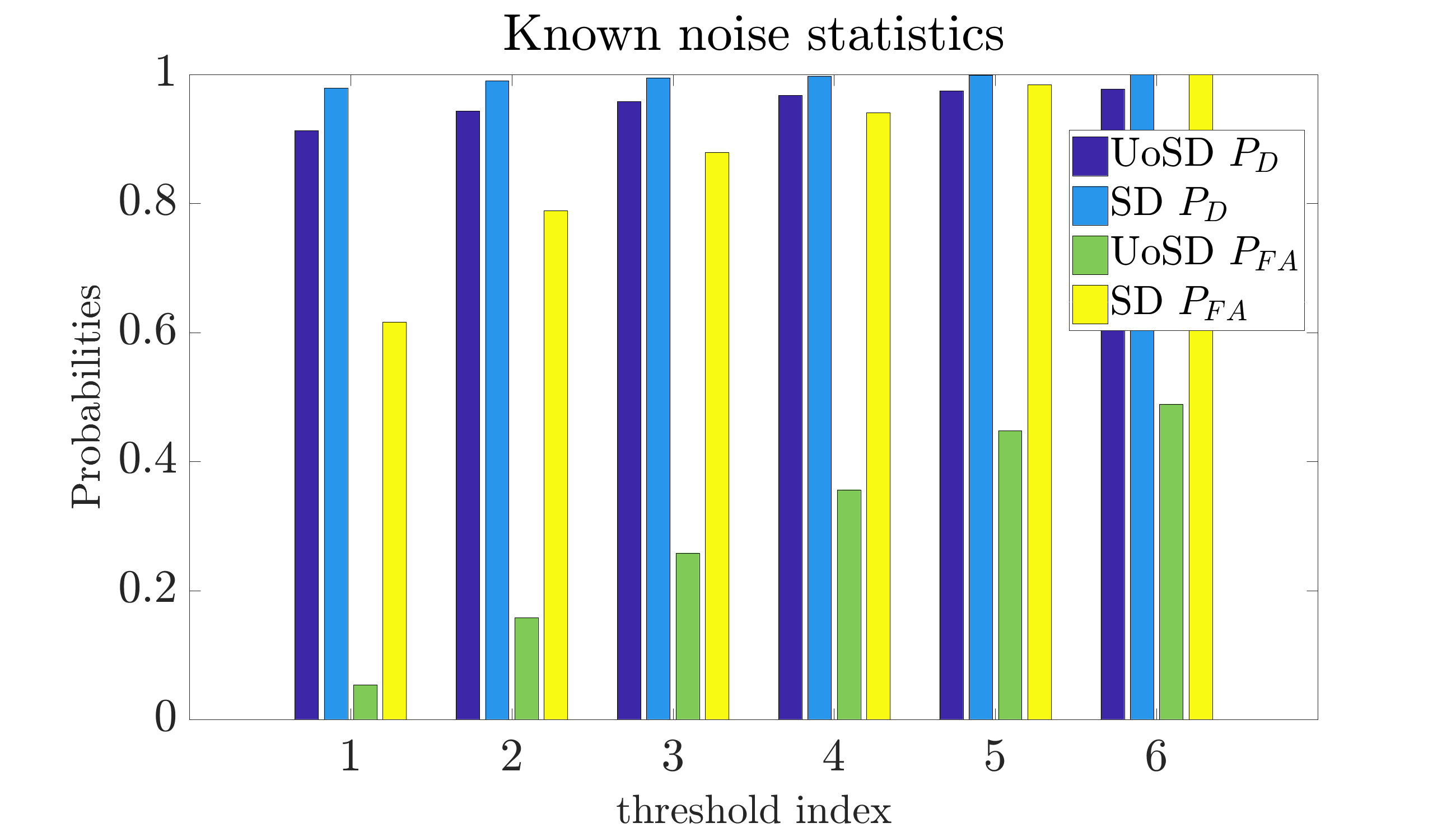}}
			{\includegraphics[height=3.43cm] {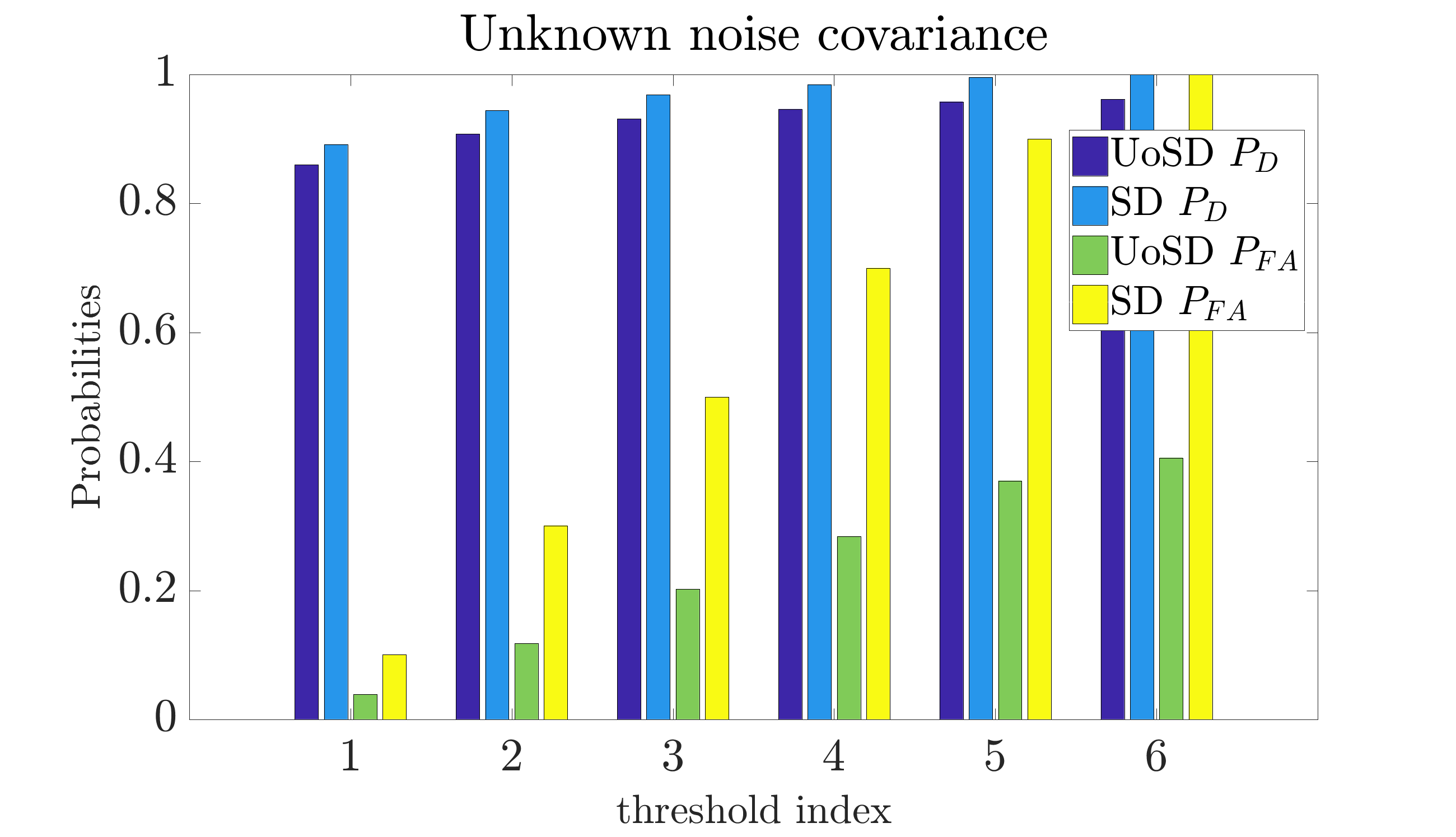}}
			{\includegraphics[height=3.43cm] {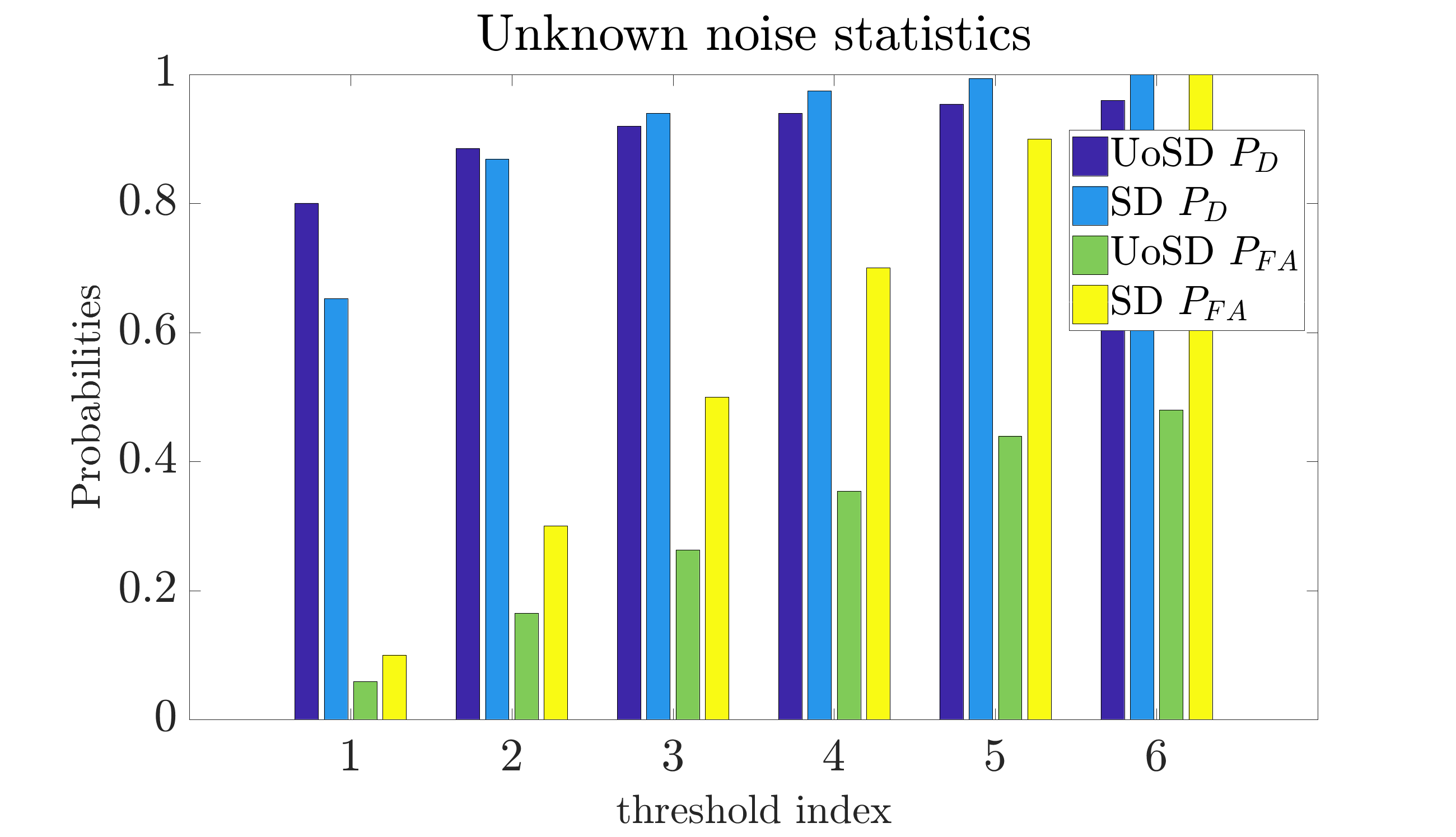}}
		\end{tabular}
	\end{center}
	\caption{\label{fig:num_sims:UoS_vs_Sub}\revise{Performance comparison of UoS-based and subspace-based detection of signals generated under the UoS model. Under all noise conditions, classical subspace detection incurs a significantly higher false alarm rate than UoS-based signal detection.}}
\end{figure*}

\revise{\subsubsection{Comparison with existing approaches}
Of the existing methods, we can only compare the signal detection performance of the GLRTs derived in this paper with that of the subspace-based GLRTs. Indeed, the active subspace detection problem under the UoS model has no counterpart in the classical subspace model. Likewise, comparison with a simple GLRT (for signal detection) is also infeasible as a simple GLRT requires knowledge of the signal template, whereas we only assume access to the subspaces that generate the signal. In addition, as noted earlier in the paper, reliance of existing compressive detection frameworks on the use of measurement matrices and (exponentially many, equiangular) random subspaces renders them impractical for UoS-based detection involving finitely many, arbitrary subspaces. In order to compare UoS-based detection with the classical subspace detection, we consider three 2-dimensional subspaces in an 8-dimensional space. Subspace detection in this setting requires projection of the observed signal onto the direct sum of the three subspaces. We compare the probability of signal detection and probability of false alarm for both UoS and subspace methods under the same SNR (5 dB) and the same detection threshold. The results, provided in Fig.~\ref{fig:num_sims:UoS_vs_Sub} for six different threshold values, show that the probability of detection of the classical subspace method is slightly higher than the detection probability under the UoS model. This is because the classical subspaces model considers the direct sum of the subspaces instead of the union and ends up declaring irrelevant signals as detections. However, this in turn significantly increases the false alarm rate of signal detection under the subspace model. In particular, it can be seen from Fig.~\ref{fig:num_sims:UoS_vs_Sub} that the probability of false alarm for the classical subspace detection far exceeds that of UoS-based detection.}
\subsubsection{Other observations}
\begin{figure*} [!ht]
	\begin{center}
		\begin{tabular}{c c c}
			{\includegraphics[height=3.43cm] {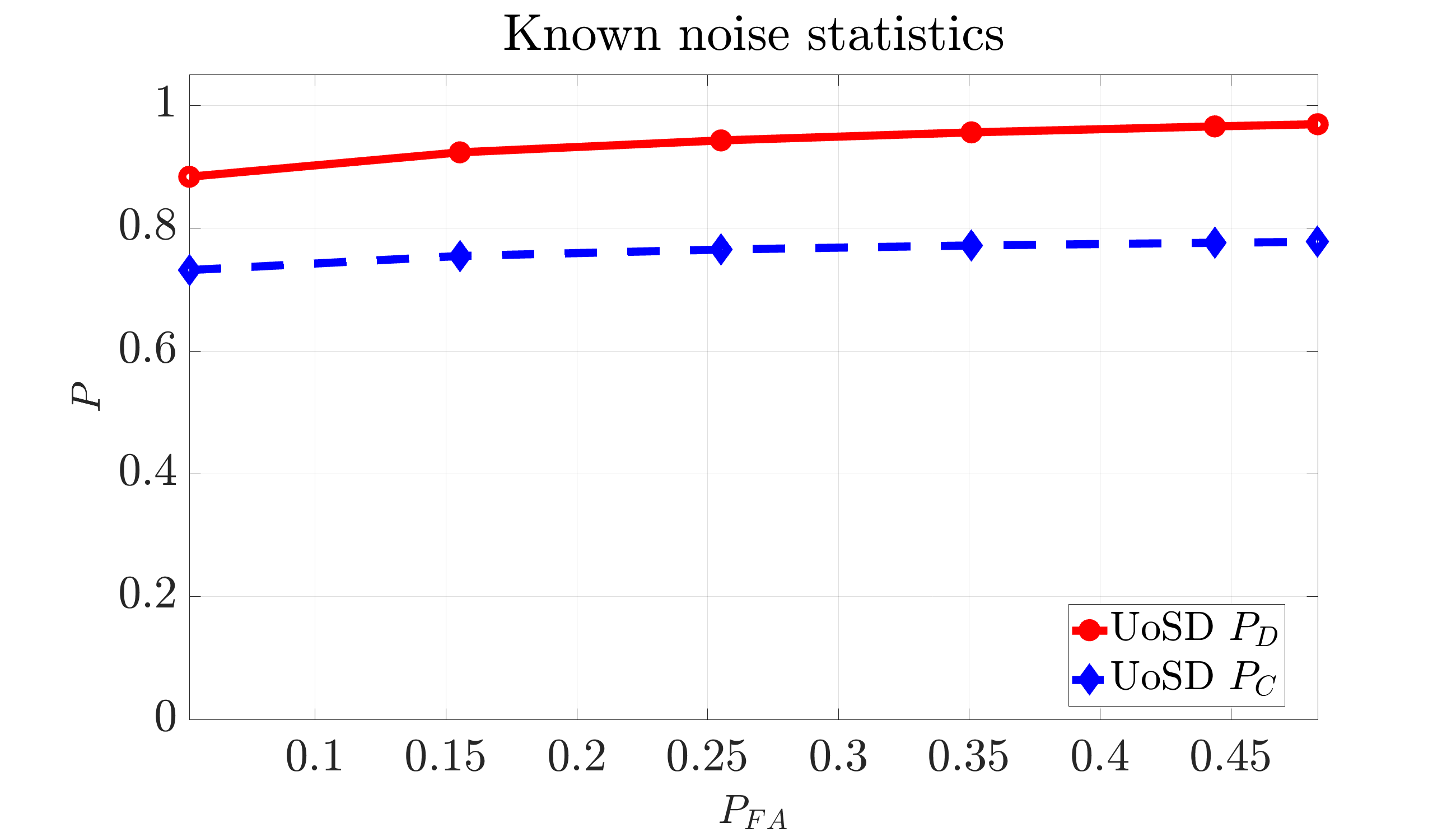}}
			{\includegraphics[height=3.43cm] {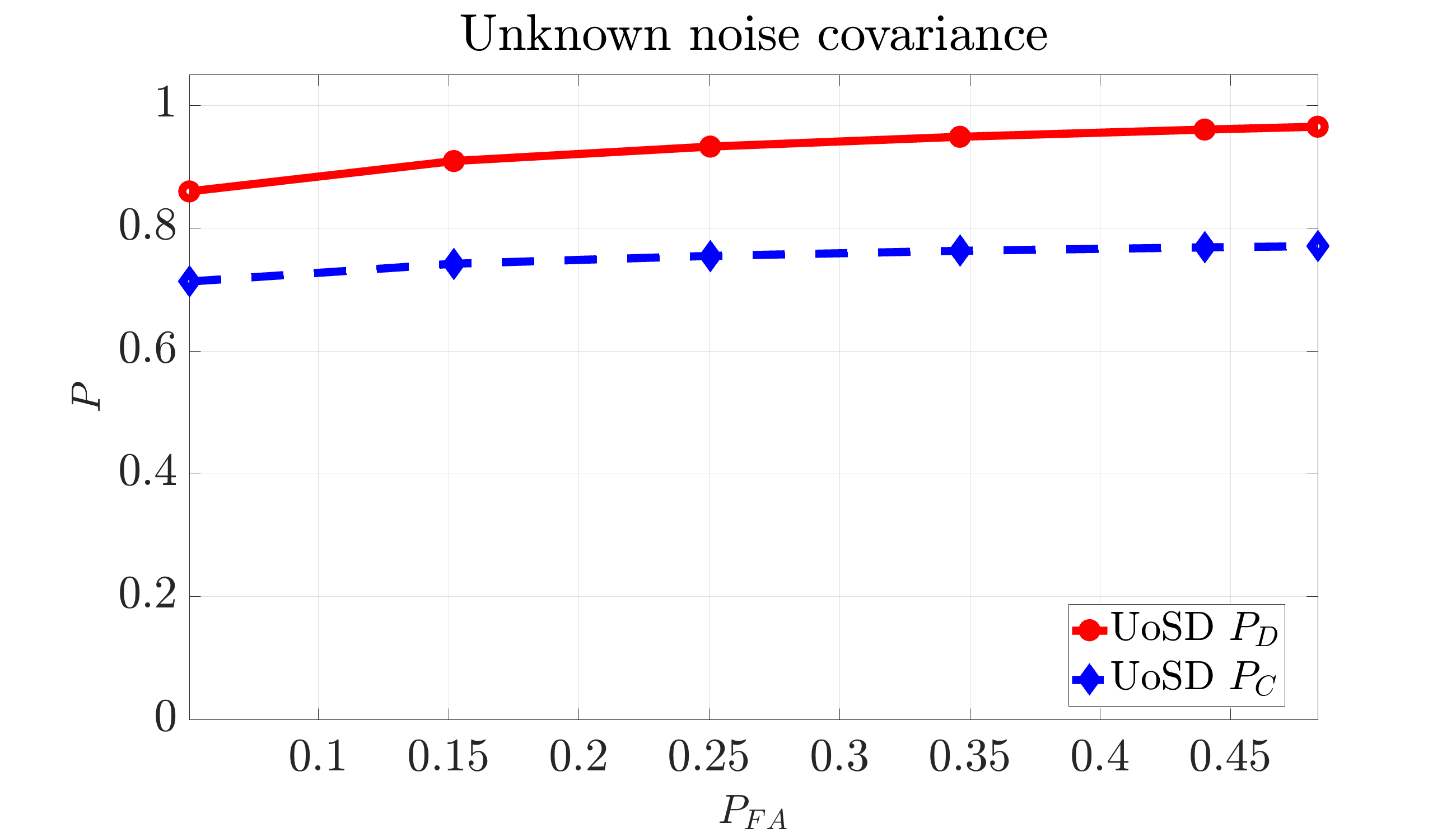}}
			{\includegraphics[height=3.43cm] {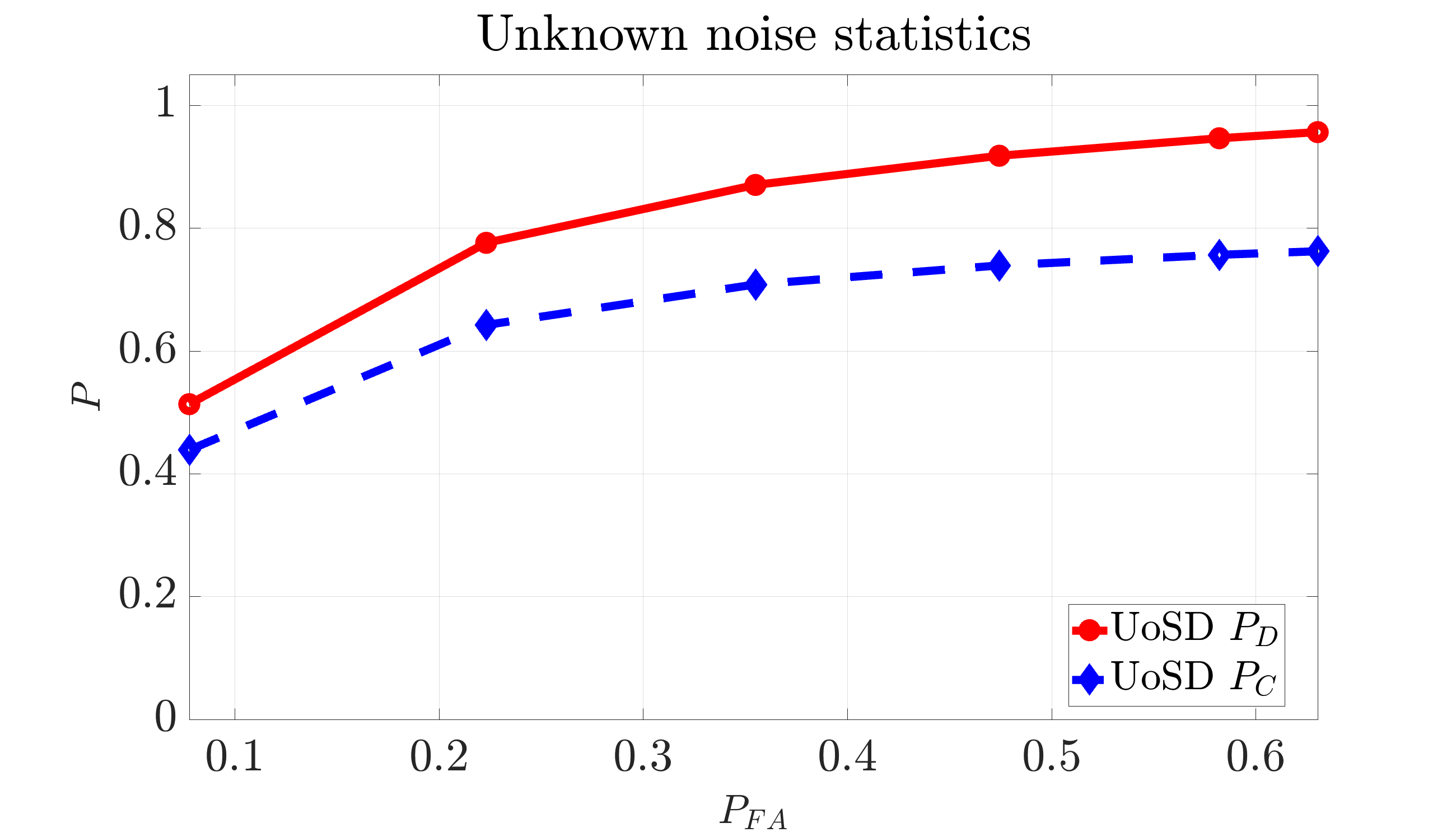}}
			\\
			{\includegraphics[height=3.43cm] {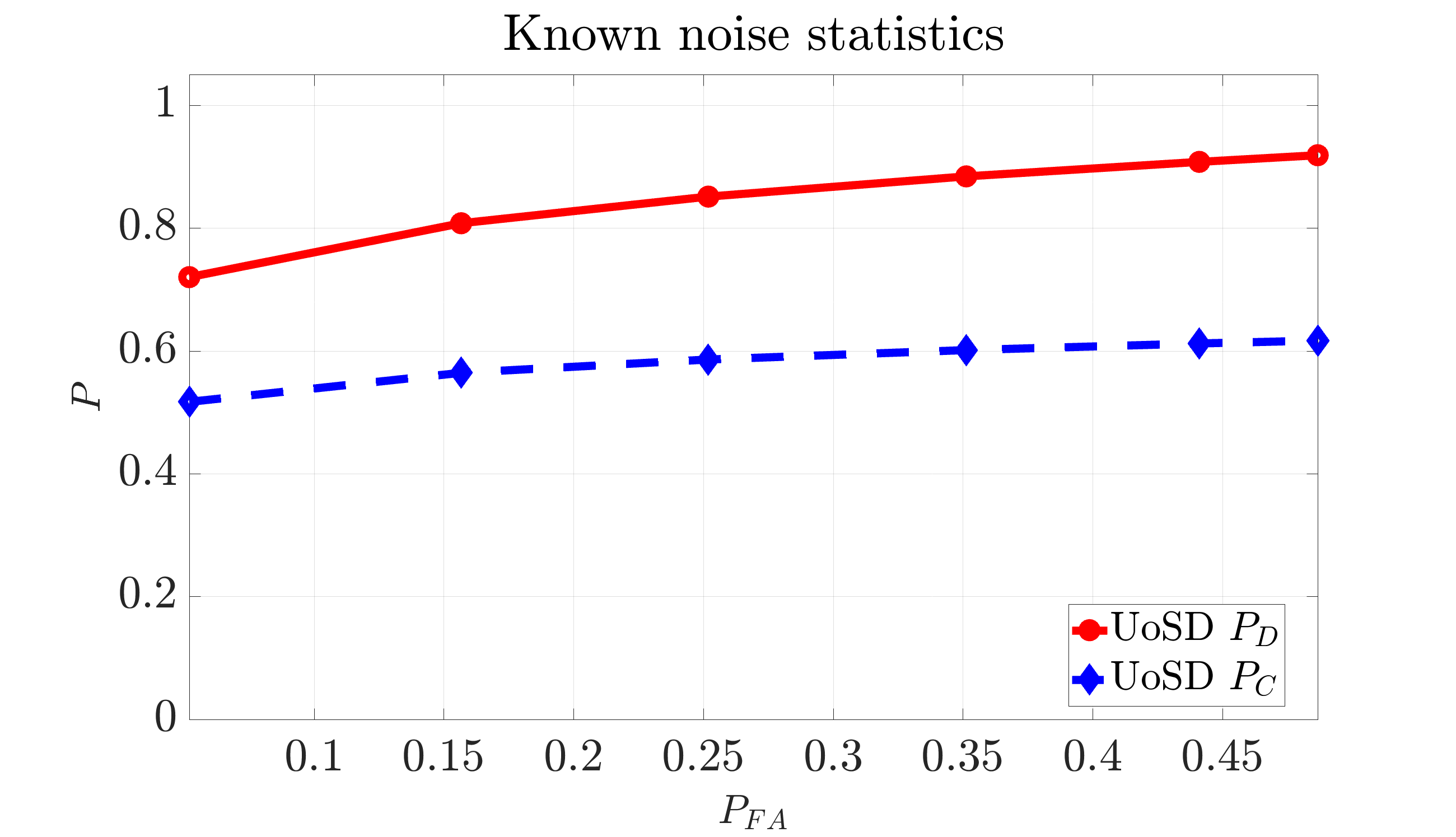}}
			{\includegraphics[height=3.43cm] {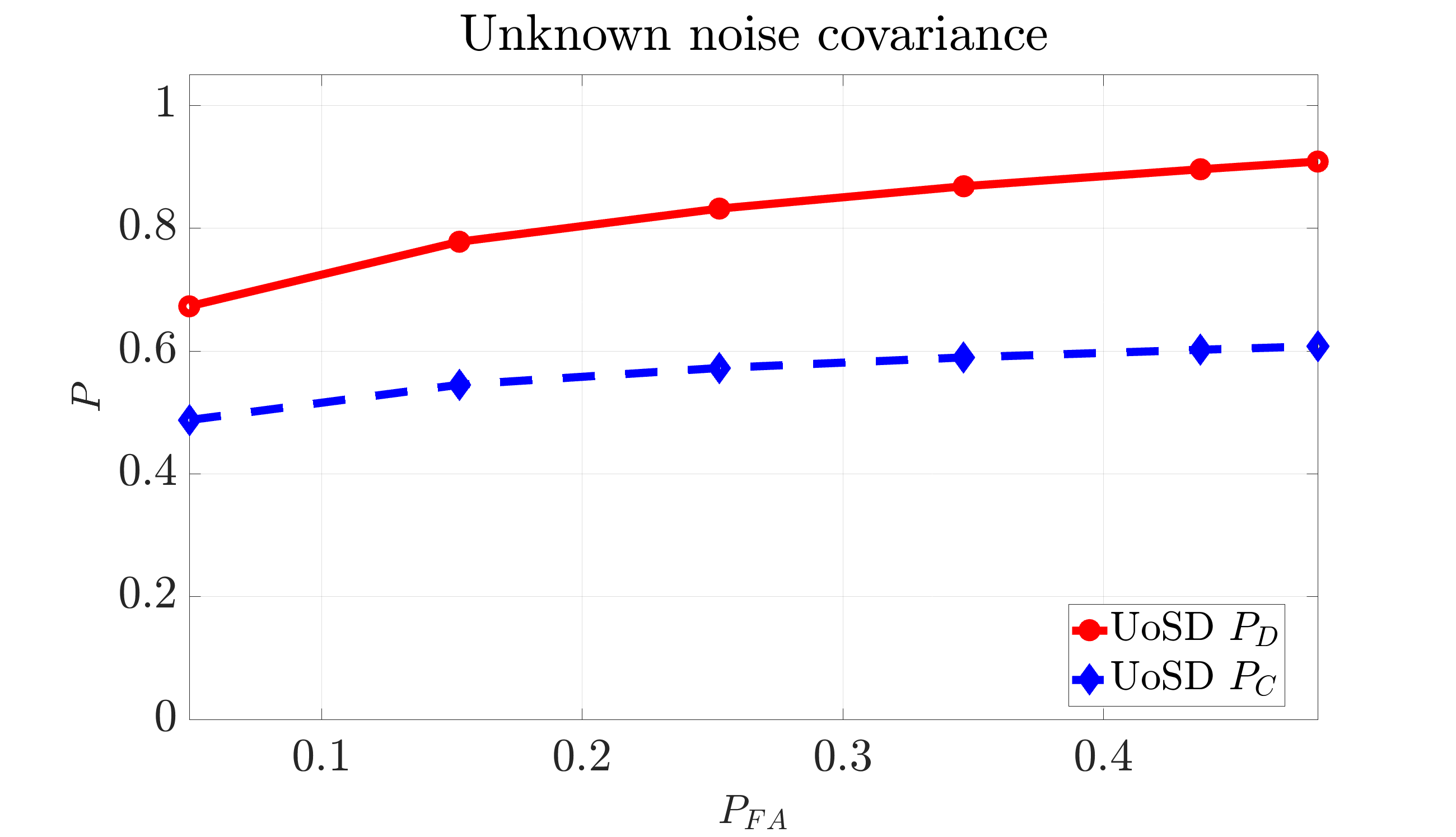}}
			{\includegraphics[height=3.43cm] {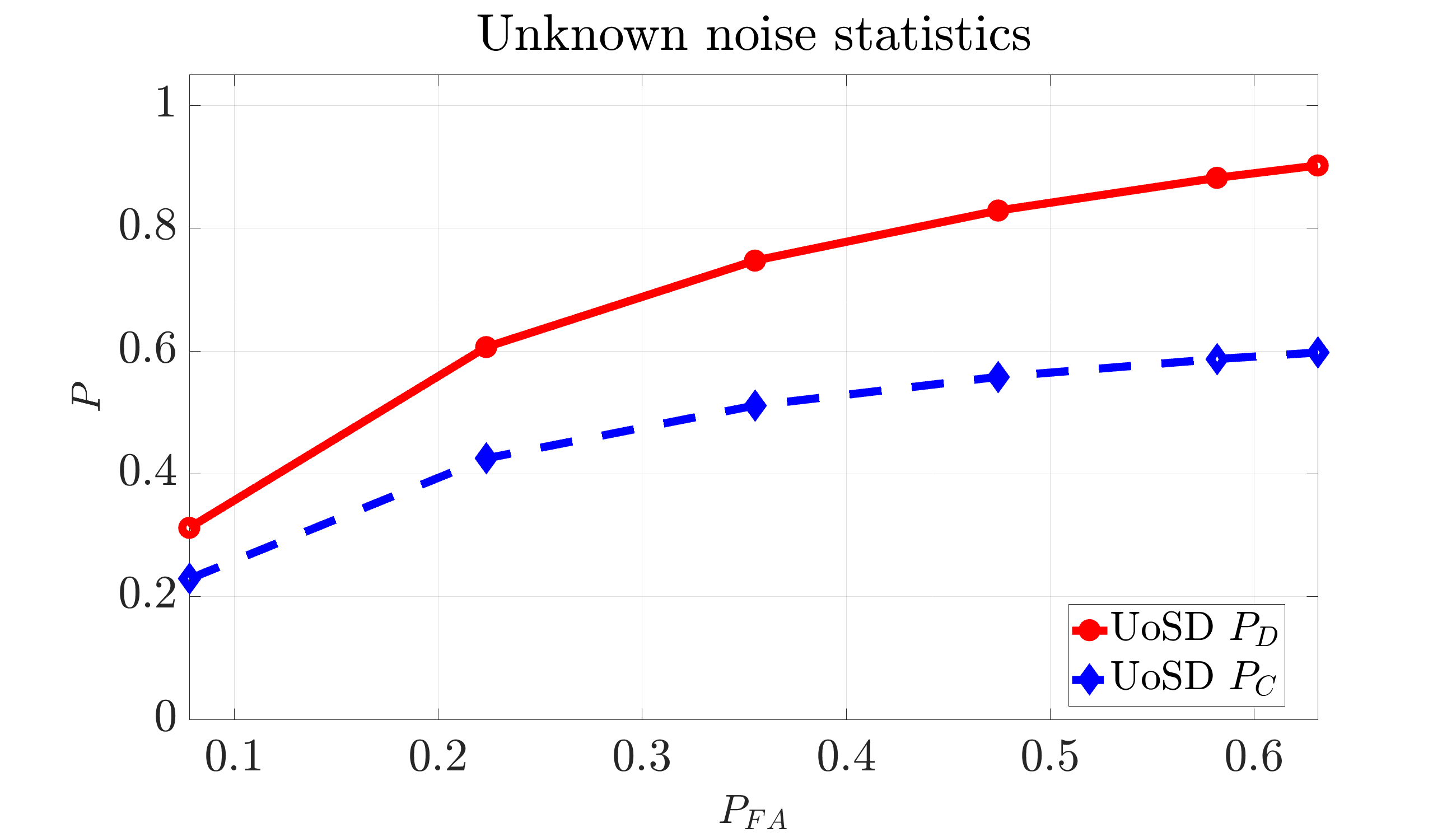}}
		\end{tabular}
	\end{center}
	\caption{\label{fig:num_sims:ROC:P_D_P_C_comparison}
		Gap between the probability of detection and the probability of correct classification under various noise settings. The two rows have SNR levels 10 dB and 5 dB, respectively. We can see that higher SNR results in a lower gap.}
\end{figure*}

Fig.~\ref{fig:num_sims:ROC:P_D_P_C_comparison} shows the gap between detection and classification probabilities for different noise settings and different SNR levels. We can see that the gap decreases for higher SNR levels.

\begin{figure} [!ht]
	\begin{center}
		\begin{tabular}{c c}
			{\includegraphics[height=2.5cm] {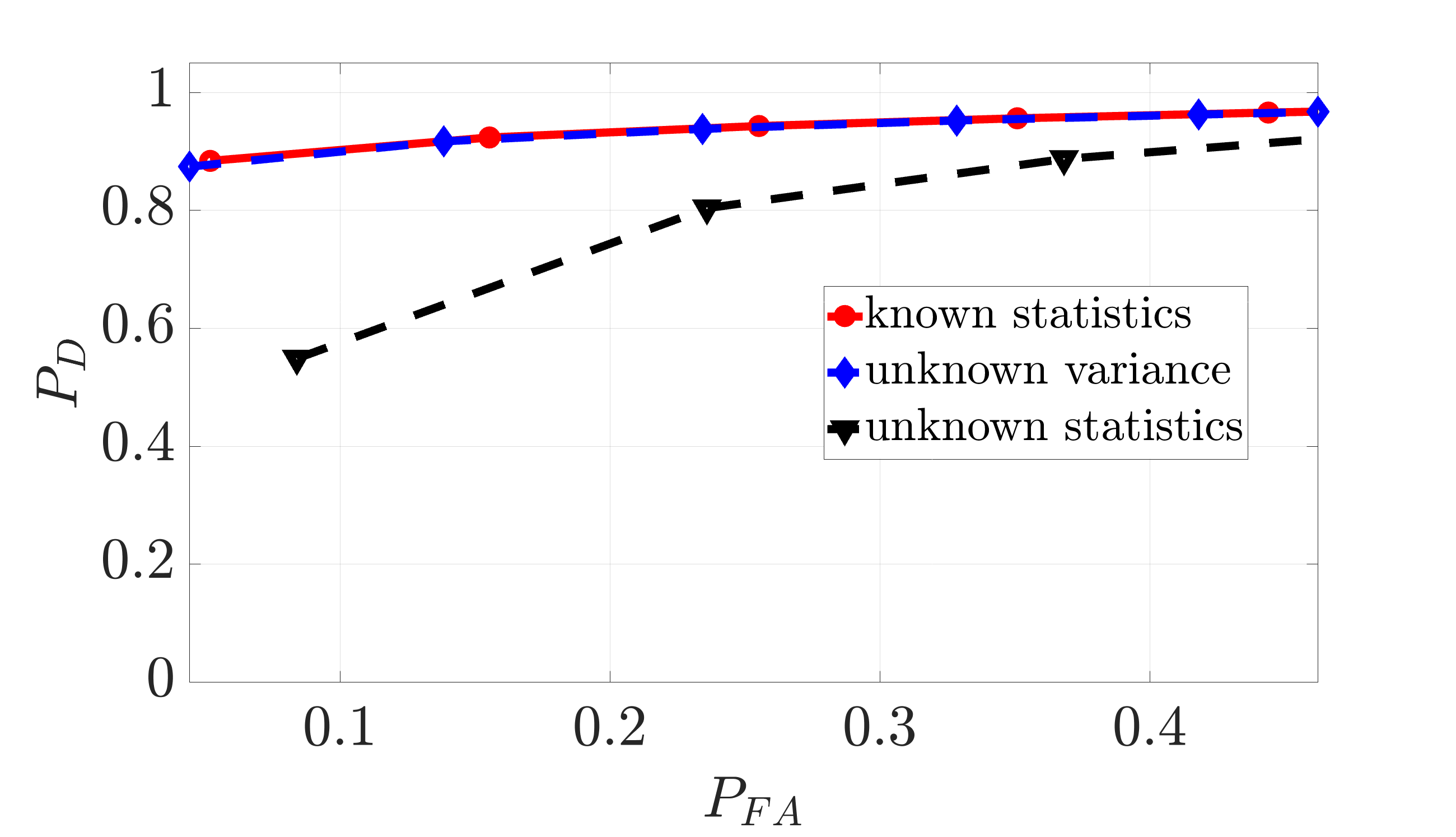}}
			{\includegraphics[height=2.5cm] {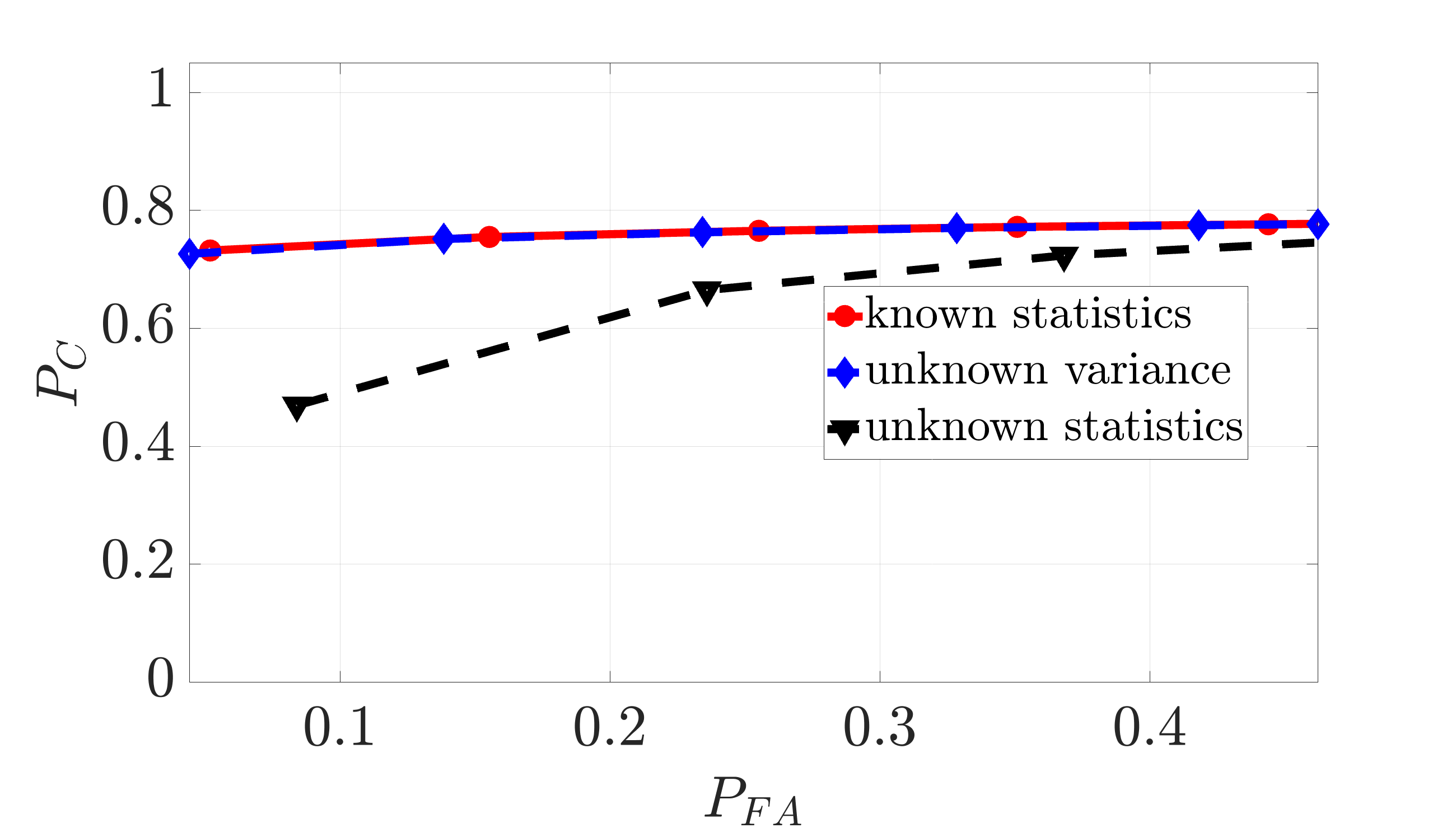}}
			\\
			{\includegraphics[height=2.5cm] {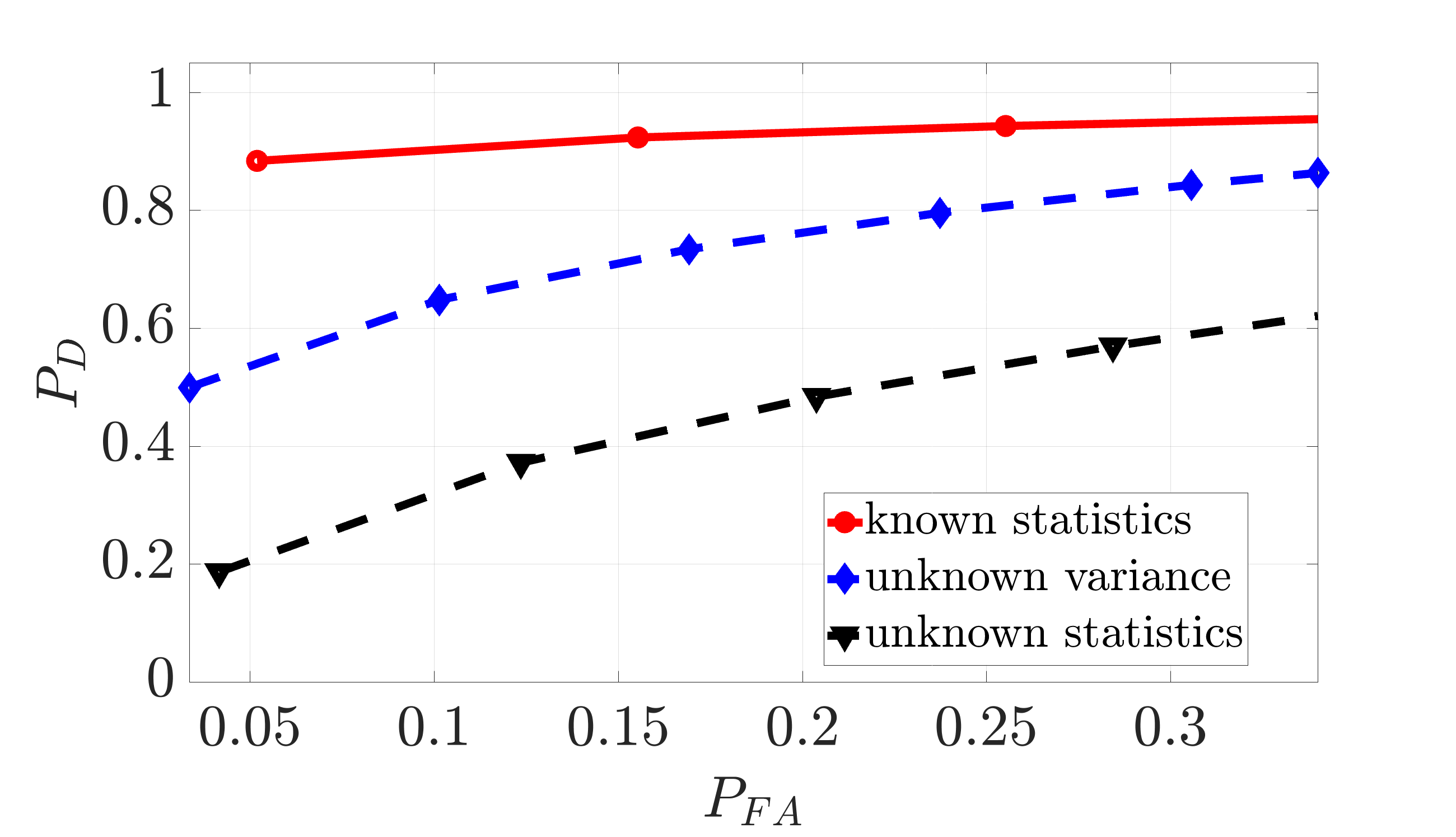}}
			{\includegraphics[height=2.5cm] {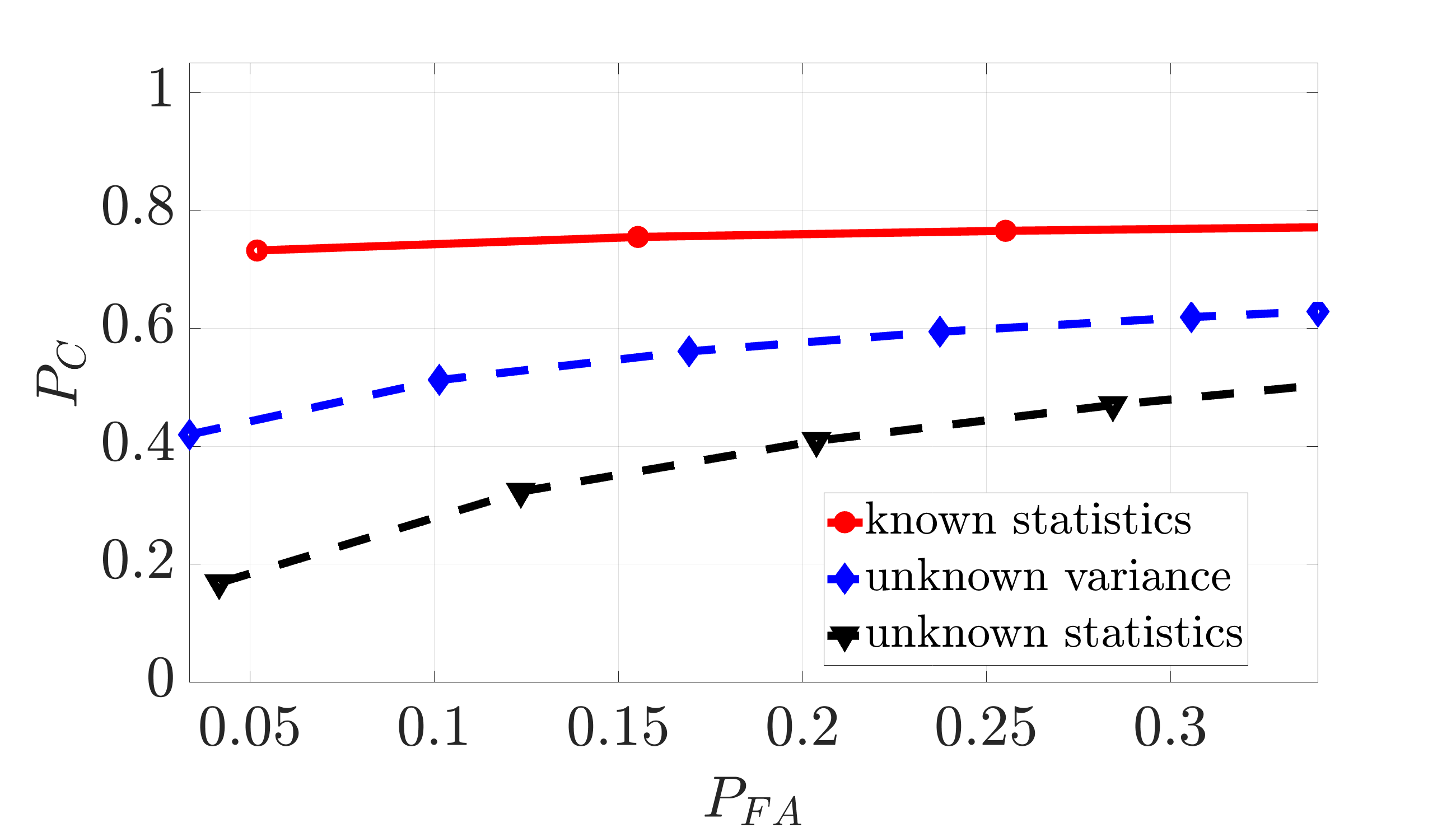}}
		\end{tabular}
	\end{center}
	\caption{ \label{fig:num_sims:ROC:P_noise_comparison}
		Gap between the ROC curves under various noise settings for different number of noise samples. Figures in the first row use $200$ noise samples whereas the ones in the second row use $8$ noise samples.}
\end{figure}

We make a final observation by plotting the ROC curves under various noise settings for different number of noise samples. From Fig.~\ref{fig:num_sims:ROC:P_noise_comparison}, we see that the gap between probabilities for known noise settings and unknown noise covariance decreases as the number of noise samples increases. This is since with increasing number of noise samples, our estimates of noise statistics get better and we move closer to the regime of known noise statistics. 

\subsection{Real-world datasets} \label{num_sims:real}
\begin{figure} [!ht]
   	\begin{center}
   	\begin{tabular}{c c}
   	{\includegraphics[height=4cm] {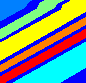}}
   	{\includegraphics[height=4cm] {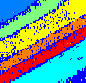}}
   	\end{tabular}
   	\end{center}
   	\caption{ \label{fig:num_sims:salinas}
	This figure shows the ground truth (left) for different classes in Salinas A scene and the detected targets (right) using the UoS detector under unknown noise statistics. The targets were detected with the classification accuracy of $91.16\%$ when upper bounding the false alarm rate at $5\times 10^{-4}$.}
   	\end{figure}

In this subsection, we report results on some real-world datasets that potentially conform to the UoS model. The first dataset we consider is the Salinas `A' Scene Hyperspectral Data \cite{green1998imaging}. This data was acquired by a 224-band AVIRIS sensor over Salinas Valley (California). There are six target classes in the data. We assume each target class is lying in a different subspace, thus modeling the set of targets as belonging to a union of subspaces. To obtain the bases for the subspaces, we randomly select 20 pixels belonging to each target and use singular value decomposition (SVD) to get the bases for 10-dimensional target subsapces. For the Salinas `A' Scene, the ground truth and the detected targets are shown in Fig.~\ref{fig:num_sims:salinas}. Assuming noise with unknown statistics and false alarm probability upper bounded at $5 \times 10^{-4}$, the targets are classified with the overall probability of correct classification 0.9116.

Next, the face of a subject with varying illumination conditions has been shown to lie near a 9-dimensional subspace \cite{basri2003lambertian}. Thus a set of subjects can be assumed to lie near a union of subspaces. Using this assumption, for the Yale Database B \cite{georghiades2001few}, we first obtain subspace bases for each subject by using SVD on 18 randomly selected subject images. With these bases and assuming unknown noise statistics, we correctly identify subjects with probability 0.76 while upper bounding the false alarm rate at $1\times 10^{-3}$.

The third dataset in consideration is the Hopkins 155 motion segmentation dataset \cite{tron2007benchmark}, which consists of sequences of two and three motions extracted from several videos. It has been argued that different motion sequences extracted from tracking a set of points in a video lie in 3-dimensional subspaces \cite{tron2007benchmark}. We again use SVD on randomly selected sequences to learn the subspace bases. Using the UoS model with unknown noise statistics, the probability of correct classification over all sequences comes out to be 0.7664 by upper bounding the false alarm rate at $5\times 10^{-2}$. 

\begin{figure} [!ht]
	\begin{center}
		\begin{tabular}{c}
			{\includegraphics[width=\columnwidth] {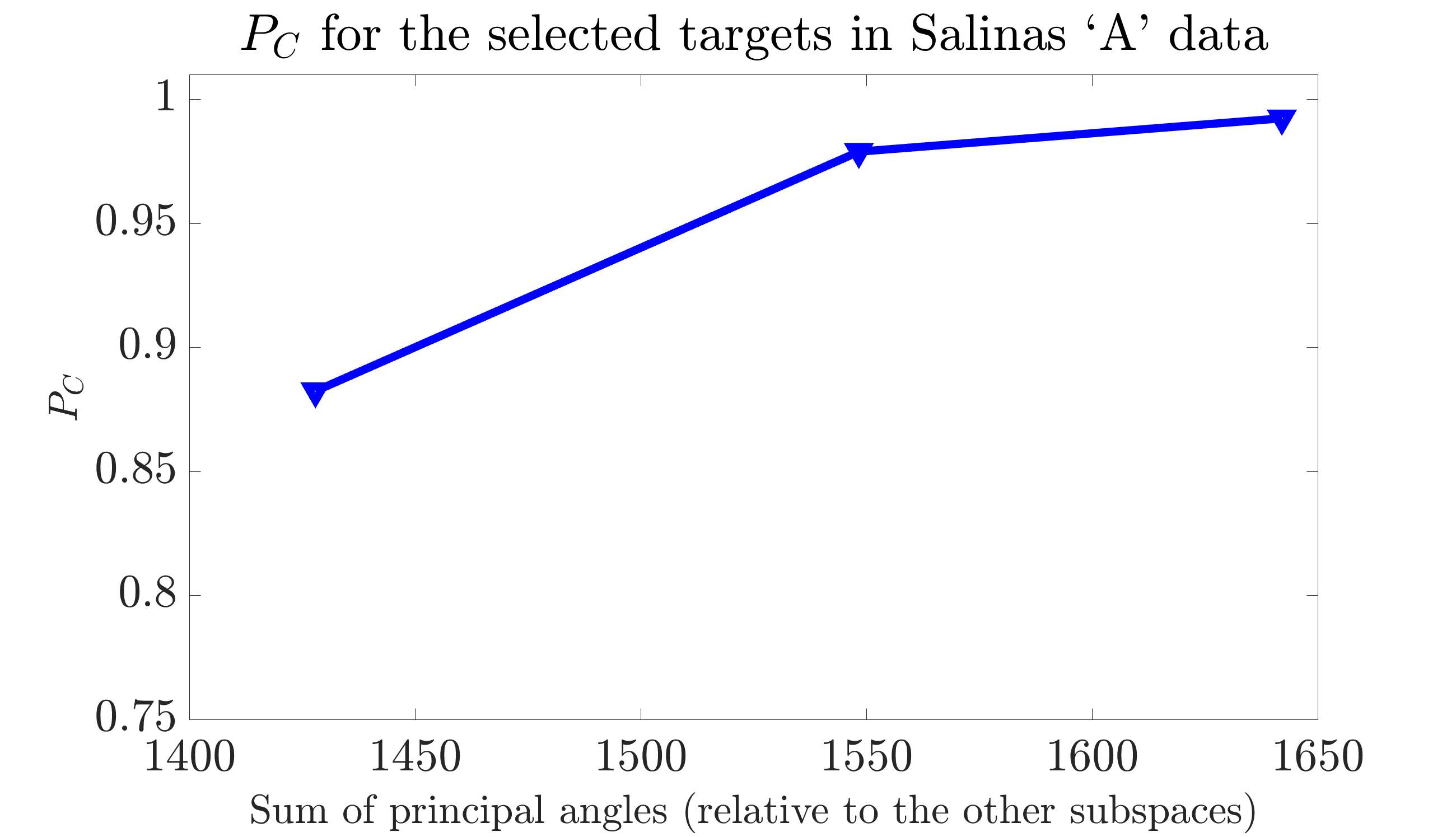}}\\
			{\includegraphics[width=\columnwidth] {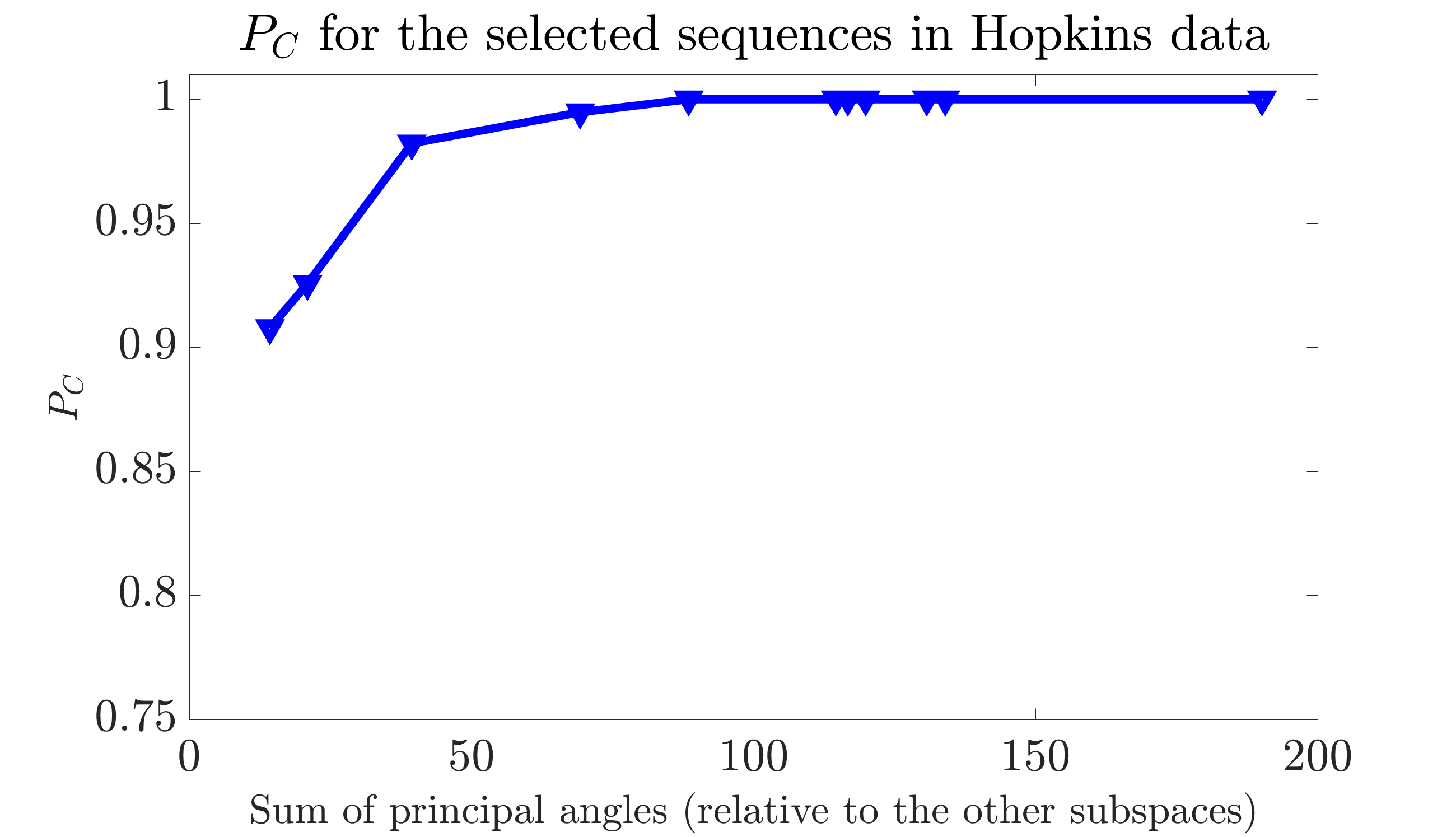}}
		\end{tabular}
	\end{center}
	\caption{\label{fig:num_sims:geometry}
		\revise{This figure shows the effects of geometry between subspaces for the Salinas `A' hyperspectral and the Hopkins motion datasets. Three targets from Salinas `A' data and 11 sequences from Hopkins motion data are selected such that they have increasing minimum and increasing cumulative principal angles with respect to the subspaces of other selected targets/sequences. One can see from the plots that target/subspaces (indicated with markers) having larger (cumulative) principal angles result in higher probabilities of correct classification (and vice versa).}}
\end{figure}

\revise{Next, recall that one of the main theses of this paper is that the geometry of subspaces underlying a union impact the performance of active subspace detection. We now validate this claim on real-world data using the Salinas `A' hyperspectral and the Hopkins motion datasets. In the case of Salinas `A' data, we select three targets whose underlying subspaces, when compared to other targets in the data, have increasing minimum principal angle and an increasing sum of principal angles (relative to the other subspaces). In the case of the Hopkins motion dataset, we select 11 sequences from the data in a similar fashion. We then carry out active subspace detection using the GLRTs derived in this paper and report the results in Fig.~\ref{fig:num_sims:geometry} for the selected targets and sequences under the same SNR and detection thresholds. It can be seen from the figure that, even though the detection of the selected targets/sequences is carried out under identical conditions, the probability of correct classification of different targets/sequences varies as a function of the geometry of subspaces in the union. In particular, targets/sequences whose cumulative principal angles (relative to the subspaces of other targets/sequences) are larger result in higher probabilities of correct classification and vice versa. These results, coupled with the ones reported for synthetic data, confirm that geometry of subspaces play an integral role in the problem of active subspace detection under the UoS model.}

\subsection{Discussion}
The experiments performed in Sec.~\ref{num_sims:monte_carlo} suggest that even though the bounds we obtain for probabilities of detection and correct classification are loose, they still predict the effect of subspace geometry on these probabilities correctly. In particular, we correctly predict that as the angles between whitened subspaces increase, the probabilities of detection and correct classification get higher, and vice-versa.

The results obtained in Sec.~\ref{num_sims:real} for real-world datasets are not as good as some state-of-the-art algorithms (e.g., see \cite{tron2007benchmark}). However, there are certain advantages that our approach enjoys over the state-of-the-art methods. The first advantage is that our detection and classification methods allow control over the false alarm rate, which is not an option for other methods. Secondly, our method can work with just enough data, i.e., we just need enough samples to get good estimates of subspace bases and noise statistics. The third advantage is that our results explicitly cater to different levels of knowledge about the noise statistics and include that information in the detection and classification processes. 

\section{Conclusion} \label{summary}
We introduced GLRTs for signal and active subspace detection under the UoS model. We analyzed the performance of the derived test statistics under various levels of knowledge about noise and explained the effect of colored noise geometry and geometry between subspaces on the detection and classification capabilities of these statistics. This was achieved by obtaining bounds on detection and classification probabilities in terms of the angles between subspaces and the angles that subspaces make with the noise eigenvectors. We also validated the insights of our analysis through Monte-Carlo experiments and experiments with real-world datasets.

\begin{appendices}
\section{Proof of Theorem \ref{th:MSD:KN:test}} \label{th:MSD:KN:test:proof}
In the case of the signal detection problem, the likelihoods under the two hypotheses are given by:
\begin{align} \label{MSD:KN:likelihoods}
l_0(\mathbf{y}) &\propto \exp \Big( - \frac{ \mathbf{y}^T \mathbf{R}^{-1} \mathbf{y} }{2\sigma^2} \Big) \nonumber,  \ \text{and} \\
l_1(\mathbf{y}) &\propto \exp \Big( - \frac{ ({\mathbf{y}-{\mathbf{x}}})^T \mathbf{R}^{-1} ({\mathbf{y}-{\mathbf{x}}}) }{2\sigma^2} \Big).
\end{align}
Since $\mathbf{x}$ is unknown in~\eqref{MSD:KN:likelihoods}, we replace it with its \emph{maximum likelihood} (ML) estimate $\mathbf{\widehat{\mathbf{x}}}$, which is given by $\argmin_k (\mathbf{y}-\mathbf{H}_k \boldsymbol{\theta})^T {\mathbf{R}}^{-1}(\mathbf{y}-\mathbf{H}_k \boldsymbol{\theta})$,
where $\mathbf{P}_{{S}_k} = \mathbf{H}_{{k}}(\mathbf{H}_{{k}}^T {\mathbf{R}}^{-1} \mathbf{H}_{{k}})^{-1} \mathbf{H}_{{k}}^T {\mathbf{R}}^{-1}$~\cite{scharf1994matched}. 
Consequently, the GLRT for this problem leads to the decision rule
\begin{align}
\frac{l_{1}(\mathbf{y})}{l_0(\mathbf{y})} &\underset{\mathcal{H}_{0}}{\overset{\mathcal{H}_{1}}{\gtrless}} \gamma \ \Leftrightarrow \
\frac{\mathbf{y}^T {\mathbf{R}}^{-1} \mathbf{P}_{{S}_{\widehat{k}}} \mathbf{y}}{2 \sigma^2} \underset{\mathcal{H}_{0}}{\overset{\mathcal{H}_{1}}{\gtrless}} \bar{\gamma},
\end{align}
where $\widehat{k} = \argmax_k (\mathbf{y}^T {\mathbf{R}}^{-1} \mathbf{P}_{{S}_k} \mathbf{y})$, and $\bar{\gamma} = \log \gamma$ is the threshold used to control the probability of false alarm. Now, with appropriate substitutions, we can rewrite the final decision rule as:
$T_{\mathbf{z}}^{2 \sigma^2}\left(\mathbf{P}_{\bar{S}_{\widehat{k}}}\right) \underset{\mathcal{H}_{0}}{\overset{\mathcal{H}_{1}}{\gtrless}} \bar{\gamma}$ with $\widehat{k} = \argmax_k \; \mathbf{z}^T \mathbf{P}_{\bar{S}_k} \mathbf{z}$.

Similarly, the likelihoods under different hypotheses for the active subspace detection problem are given by:
\begin{align} \label{MSuD:KN:likelihoods}
l_0(\mathbf{y}) &\propto \exp \Big( - \frac{ \mathbf{y}^T \mathbf{R}^{-1} \mathbf{y} }{2\sigma^2} \Big) , \ \text{and} \nonumber \\
l_k(\mathbf{y}) &\propto \exp \Big( - \frac{ ({\mathbf{y} - \mathbf{H}_k \boldsymbol{\theta}_k })^T \mathbf{R}^{-1} ({\mathbf{y} - \mathbf{H}_k \boldsymbol{\theta}_k }) }{2\sigma^2} \Big),
\end{align}
where $k=1,\dots,K_0$. Replacing the unknown $\boldsymbol{\theta}_k$'s in \eqref{MSuD:KN:likelihoods} with their ML estimates 
$\widehat{\boldsymbol{\theta}}_k = (\mathbf{H}_k^T {\mathbf{R}}^{-1} \mathbf{H}_k)^{-1} \mathbf{H}_k^T {\mathbf{R}}^{-1} \mathbf{y}$~\cite{scharf1994matched} and comparing the generalized likelihoods lead to the rule
\begin{align} \label{eq:MSuD}
\frac{l_{\widehat{k}}(\mathbf{y})}{l_0(\mathbf{y})} \underset{\mathcal{H}_{0}}{\overset{\mathcal{H}_{\widehat{k}}}{\gtrless}} \gamma  \ \Leftrightarrow \
\frac{\mathbf{y}^T \mathbf{R}^{-1} \mathbf{P}_{S_{\widehat{k}}} \mathbf{y}}{2 \sigma^2} \underset{\mathcal{H}_{0}}{\overset{\mathcal{H}_{\widehat{k}}}{\gtrless}} \bar{\gamma}.
\end{align}
Making the same substitutions as before, the final decision rule becomes: $T_{\mathbf{z}}^{2 \sigma^2}\left(\mathbf{P}_{\bar{S}_{\widehat{k}}}\right) \underset{\mathcal{H}_{0}}{\overset{\mathcal{H}_{\widehat{k}}}{\gtrless}} \bar{\gamma}$.
\qed

\section{Proof of Theorem \ref{th:MSD:KN:prob}} \label{th:MSD:KN:prob:proof}
The probability of false alarm in the case of signal detection is given by:
\begin{align} \label{eq:MSD:P_FA}
P_{FA} &= {P}_{\mathcal{H}_0} \Big( \widehat{\mathcal{H}}_1 \Big)
	   = {P}_{\mathcal{H}_0} \Big( T_{\mathbf{z}}^{2 \sigma^2}(\mathbf{P}_{\bar{S}_{\widehat{k}}}) > \bar{\gamma} \Big) \nonumber \\
	   &\overset{(a)}= {\Pr}\Big( T_{\mathbf{w}}^{2 \sigma^2}(\mathbf{P}_{\bar{S}_{\widehat{k}}}) > \bar{\gamma} \Big) \nonumber
	   = {\Pr}\Big( \underset{k = 1}{\overset{K_0} \bigcup} \Big\{ T_{\mathbf{w}}^{2 \sigma^2}(\mathbf{P}_{\bar{S}_k}) > \bar{\gamma} \Big\} \Big) \nonumber \\
	   &= \underset{k = 1}{\overset{K_0} \sum} {\Pr}\Big( T_{\mathbf{w}}^{2 \sigma^2}(\mathbf{P}_{\bar{S}_k}) > {\bar{\gamma}} \Big) - \nonumber \\
	   &\underset{k < j}{\overset{K_0} \sum} {\Pr}\Bigg( \Big\{ T_{\mathbf{w}}^{2 \sigma^2}(\mathbf{P}_{\bar{S}_k}) > {\bar{\gamma}} \Big\} \bigcap \Big\{ T_{\mathbf{w}}^{2 \sigma^2}(\mathbf{P}_{\bar{S}_j}) > {\bar{\gamma}} \Big\} \Bigg) + \nonumber \\
	   & + {\ldots} + (-1)^{K_0 - 1} {\Pr}\Big( \underset{k = 1}{\overset{K_0} \bigcap} \Big\{ T_{\mathbf{w}}^{2 \sigma^2}(\mathbf{P}_{\bar{S}_k}) > {\bar{\gamma}} \Big\} \Big),
\end{align}
where (a) follows because $\mathbf{y} | \mathcal{H}_{0} = \mathbf{n}$. We cannot evaluate \eqref{eq:MSD:P_FA} explicitly since it contains tail probabilities of $k$-tuples $\underset{j = 1}{\overset{k} \bigcap} \Big\{ \frac{ \mathbf{w}^T  \mathbf{P}_{\bar{S}_j}  {\mathbf{w}} }{2 \sigma^2} > \bar{\gamma} \Big\}$, $k = 1,\dots,K_0$. In particular, notice that $\mathbf{w}^T  \mathbf{P}_{\bar{S}_j}  {\mathbf{w}}$ is a quadratic form of the variable $\mathbf{P}_{\bar{S}_j}  {\mathbf{w}}$ 
and has a centered chi-squared distribution. This means that the distribution of the $k$-tuple is the joint distribution of $k$ dependent chi-squared variables. These distributions exist in the literature for either independent quadratic forms or dependent quadratic forms under particular settings \cite{khatri1977note,jensen1970joint,jensen1994approximations, al2016distribution}. However, the quadratic forms in \eqref{eq:MSD:P_FA} are neither independent nor fall under these settings. 
We instead resort to upper bounding \eqref{eq:MSD:P_FA} by the union bound, i.e.,
\begin{align}
P_{FA} &= {\Pr}\Big( \underset{k = 1}{\overset{K_0} \bigcup}  \Big\{ T_{\mathbf{w}}^{2 \sigma^2}(\mathbf{P}_{\bar{S}_k}) > \bar{\gamma} \Big\} \Big) \nonumber \\
	   &\leq \min \Big\{ 1 \;,\; \underset{k = 1}{\overset{K_0} \sum} {\Pr} \Big( T_{\mathbf{w}}^{2 \sigma^2}(\mathbf{P}_{\bar{S}_k}) > \bar{\gamma} \Big) \Big\}.
\end{align}
Finally since, the null hypotheses for both signal and active subspace detection problems are the same, they end up having the same probability of false alarm.

Next, for the {probability of detection} $P_{D}$, note that 
\begin{align} \label{eq:MSD:P_D_k}
P_{S_{k}} \Big( \widehat{\mathcal{H}}_1 \Big)	
											&= {P}_{S_{k}} \Big( \underset{i = 1}{\overset{K_0} \bigcup} \Big\{ T_{\mathbf{z}}^{2 \sigma^2}(\mathbf{P}_{\bar{S}_i}) > \bar{\gamma} \Big\} \Big) \nonumber \\
											&\overset{(b)}= \underset{i = 1}{\overset{K_0} \sum} {P}_{S_{k}} \Big( T_{\mathbf{z}}^{2 \sigma^2}(\mathbf{P}_{\bar{S}_i}) > \bar{\gamma} \Big) \nonumber \\
											&- \underset{i < j}{\overset{K_0} \sum} {P}_{S_{k}} \Big( \Big\{ T_{\mathbf{z}}^{2 \sigma^2}(\mathbf{P}_{\bar{S}_i}) > \bar{\gamma} \Big\} , \Big\{ T_{\mathbf{z}}^{2 \sigma^2}(\mathbf{P}_{\bar{S}_j}) > \bar{\gamma} \Big\} \Big) \nonumber \\
	   										&- \dots + (-1)^{K_0 - 1} {P}_{S_{k}} \Big( \underset{i = 1}{\overset{K_0} \bigcap} \Big\{ T_{\mathbf{z}}^{2 \sigma^2}(\mathbf{P}_{\bar{S}_i}) > \bar{\gamma} \Big\} \Big) \nonumber \\
	   										&\overset{(c)}\leq \min \Big\{ 1 \;,\; {\underset{i = 1}{\overset{K_0} \sum}} {P}_{S_{k}} \Big( T_{\mathbf{z}}^{2 \sigma^2}(\mathbf{P}_{\bar{S}_i}) > \bar{\gamma} \Big) \Big\},
\end{align}
%
where $(c)$ is again obtained using the union bound since the $k$-tuples in $(b)$ cannot be expressed in closed form. Further, the lower bound in \eqref{eq:MSD:KN:prob:detect} follows from \cite[Theorem~1]{de1997lower}.
%

Finally for the probability of classification $P_{C}$, we have:
\begin{align} \label{eq:MSuD:P_C_}
&P_{\mathcal{H}_k}(\widehat{\mathcal{H}}_k) 
 = {P}_{S_k} \big( \{ T_{\mathbf{z}}^{2 \sigma^2}(\mathbf{P}_{\bar{S}_k}) > \bar{\gamma} \} , \nonumber \\
& \quad \quad \quad \quad {{\underset{j = 1, j \neq k}{\overset{K_0} \bigcap}}} \{ T_{\mathbf{z}}( \mathbf{P}_{\bar{S}_k} , \mathbf{P}_{\bar{S}_j} ) > 1 \} \big).
\end{align}
Since \eqref{eq:MSuD:P_C_} cannot be evaluated explicitly as it involves \emph{dependent definite and indefinite quadratic forms}, 
we lower bound it by using the  Fr{\'e}chet inequalities \cite{frechet1935generalisation}:
\begin{align} \label{eq:MSuD:P_C}
&P_{\mathcal{H}_k}(\widehat{\mathcal{H}}_k) 
\geq  \max \Big\{ 0 \;,\; {P}_{S_k} (  T_{\mathbf{z}}^{2 \sigma^2}(\mathbf{P}_{\bar{S}_k}) > \bar{\gamma} ) + \nonumber \\
&\quad {{\underset{j = 1,j \neq k}{\overset{K_0} \sum}}} {P}_{S_k} ( T_{\mathbf{z}}( \mathbf{P}_{\bar{S}_k} , \mathbf{P}_{\bar{S}_j} ) > 1) - (K_0 - 1) \Big\}.
\end{align}

We conclude by noting that one could use \cite[Lemma~1]{wimalajeewa2015subspace} to further lower bound \eqref{eq:MSuD:P_C}. Specifically, 
\begin{align} \label{eq:MSuD:P_C:quadratic_form}
{P}_{S_k} &( T_{\mathbf{z}}( \mathbf{P}_{\bar{S}_k} , \mathbf{P}_{\bar{S}_j} ) > 1 ) = {P}_{S_k} ( \mathbf{z}^T \mathbf{P}_{\bar{S}_j}^{\bot} \mathbf{z} - \mathbf{z}^T \mathbf{P}_{\bar{S}_k}^{\bot} \mathbf{z} > 0) \nonumber \\
&= 1 - {P}_{S_k} ( \mathbf{z}^T \mathbf{P}_{\bar{S}_j}^{\bot} \mathbf{z} - \mathbf{z}^T \mathbf{P}_{\bar{S}_k}^{\bot} \mathbf{z} < 0) \nonumber \\
&\geq 1 - Q \big( \frac{1}{2} (1-2\eta_{0}) \sqrt{\lambda_{j \backslash k}}\big) - \Psi(n,\lambda_{j \backslash k}),
\end{align}
where $\lambda_{j \backslash k} = \frac{1}{\sigma^2} \mathbf{z}^T \mathbf{P}_{\bar{S}_j}^{\bot} \mathbf{z}$ when $\mathbf{z} \in \bar{S}_k$. This leads to $P_{\mathcal{H}_k}(\widehat{\mathcal{H}}_k) \geq \max \big\{0 , {P}_{S_k}(T_{\mathbf{z}}^{2 \sigma^2}(\mathbf{P}_{\bar{S}_k}) > \bar{\gamma}) - \sum\limits_{j:j\not=k} Q \big( \frac{1}{2} (1-2\eta_{0}) \sqrt{\lambda_{j \backslash k}}\big) - \sum\limits_{j:j\not=k} \Psi(\eta_{0},\lambda_{j \backslash k}) \big\}$.
$\hfill \blacksquare$

\section{Proof of Theorem \ref{th:ASD:UC:test}} \label{th:ASD:UC:test:proof}
The results derived in this appendix closely follow the derivations in \cite{kelly1986adaptive}. The likelihood of $\boldsymbol{\xi}_p$ is given by: 
\begin{align} \label{training:likelihoods}
&l(\boldsymbol{\xi}_p) = \frac{1}{\sqrt{(2 \pi)^m |\mathbf{R}|}} \exp \Big \{ \frac{-1}{2} {\boldsymbol{\xi}_p}^T \mathbf{R}^{-1} {\boldsymbol{\xi}_p} \Big \},
\end{align}
which is used to get the joint likelihoods under each hypothesis $\mathcal{H}_{1}$ and $\mathcal{H}_{0}$: $l_0(\mathbf{y},\boldsymbol{\Xi})$ and $l_1(\mathbf{y},\boldsymbol{\Xi})$, where $\boldsymbol{\Xi} = [\boldsymbol{\xi}_1, \boldsymbol{\xi}_2, \cdots, \boldsymbol{\xi}_{N_0}]$.
From these joint likelihoods, the ML estimate of $\mathbf{R}$ under $\mathcal{H}_{1}$ and $\mathcal{H}_{0}$ can be computed as $\widehat{\mathbf{R}}_1 = \frac{N_0}{N_0+1} \boldsymbol{\Sigma} + \frac{(\mathbf{y}-\mathbf{x}) (\mathbf{y}-\mathbf{x})^T }{\sigma^2(N_0+1)}$ and $\widehat{\mathbf{R}}_0 = \widehat{\mathbf{R}}_1|_{\mathbf{x} = 0}$, respectively.

Now, following the same steps as in the proof of Theorem \ref{th:MSD:KN:test}, we can proceed to calculate the final decision rule for signal detection as
$\overline{T}_{\widehat{\mathbf{z}}}^{N_0 \sigma^2}(\widehat{\mathbf{P}}_{\bar{S}_{\widehat{k}}}) \underset{\mathcal{H}_{0}}{\overset{\mathcal{H}_{1}}{\gtrless}} \bar{\gamma}$,
where $\widehat{k} = \argmax_k (\widehat{\mathbf{z}}^T \widehat{\mathbf{P}}_{\bar{S}_k} \widehat{\mathbf{z}})$ and $\bar{\gamma} = \log \gamma$.

Next, note that the likelihood in \eqref{training:likelihoods} combined with the likelihoods in \eqref{MSuD:KN:likelihoods} also provide the joint likelihoods under each hypothesis for the active subspace detection problem.
With trivial algebraic manipulations, the ML estimates of $\mathbf{R}$ in this case can be expressed as:
\begin{align}
&\mathcal{H}_{0}: \widehat{\mathbf{R}}_0 = \frac{N_0}{N_0+1} \boldsymbol{\Sigma} + \frac{\mathbf{y} \mathbf{y}^T }{\sigma^2(N_0+1)}, \text{ and} \nonumber \\
&\mathcal{H}_{k}: \widehat{\mathbf{R}}_k = \frac{N_0}{N_0+1} \boldsymbol{\Sigma} + \frac{(\mathbf{y}-\mathbf{x}) (\mathbf{y}-\mathbf{x})^T }{\sigma^2(N_0+1)},
\end{align}
where $\mathbf{x} | \mathcal{H}_{k} = \mathbf{H}_{k} \boldsymbol{\theta}_{k}$.
%
Using the ML estimates of $\mathbf{R}$ and the joint likelihoods, we can calculate the decision rule (similar to the proof of Theorem \ref{th:MSD:KN:test}) as
$\overline{T}_{\widehat{\mathbf{z}}}^{N_0 \sigma^2}(\widehat{\mathbf{P}}_{\bar{S}_{\widehat{k}}}) \underset{\mathcal{H}_{0}}{\overset{\mathcal{H}_{\widehat{k}}}{\gtrless}} \bar{\gamma}$.
\qed

\section{Proof of Theorem \ref{th:ASD:UN:test}} \label{th:ASD:UN:test:proof}
This proof uses derivations from the proof of Theorem~\ref{th:ASD:UC:test}. The only additional estimate we need is for the variance $\sigma^{2}$ which can be found from the joint likelihoods with the estimate $\widehat{\mathbf{R}}$ substituted in them.
This results in:
\begin{align}
\widehat{\sigma^2} | \mathcal{H}_{1} &= \frac{N_0 - m + 1}{N_0 m} {(\mathbf{y}-\mathbf{x})^T {\boldsymbol{\Sigma}}^{-1}(\mathbf{y}-\mathbf{x})}, \text{and} \nonumber \\
\widehat{\sigma^2} | \mathcal{H}_{0} &= \frac{N_0 - m + 1}{N_0 m} {\mathbf{y}^T {\boldsymbol{\Sigma}}^{-1}\mathbf{y}}.
\end{align}
%
The ML estimate of $\mathbf{x}$ in this case is the same as in the proof of Theorem~\ref{th:MSD:KN:test}. Putting these estimates together, we arrive at the final decision rule 
$T_{\mathbf{z}}(\mathbf{P}_{\bar{S}_{\widehat{k}}}) \underset{\mathcal{H}_{0}}{\overset{\mathcal{H}_{1}}{\gtrless}} \bar{\gamma}$,
where $\widehat{k} = \argmax_k (\widehat{\mathbf{z}}^T \widehat{\mathbf{P}}_{\bar{S}_k} \widehat{\mathbf{z}})$ and $\bar{\gamma} = \log \gamma$.

Similarly, the active subspace detection problem takes the same from as in Theorem~\ref{th:ASD:UC:test} with an additional unknown variable $\sigma^2$. However, we can use the previously calculated ML estimates of $\sigma^2$, $\mathbf{R}$, and $\mathbf{x}$ to arrive at the final decision rule of
$T_{\widehat{\mathbf{z}}}(\widehat{\mathbf{P}}_{\bar{S}_{\widehat{k}}}) \underset{\mathcal{H}_{0}}{\overset{\mathcal{H}_{\widehat{k}}}{\gtrless}} \bar{\gamma}.$
\qed

\section{Proof of Theorem \ref{th:MSuD:KN:angles}} \label{th:MSuD:KN:angles:proof}
To get a better understanding of the parameters that effect the probability of correct classification, we analyze the terms ${P}_{S_k} ( T_{\mathbf{z}}( \mathbf{P}_{\bar{S}_k} , \mathbf{P}_{\bar{S}_j} ) > 1 )$ in \eqref{eq:MSuD:KN:prob} since these terms characterize the interactions between the whitened subspaces.
Assuming $\mathbf{x} \in S_k$, notice that:
\begin{align} \label{eq:MSuD:influence_eq_1}
&T_{\mathbf{z}}( \mathbf{P}_{\bar{S}_k} , \mathbf{P}_{\bar{S}_j} ) > 1 \Leftrightarrow \mathbf{z}^T \mathbf{P}_{\bar{S}_k} \mathbf{z} > \mathbf{z}^T \mathbf{P}_{\bar{S}_j} \mathbf{z} \nonumber \\
&\Leftrightarrow {(\bar{\mathbf{x}}+\mathbf{w})^T \mathbf{P}_{\bar{S}_k} (\bar{\mathbf{x}}+\mathbf{w})} > {(\bar{\mathbf{x}}+\mathbf{w})^T \mathbf{P}_{\bar{S}_j} (\bar{\mathbf{x}}+\mathbf{w})} \nonumber \\
&\overset{(a)}{\Leftrightarrow} \mathbf{w}^T \mathbf{P}_{\bar{S}_k}\mathbf{w} -\mathbf{w}^T \mathbf{P}_{\bar{S}_j}\mathbf{w} > \nonumber \\ 
&\quad \quad - \bar{\mathbf{x}}^T \bar{\mathbf{x}} + \bar{\mathbf{x}}^T \mathbf{P}_{\bar{S}_j} \bar{\mathbf{x}} - 2\mathbf{w}^T \bar{\mathbf{x}} + 2\mathbf{w}^T \mathbf{P}_{\bar{S}_j} \bar{\mathbf{x}},
\end{align}
where $\bar{\mathbf{x}} = \mathbf{R}^{-\frac{1}{2}} \mathbf{x}$ is the whitened signal.
We now focus on the quadratic forms $\bar{\mathbf{x}}^T \mathbf{P}_{\bar{S}_j} \bar{\mathbf{x}}$ and $\mathbf{w}^T \mathbf{P}_{\bar{S}_j} \bar{\mathbf{x}}$ in \eqref{eq:MSuD:influence_eq_1} because these are the terms where different subspaces interact with each other and that can be expressed in terms of the principal angles between whitened subspaces. Using the derivation provided in Appendix \ref{lower_bound_quadratic}, we can bound ${P}_{S_k} ( T_{\mathbf{z}}( \mathbf{P}_{\bar{S}_k} , \mathbf{P}_{\bar{S}_j} ) > 1 )$ as:
\begin{align} \label{eq:MSuD:influence:all:final}
&{P}_{S_k} (\mathbf{w}^T \mathbf{P}_{\bar{S}_k}\mathbf{w} -\mathbf{w}^T \mathbf{P}_{\bar{S}_j}\mathbf{w} > - \bar{\mathbf{x}}^T \bar{\mathbf{x}} + \bar{\mathbf{x}}^T \mathbf{P}_{\bar{S}_j} \bar{\mathbf{x}} \nonumber \\ 
& \quad \quad \quad \quad \quad \quad \quad \quad \quad \quad \quad\quad \quad - 2\mathbf{w}^T \bar{\mathbf{x}} + 2\mathbf{w}^T \mathbf{P}_{\bar{S}_j} \bar{\mathbf{x}}) \nonumber \\ 
											  &\geq  {P}_{S_k} \Big( \|\mathbf{n}\|_2^2 ( \cos^2 \psi_{k} - \cos^2 \psi_{j} ) > \nonumber \\
											  &- \underset{i = 1}{\overset{n}{\sum}} \theta_{ki}^2 \sin^2 \varphi_{i}^{(k,j)} + 2 \underset{i < p}{\overset{n}{\sum}} |\theta_{ki} \theta_{kp}| \cos \varphi_{i}^{(k,j)} \cos \varphi_{p}^{(k,j)} \nonumber \\
															&+ \|\mathbf{n} \|_2 \cos \psi_j \Big( \underset{i = 1}{\overset{n}{\sum}} \theta_{ki}^2 \cos^2 \varphi_{i}^{(k,j)} \Big)^{\frac{1}{2}} - \|\mathbf{n} \|_2 \cos \psi_k \Big( \underset{i = 1}{\overset{n}{\sum}} \theta_{ki}^2 \Big)^{\frac{1}{2}} \nonumber \\
																		&+ \|\mathbf{n} \|_2 \cos \psi_j \Big( 2 \underset{i < p}{\overset{n}{\sum}} |\theta_{ki} \theta_{kp}| \cos \varphi_{i}^{(k,j)} \cos \varphi_{p}^{(k,j)} \Big)^{\frac{1}{2}} \Big),
\end{align}
where $\varphi_{i}^{(k,j)}$ is the angle that $\mathbf{g}_{i}^{k}$ ($i$-th basis vector of whitened subspace $\bar{S}_k$, i.e., $i$-th column of $\mathbf{G}_{k}$) makes with the whitened subspace $\bar{S}_j$, $\varphi_{ip}^{(k,j)}$ is the angle between $\mathbf{g}_{i}^{k \rightarrow j}$ and $\mathbf{g}_{p}^{k \rightarrow j}$ (i.e., the angles between the $i$-th and $p$-th basis vectors of whitened subspace $\bar{S}_k$ after projection onto the whitened subspace $\bar{S}_j$) and $\psi_j$ is the angle between $\mathbf{w}$ and the whitened subspace $\bar{S}_j$.

This lower bound on ${P}_{S_k} ( T_{\mathbf{z}}( \mathbf{P}_{\bar{S}_k} , \mathbf{P}_{\bar{S}_j} ) > 1)$ is dependent on the principal angles $\varphi_{i}^{(k,j)}$ between the whitened subspace $\bar{S}_k$ and $\bar{S}_j$.
In particular, we can see that as the principal angles $\varphi_{i}^{(k,j)}$ increase, the bound on the right hand side of the inequality $(a)$ in \eqref{eq:MSuD:influence_eq_1} becomes smaller. This implies that lower bound on the tail probability in \eqref{eq:MSuD:influence:all:final} becomes larger as the principal angles increase. This trend holds for all pairs of whitened subspaces $\bar{S}_j$ and $\bar{S}_k$ (for $j,k = 1,\dots,K_0$ and $j \neq k$). This means that the lower bound for $P_{\mathcal{H}_k}(\widehat{\mathcal{H}}_k)$ in \eqref{eq:MSuD:KN:prob} also increases with increasing principal angles between the whitened subspaces. 

We conclude by noting that this trend can also be derived from the lower bound expression in Remark \ref{remark:MSuD:KN}. The quantities $Q()$ and $\Psi()$ in that expression are functions of $\lambda_{j \backslash i}$ and decrease monotonically as $\lambda_{j \backslash i}$ is increased \cite{wimalajeewa2015subspace}. This means that an increase in $\lambda_{j \backslash i}$ will result in an increase in the probability of correct classification. Since $\lambda_{j \backslash i}$ can be expressed as $\lambda_{j \backslash i} = \frac{1}{\sigma^2} \mathbf{z}^T \mathbf{P}_{\bar{S}_j}^{\bot} \mathbf{z} = \frac{1}{\sigma^2} \big( \mathbf{z}^T \mathbf{z} - \mathbf{z}^T \mathbf{P}_{\bar{S}_j} \mathbf{z} \big) = \frac{1}{\sigma^2} \big( \mathbf{z}^T \mathbf{z} - \bar{\mathbf{x}}^T \mathbf{P}_{\bar{S}_j} \bar{\mathbf{x}} - 2\mathbf{w}^T \mathbf{P}_{\bar{S}_j} \bar{\mathbf{x}} - \mathbf{w}^T \mathbf{P}_{\bar{S}_j}\mathbf{w} \big)$, one can use results from Appendix~\ref{lower_bound_quadratic} to once again argue that as the angles between whitened subspaces increase, the lower bound on $\lambda_{j \backslash i}$ increases which in turn results in larger (lower) bound on the probability of correct classification.
\qed
\section{Probability bound on ratio of quadratic forms} \label{lower_bound_quadratic}
The outline of our procedure for deriving a lower bound on the probability of the comparison of quadratic forms is as follows: we first express $\bar{\mathbf{x}}^T \mathbf{P}_{\bar{S}_j} \bar{\mathbf{x}}$ and $\mathbf{w}^T \mathbf{P}_{\bar{S}_j} \bar{\mathbf{x}}$ in terms of the principal angles between whitened subspaces. We then obtain upper bounds on these quadratic forms that depend on the principal angles. Next we put these upper bounds in the expression for the probability of correct classification of the individual subspaces and finally we derive a lower bound on the probability of correct classification that is dependent on the principal angles between the whitened subspaces.

Let's consider $\bar{\mathbf{x}}^T \mathbf{P}_{\bar{S}_j} \bar{\mathbf{x}}$ when $\mathbf{x} \in S_k$:
\begin{align} \label{eq:MSuD:influence:all:1}
&\bar{\mathbf{x}}^T \mathbf{P}_{\bar{S}_j} \bar{\mathbf{x}}	= \| \mathbf{P}_{\bar{S}_j} \bar{\mathbf{x}} \|_2^2 
				\overset{(a)}{=} \| \mathbf{P}_{\bar{S}_j} {\mathbf{G}}_{k} {\boldsymbol{\theta}}_{k} \|_2^2 \nonumber \\
				&\overset{(b)}{=} \| \theta_{k1} \mathbf{g}_{1}^{k \rightarrow j} + \dots + \theta_{kn} \mathbf{g}_{n}^{k \rightarrow j} \|_2^2 \nonumber \\
				&\overset{(c)}{=} \underset{i = 1}{\overset{n}{\sum}} \| \theta_{ki} \mathbf{g}_{i}^{k \rightarrow j} \|_2^2 + 2 \underset{i < p}{\overset{n}{\sum}}   \langle \theta_{ki} \mathbf{g}_{i}^{k \rightarrow j} , \theta_{kp} \mathbf{g}_{p}^{k \rightarrow j} \rangle, \nonumber \\
				&= \underset{i = 1}{\overset{n}{\sum}} \theta_{ki}^2 \| \mathbf{g}_{i}^{k} \|_2^2 \cos^2 \varphi_{i}^{(k,j)} + \nonumber \\
				&2 \underset{i < p}{\overset{n}{\sum}} |\theta_{ki}| \|\mathbf{g}_{i}^{k} \|_ 2 \cos \varphi_{i}^{(k,j)} |\theta_{kp}| \|\mathbf{g}_{p}^{k} \|_2 \cos \varphi_{p}^{(k,j)} \cos \varphi_{ip}^{(k,j)}
\end{align}
where $\varphi_{i}^{(k,j)}$ are as defined in Appendix~\ref{th:MSuD:KN:angles:proof}. Note that (a) in \eqref{eq:MSuD:influence:all:1} follows from $\bar{\mathbf{x}} = {\mathbf{G}}_{k} {\boldsymbol{\theta}}_{k}$, (b) uses the notation $\mathbf{g}_{i}^{k \rightarrow j} = \mathbf{P}_{\bar{S}_j} \mathbf{g}_{i}$ and (c) uses the identity $\|\mathbf{a} + \mathbf{b}\|_2^2 = \|\mathbf{a}\|_2^2 + \|\mathbf{b}\|_2^2 + 2 \langle \mathbf{a}, \mathbf{b} \rangle$.

Now, if we assume $\mathbf{g}_{i}^{k}$'s to be the unit-norm principal vectors of $\bar{S}_k$, we can bound \eqref{eq:MSuD:influence:all:1} as $\bar{\mathbf{x}}^T \mathbf{P}_{\bar{S}_j} \bar{\mathbf{x}} \leq \underset{i = 1}{\overset{n}{\sum}} \theta_{ki}^2 \cos^2 \varphi_{i}^{(k,j)} + 2 \underset{i < p}{\overset{n}{\sum}} |\theta_{ki}| \cos \varphi_{i}^{(k,j)} \: |\theta_{kp}| \cos \varphi_{p}^{(k,j)}.$
Similarly we have $\mathbf{w}^T \mathbf{P}_{\bar{S}_j} \bar{\mathbf{x}} \leq \|\mathbf{n} \|_2 \cos \psi_j \Big( \underset{i = 1}{\overset{n}{\sum}} \theta_{ki}^2 \cos^2 \varphi_{i}^{(k,j)} \Big)^{\frac{1}{2}} + \|\mathbf{n} \|_2 \cos \psi_j \Big( 2 \underset{i < p}{\overset{n}{\sum}} |\theta_{ki}| \cos \varphi_{i}^{(k,j)} \: |\theta_{kp}| \cos \varphi_{p}^{(k,j)} \Big)^{\frac{1}{2}}$
%
, where we have used the fact that $\sqrt{a + b} \leq \sqrt{a} + \sqrt{b}$ and $\psi_j$ is the angle between $\mathbf{w}$ and the whitened subspace $\bar{S}_j$.
Substituting these upper bounds in \eqref{eq:MSuD:influence_eq_1} we get:
\begin{align} \label{eq:MSuD:influence:all:inequality_combined}
\|&\mathbf{n}\|_2^2 ( \cos^2 \psi_{k} - \cos^2 \psi_{j} ) 	> - \underset{i = 1}{\overset{n}{\sum}} \theta_{ki}^2 \sin^2 \varphi_{i}^{(k,j)} + \nonumber \\
&2 \underset{i < p}{\overset{n}{\sum}} |\theta_{ki} \theta_{kp}| \cos \varphi_{i}^{(k,j)} \cos \varphi_{p}^{(k,j)} + \nonumber \\
															&\|\mathbf{n} \|_2 \cos \psi_j \Big( \underset{i = 1}{\overset{n}{\sum}} \theta_{ki}^2 \cos^2 \varphi_{i}^{(k,j)} \Big)^{\frac{1}{2}} - \|\mathbf{n} \|_2 \cos \psi_k \Big( \underset{i = 1}{\overset{n}{\sum}} \theta_{ki}^2 \Big)^{\frac{1}{2}} \nonumber \\
															&+ \|\mathbf{n} \|_2 \cos \psi_j \Big( 2 \underset{i < p}{\overset{n}{\sum}} |\theta_{ki} \theta_{kp}| \cos \varphi_{i}^{(k,j)} \cos \varphi_{p}^{(k,j)} \Big)^{\frac{1}{2}},
\end{align}
which can be used to obtain \eqref{eq:MSuD:influence:all:final}.
\qed
\end{appendices} 

\balance


\begin{thebibliography}{10}
\providecommand{\url}[1]{#1}
\csname url@samestyle\endcsname
\providecommand{\newblock}{\relax}
\providecommand{\bibinfo}[2]{#2}
\providecommand{\BIBentrySTDinterwordspacing}{\spaceskip=0pt\relax}
\providecommand{\BIBentryALTinterwordstretchfactor}{4}
\providecommand{\BIBentryALTinterwordspacing}{\spaceskip=\fontdimen2\font plus
\BIBentryALTinterwordstretchfactor\fontdimen3\font minus
  \fontdimen4\font\relax}
\providecommand{\BIBforeignlanguage}[2]{{%
\expandafter\ifx\csname l@#1\endcsname\relax
\typeout{** WARNING: IEEEtran.bst: No hyphenation pattern has been}%
\typeout{** loaded for the language `#1'. Using the pattern for}%
\typeout{** the default language instead.}%
\else
\language=\csname l@#1\endcsname
\fi
#2}}
\providecommand{\BIBdecl}{\relax}
\BIBdecl

\bibitem{LodhiBajwa.ConfSSP18}
M.~A. Lodhi and W.~U. Bajwa, ``Union of subspaces signal detection in subspace
  interference,'' in \emph{Proc. IEEE Workshop Statistical Signal Processing
  (SSP'18)}, Freiburg, Germany, Jun. 2018, pp. 548--552.

\bibitem{louis1991statistical}
L.~S. Louis, ``Statistical signal processing: Detection, estimation, and time
  series analysis,'' \emph{Addision-Wesley Publishing Company}, 1991.

\bibitem{Kay.Book1998}
S.~M. Kay, \emph{Fundamentals of Statistical Signal Processing: Detection
  Theory}.\hskip 1em plus 0.5em minus 0.4em\relax Upper Saddle River, NJ:
  Prentice Hall, 1998.

\bibitem{lu2008theory}
Y.~M. Lu and M.~N. Do, ``A theory for sampling signals from a union of
  subspaces,'' \emph{IEEE Trans. Sig. Proc.}, vol.~56, no.~6, pp. 2334--2345,
  2008.

\bibitem{WuBajwa.ConfICASSP14}
T.~Wu and W.~U. Bajwa, ``Revisiting robustness of the union-of-subspaces model
  for data-adaptive learning of nonlinear signal models,'' in \emph{Proc. IEEE
  Intl. Conf. Acoustics, Speech, and Signal Processing (ICASSP'14)}, Florence,
  Italy, May 2014, pp. 3390--3394.

\bibitem{WuBajwa.ConfICASSP15}
------, ``Metric-constrained kernel union of subspaces,'' in \emph{Proc. IEEE
  Intl. Conf. Acoustics, Speech, and Signal Processing (ICASSP'15)}, Brisbane,
  Australia, Apr. 2015, pp. 5778--5782.

\bibitem{WuBajwa.ITSP15}
------, ``Learning the nonlinear geometry of high-dimensional data: {M}odels
  and algorithms,'' \emph{IEEE Trans. Signal Processing}, vol.~63, no.~23, pp.
  6229--6244, Dec. 2015.

\bibitem{ho2003clustering}
J.~Ho, M.-H. Yang, J.~Lim, K.-C. Lee, and D.~Kriegman, ``Clustering appearances
  of objects under varying illumination conditions,'' in \emph{IEEE Computer
  Vision and Pattern Recognition}, vol.~1.\hskip 1em plus 0.5em minus
  0.4em\relax IEEE, 2003, pp. 1--11.

\bibitem{hong2006multiscale}
W.~Hong, J.~Wright, K.~Huang, and Y.~Ma, ``Multiscale hybrid linear models for
  lossy image representation,'' \emph{IEEE Trans. Image Proc.}, vol.~15,
  no.~12, pp. 3655--3671, 2006.

\bibitem{yang2008unsupervised}
A.~Y. Yang, J.~Wright, Y.~Ma, and S.~S. Sastry, ``Unsupervised segmentation of
  natural images via lossy data compression,'' \emph{Computer Vision and Image
  Understanding}, vol. 110, no.~2, pp. 212--225, 2008.

\bibitem{Elhamifar.Vidal.ITPAMI2013}
E.~Elhamifar and R.~Vidal, ``Sparse subspace clustering: {A}lgorithm, theory,
  and applications,'' \emph{{IEEE} Trans. Pattern Anal. Mach. Intel.}, vol.~35,
  no.~11, pp. 2765--2781, Nov. 2013.

\bibitem{bajwa2014multiple}
W.~U. Bajwa and D.~G. Mixon, ``A multiple hypothesis testing approach to
  low-complexity subspace unmixing,'' \emph{arXiv preprint arXiv:1408.1469},
  2014.

\bibitem{gini2004radar}
F.~Gini, M.~Greco, and F.~Farina, ``Radar detection and preclassification based
  on multiple hypothesis,'' \emph{IEEE Trans. Aero. Elec. Sys.}, vol.~40,
  no.~3, pp. 1046--1059, 2004.

\bibitem{wu2015hierarchical}
T.~Wu, P.~Gurram, R.~M. Rao, and W.~U. Bajwa, ``Hierarchical union-of-subspaces
  model for human activity summarization,'' in \emph{Proc. IEEE Intl. Conf.
  Comp. Vision}, 2015, pp. 1--9.

\bibitem{davenport2010signal}
M.~A. Davenport, P.~T. Boufounos, M.~B. Wakin, and R.~G. Baraniuk, ``Signal
  processing with compressive measurements,'' \emph{IEEE J. Sel. Topics Signal
  Process}, vol.~4, no.~2, pp. 445--460, 2010.

\bibitem{yap2014false}
H.~L. Yap and R.~Pribi{\'c}, ``False alarms in multi-target radar detection
  within a sparsity framework,'' in \emph{Proc. IEEE Intl. Radar Conf.}, 2014,
  pp. 1--6.

\bibitem{joneidi2016union}
M.~Joneidi, P.~Ahmadi, M.~Sadeghi, and N.~Rahnavard, ``Union of low-rank
  subspaces detector,'' \emph{IET Signal Processing}, vol.~10, no.~1, pp.
  55--62, 2016.

\bibitem{wimalajeewa2015subspace}
T.~Wimalajeewa, Y.~C. Eldar, and P.~K. Varshney, ``Subspace recovery from
  structured union of subspaces,'' \emph{IEEE Trans. Info. Th.}, vol.~61,
  no.~4, pp. 2101--2114, 2015.

\bibitem{scharf1994matched}
L.~L. Scharf and B.~Friedlander, ``Matched subspace detectors,'' \emph{IEEE
  Trans. Sig. Proc.}, vol.~42, no.~8, pp. 2146--2157, 1994.

\bibitem{kraut2001adaptive}
S.~Kraut, L.~L. Scharf, and L.~T. McWhorter, ``Adaptive subspace detectors,''
  \emph{IEEE Trans. Sig. Proc.}, vol.~49, no.~1, pp. 1--16, 2001.

\bibitem{kraut1999cfar}
S.~Kraut and L.~L. Scharf, ``The {CFAR} adaptive subspace detector is a
  scale-invariant {GLRT},'' \emph{IEEE Trans. Sig. Proc.}, vol.~47, no.~9, pp.
  2538--2541, 1999.

\bibitem{kelly1986adaptive}
E.~J. Kelly, ``An adaptive detection algorithm,'' \emph{IEEE Trans. Aero. Elec.
  Sys.}, no.~2, pp. 115--127, 1986.

\bibitem{johnson1984extensions}
W.~B. Johnson and J.~Lindenstrauss, ``Extensions of {Lipschitz} mappings into a
  {Hilbert} space,'' \emph{Contemporary Mathematics}, vol.~26, no. 189-206,
  p.~1, 1984.

\bibitem{tibshirani1996regression}
R.~Tibshirani, ``Regression shrinkage and selection via the lasso,'' \emph{J.
  Royal Stat. Soc.}, pp. 267--288, 1996.

\bibitem{afriat1957orthogonal}
S.~N. Afriat, ``Orthogonal and oblique projectors and the characteristics of
  pairs of vector spaces,'' in \emph{Mathematical Proceedings of the Cambridge
  Philosophical Society}, 1957, pp. 800--816.

\bibitem{green1998imaging}
R.~O. Green, M.~L. Eastwood, C.~M. Sarture, T.~G. Chrien, M.~Aronsson, B.~J.
  Chippendale, J.~A. Faust, B.~E. Pavri, C.~J. Chovit, M.~Solis \emph{et~al.},
  ``Imaging spectroscopy and the airborne visible/infrared imaging spectrometer
  ({AVIRIS}),'' \emph{Remote sensing of environment}, vol.~65, no.~3, pp.
  227--248, 1998.

\bibitem{basri2003lambertian}
R.~Basri and D.~W. Jacobs, ``Lambertian reflectance and linear subspaces,''
  \emph{IEEE Trans. Patt. Anal. Mach. Intllg.}, vol.~25, no.~2, pp. 218--233,
  2003.

\bibitem{georghiades2001few}
A.~S. Georghiades, P.~N. Belhumeur, and D.~J. Kriegman, ``From few to many:
  Illumination cone models for face recognition under variable lighting and
  pose,'' \emph{IEEE Trans. Patt. Anal. Mach. Intllg.}, vol.~23, no.~6, pp.
  643--660, 2001.

\bibitem{tron2007benchmark}
R.~Tron and R.~Vidal, ``A benchmark for the comparison of 3-d motion
  segmentation algorithms,'' in \emph{IEEE Comp. Vision and Pattern
  Recognition}.\hskip 1em plus 0.5em minus 0.4em\relax IEEE, 2007, pp. 1--8.

\bibitem{khatri1977note}
C.~Khatri, P.~Krishnaiah, and P.~K. Sen, ``A note on the joint distribution of
  correlated quadratic forms,'' \emph{J. Stat. Planning Inference}, vol.~1,
  no.~3, pp. 299--307, 1977.

\bibitem{jensen1970joint}
D.~Jensen, ``The joint distribution of quadratic forms and related
  distributions1,'' \emph{Australian J. Statistics}, vol.~12, no.~1, pp.
  13--22, 1970.

\bibitem{jensen1994approximations}
D.~Jensen and H.~Solomon, ``Approximations to joint distributions of definite
  quadratic forms,'' \emph{J. Amer. Statistical Assoc.}, vol.~89, no. 426, pp.
  480--486, 1994.

\bibitem{al2016distribution}
T.~Y. Al-Naffouri, M.~Moinuddin, N.~Ajeeb, B.~Hassibi, and A.~L. Moustakas,
  ``On the distribution of indefinite quadratic forms in {Gaussian} random
  variables,'' \emph{IEEE Trans. Commun.}, vol.~64, no.~1, pp. 153--165, 2016.

\bibitem{de1997lower}
D.~De~Caen, ``A lower bound on the probability of a union,'' \emph{Discrete
  Mathematics}, vol. 169, no. 1-3, pp. 217--220, 1997.

\bibitem{frechet1935generalisation}
M.~Fr{\'e}chet, ``G{\'e}n{\'e}ralisation du th{\'e}oreme des probabilit{\'e}s
  totales,'' \emph{Fundamenta Mathematicae}, vol.~1, no.~25, pp. 379--387,
  1935.

\end{thebibliography}
\end{document}